\documentclass{jfm}
\usepackage{graphicx}
\usepackage{epstopdf, epsfig}
\usepackage{bm,amsmath}

\renewcommand{\vec}[1]{{\bm #1}}

\newcommand{\e}{{\mathrm e}}
\renewcommand{\d}{{\mathrm d}}
\renewcommand{\i}{{\mathrm i}}

\title{Correlation function and linear response function of homogeneous isotropic turbulence in the Eulerian and Lagrangian coordinates}

\shorttitle{Eulerian and Lagrangian correlations and response functions}
\shortauthor{Matsumoto,  Otsuki, Ooshida and Goto}

\author{Takeshi Matsumoto \aff{1}
  \corresp{\email{takeshi@kyoryu.scphys.kyoto-u.ac.jp}},
  Michio Otsuki\aff{2},
  Takeshi Ooshida\aff{3}
 \and Susumu Goto\aff{2}}

\affiliation{%
\aff{1} Division of Physics and Astronomy, Graduate School of Science, Kyoto University, Kyoto, 606-8502, Japan %
\aff{2} Graduate School of Engineering Science, Osaka University, Toyonaka, 560-8531, Japan %
\aff{3} Department of Mechanical and Aerospace Engineering, Tottori University, Tottori, 680-8552, Japan %
}

\begin{document}

\maketitle

\begin{abstract}
We study the correlation function and mean linear response function of the velocity Fourier mode
of statistically steady-state, homogeneous and isotropic turbulence
in the Eulerian and Lagrangian coordinates through direct numerical simulation (DNS).
As the Lagrangian velocity, we here adopt Kraichnan's Lagrangian history framework where
Lagrangian particles are labelled with current positions and their velocity are measured at 
some time before. This Lagrangian velocity is numerically calculated with a method known
as passive vector method.
Our first goal is to study relation between the correlation function and 
the mean linear response function in the Eulerian and Lagrangian coordinates.
Such a relation is known to be important in analysing the closed set of equations for the
two functions, which are obtained by direct-interaction-approximation type closures. 
We demonstrate numerically that the fluctuation-dissipation theorem (proportionality between
the two functions)
does not hold. The relation is further investigated with general analytical expressions
of the mean linear response function under stochastic settings, which are known as the
fluctuation-response relations in non-equilibrium statistical mechanics. 
Our second goal is to identify characteristic times associated with the two functions and
to compare the times between the Eulerian and Lagrangian coordinates.
Our DNS result supports the common view that
the Eulerian characteristic times have the sweeping-time scaling ($\propto k^{-1}$, where $k$ is the wavenumber)
for both functions
and the Lagrangian characteristic times in the inertial range have the Kolmogorov-time scaling 
($\propto k^{-2/3}$) for both functions.
\end{abstract}


\maketitle

\section{\label{intro}Introduction}

Two-point correlation function of the velocity in turbulence
has been the central object in statistical theory of homogeneous and isotropic 
turbulence. In particular, one goal of the theory is to derive the functional form
of the energy spectrum from the incompressible Navier-Stokes equations in the Fourier
space.
However, due to the quadratic nonlinearity, 
an equation for the correlation function cannot be obtained rigorously in a closed form, which is
known as the closure problem \citep[see, e.g.,][]{Leslie, Pope00, Davidson}.

To overcome this intrinsic problem, various approximations have been
proposed to close the equation for the correlation function,
as described, for example, critically in \citet{Davidson}.
Among those approximations, there is an exceptional one:
the direct interaction approximation (DIA) proposed by \citet{K59},
although the first DIA in the Eulerian coordinates failed to recover 
the Kolmogorov spectrum $k^{-5/3}$ in the inertial range
\citep[see, e.g.,][]{Leslie}.
By exceptional, it is understood that the DIA does not have any adjustable
parameter and that the mean linear response function was introduced
for the first time in the closure approximations of the Navier-Stokes equations \citep[see, e.g.,][]{MPRV08, EF11}.
The mean linear response function, or the Green function, which many physicists started to use in 1950s,
is now a standard theoretical device of closure approximation of the correlation function 
in nonlinear statistical problems \citep[see, e.g., ][]{UF,MPRV08}. 
Specifically, the reason to utilise
the mean linear response function (\emph{linear response function} for short) 
is to describe the nonlinear effect in a perturbative manner. 
In this closure framework, the linear response function and the correlation function are
considered on an equal footing.

Motivated by the framework, we study several important aspects of the linear 
response function via direct numerical simulation (DNS) 
in the Eulerian and Lagrangian coordinates. These aspects are described in the next subsections.
In particular, to our knowledge, 
a DNS study of Lagrangian linear response function is reported for the first time.

\subsection{Relation between the linear response function and the correlation function: fluctuation-response relation (FRR)}

In the DIA-type closures, one of the crucial elements is
the relation between the linear response function and the correlation function.
As the result of the approximations,
we end up typically with a set of two closed integro-differential equations
for the linear response function and the two-point correlation function. 
We then need to solve the set of the equations numerically.
In practice, we solve them simultaneously by assuming that the linear response function and 
the correlation function are self-similar.
In this process, we often encounter difficulty such as infra-red or ultra-violet divergence of the integrals,
see, e.g., a discussion concerning the mode-coupling theory of colloidal suspensions 
in \citet{miyazaki} (this is not always the case for turbulence, though).

One way to circumvent this problem is to utilise an expression of the linear response function 
in terms of suitable correlation functions, which is called fluctuation-response relation (FRR).
The special case of FRR is the fluctuation-dissipation theorem (FDT):
in equilibrium statistical mechanics the two functions are proportional with the proportionality constant being the inverse
temperature, see e.g., \citet{MPRV08}. 
The FDT is considered to fail generally in systems out of equilibrium.
Indeed, it has been demonstrated so
for a number of non-equilibrium steady-state systems as discussed in \citet{MPRV08}. 
In particular, it was shown that the FDT is invalid
for the forced Navier-Stokes turbulence in the dissipation range in \citet{CQ}
and for the forced SABRA shell model in our previous work \citep{moogn}. The breakdown
of the FDT is surely a manifestation of out-of-equilibrium character of turbulence and
of the shell model.

There are several forms of FRRs that hold for general out-of-equilibrium cases,
as reviewed in section 3 of \citet{MPRV08} and also section 4 of \citet{PSV17}. 
Unfortunately, they are not written with the two-point or multi-point correlation functions.
The most general one is written with formal derivative of the invariant measure.
Hence they cannot be used in solving the two integro-differential equations of the correlation
function and the linear response function, which are obtained by closure approximations.

However, if we add a random noise to the system, the situation
becomes different.
In this stochastic setting, there is at least one general expression of the linear response function
in terms of multi-point correlation functions, which is obtained by \citet{hs05, hs06}.
This recent development of the non-equilibrium statistical mechanics has urged us to consider the 
correlation function and the linear response function of turbulence in a new perspective.
This Harada-Sasa relation was the basis
of our previous study \citep{moogn} to consider a similar FRR for the shell model and the Navier-Stokes
equations in the Eulerian coordinates.
With the random noise, there is yet another general expression of the linear response function
in terms of the correlation between the random noise itself and the solution. This was obtained
by \citet{novi} and was studied numerically by \citet{CQ}.
We consider these two FRRs in this paper
by adding a random forcing to the Navier-Stokes equations in addition to the
deterministic large-scale forcing to maintain the turbulence in a statistically steady state.

Of course, such random forcing or noise does not have any physical origin in turbulent flows,
whereas for the microscopic systems considered in \citet{hs05, hs06}, 
the Langevin noise therein has a definite physical origin as an effect of
thermal fluctuations in the background environment. 
We regard our random forcing as a theoretical and numerical tool to investigate the response function 
and consider the zero limit of the random forcing
(here we do not intend to regard the randomly forced Navier-Stokes equations as fluctuating
hydrodynamic description for mesoscopic systems).

In the present study, first we demonstrate numerically breakdown of the FDT. 
Second, by adding small random forcing, we check whether the two types of non-equilibrium FRRs hold for the forced
Navier-Stokes turbulence in the Eulerian coordinates for the energy-containing, inertial
and dissipation ranges.
In particular, the Harada-Sasa relation is applied to the Navier-Stokes case 
for the first time.
In the Lagrangian coordinates, numerical simulation of the FRRs 
with the random forcing is almost impossible, as we will see. Hence we only give expressions 
for the Lagrangian FRRs.

\subsection{Difference in the Eulerian and Lagrangian coordinates: time scale and FRR}

There is another well-known problem in the DIA-type closures of turbulence:
it is understood that the failure of the earliest version of the DIA, leading
to the $k^{-3/2}$ scaling of the energy spectrum in the inertial range, was
due to picking up the sweeping time scale instead of the proper Kolmogorov
time scale in the inertial range. 
This is ascribed to lack of the Galilean invariance of the velocity correlation function
in the Eulerian coordinates, see, e.g., \citet{Leslie}. 
The DIA in the Lagrangian coordinates, called Lagrangian-history DIA (LHDIA),
was later elaborated by \citet{K65} and succeeded in reproducing the Kolmogorov $k^{-5/3}$ 
spectrum \citep{K66}. 

This implies that the time scales of the correlation function and the linear response function are
critical factors in order to have a correct result.
In other words, as discussed in \citet{K65}, a correct approximation to the Kolmogorov spectrum
should be capable of distinguishing between the time scales of the internal distortion caused
by the flow of the same spatial scales and that of the sweeping motion without distortion
caused by the flow of much larger scales.
However, these time scales of the correlation function and the linear response function are
not well studied numerically nor experimentally in spite of their critical role in the closures.
In the present paper,
we show via DNS that indeed the time scale of the linear response function
in the Lagrangian coordinates is consistent with the Kolmogorov scaling $k^{-2/3}$
for the first time (we analyse the linear response function in the Lagrangian history 
framework).

Given the success of the LHDIA, more straightforward DIA-type closures in the Lagrangian coordinates
have been developed without ad-hoc assumptions. Mostly, the development was
to incorporate the forward-in-time (measuring time) evolution of the Lagrangian
velocity field.
Notable ones include the Lagrangian renormalized approximation (LRA) by \citet{LRA}
and the Lagrangian direct interaction approximation (LDIA) by \citet{LDIA}.
These developments are crucial steps to extend the application 
area of the DIA-type closures to more realistic, inhomogeneous and anisotropic turbulent flows.

Then what is the role of FRR in these DIAs in the Eulerian and Lagrangian coordinates?
In Kraichnan's Eulerian DIA and LHDIA, no FRR was used upon solving the closed 
integro-differential equations for the correlation function and the linear response function. 
Instead, the FRR was invoked to justify the DIA:
his Eulerian DIA and Lagrangian history DIA were shown to be compatible to the FDT when it is applied
to the energy-equipartitioned state (fully thermalized state) of the Galerkin truncated Euler
equations, see e.g., \citet{k64decay}, \citet{K65} and \cite{K66}. 
By contrast, in the LRA and the LDIA,
the integro-differential equation for the linear response function becomes identical
to that of the correlation function. In other words, the FDT was obtained as a consequence 
of the closure approximations and hence used in solving the the integro-differential equations.

These closures suggest that 
whether or not the FDT holds, or a more general FRR should replace the FDT,
depends on the coordinates (Eulerian or Lagrangian).
We study this point by using DNS both in the Eulerian and Lagrangian (history) coordinates.
As we mentioned in the previous subsection, to explore possible forms of FRR, we use 
two known FRRs for the randomly forced cases by \citet{hs05, hs06} and by \citet{novi} 
and \citet{CQ}. 

Finally we comment on why studies about the linear response function in experiments or
numerical simulations have not been common.
One reason can be a technical one: a long-time average between the difference of the two nearby solutions
is required in order to have a statistically converged result. 
Another one may be a conceptual one: some regard the linear response function itself
as a somewhat abstract theoretical entity, leading to no interesting 
insights. 
Nevertheless, 
there are studies of the linear response function of the velocity Fourier modes 
in the Eulerian coordinates, which include a case for homogeneous and isotropic turbulence \citep{CQ}
and a case for turbulent channel flow in the context of turbulence control, see e.g., \citet{LQZ}
and references therein.
In the Lagrangian coordinates,
the correlation function of the Lagrangian velocity Fourier modes
has not been experimentally or numerically studied much either.
The notable early numerical studies of the Lagrangian correlation functions 
include:
\citet{kpv} in two dimensions;
\citet{grhk} for the Lagrangian history velocity in three dimensions;
\citet{YP} and \citet{kig} in three dimensions for the Lagrangian velocity whose measuring time evolves 
forward in time.

\subsection{Organisation of the paper}
The organisation of the paper is as follows.
In the next two sections, we study the correlation function 
and the linear response function of the Fourier coefficients of the velocity  
in both the Eulerian coordinates (section \ref{s:e}) and the Lagrangian coordinates (section \ref{s:l})
via a direct numerical simulation (DNS)
with a moderate Taylor-scale Reynolds number, $R_\lambda = 210$. 
The Reynolds number stays rather moderate since, for our purpose, integration over
hundreds of large-scale eddy turnover times is required.

More specifically, in section \ref{s:e} for the Eulerian coordinates, we discuss
two FRRs which were the results of the randomly forced case
obtained in \citet{novi} and \citet{CQ} and in our previous work \citep{moogn}.
The latter was obtained theoretically by adopting the 
relation in non-equilibrium statistical mechanics proposed by \citet{hs05, hs06}. 
We numerically compare the two FRR expressions with a small random forcing
to the linear response function measured without the random forcing, that is,
in the deterministic case (section \ref{s:NUMFRRE}).

In section \ref{s:l} for the Lagrangian coordinates, by using 
the numerical method used in \citet{kpv} and \citet{grhk}, known as the passive vector method, 
we calculate the Lagrangian correlation and linear response functions, which are the same
correlation and response functions as those considered in the abridged LHDIA (ALHDIA) by \cite{K65, K66}.
In both coordinates, the linear response function is directly calculated by using
the numerical method proposed in \citet{bdlv}.
We derive the FRRs for the Eulerian coordinates in appendix \ref{s:FRRE}
and for the Lagrangian coordinates in appendix \ref{s:FRRL}, but the Lagrangian FRRs 
are not numerically studied since their forms are not amenable to numerical simulations.

In section \ref{s:timescale}, we demonstrate numerically that characteristic times
associated with the Eulerian correlation and response
functions have indeed the sweeping scaling, $k^{-1}$ 
and that characteristic times associated with the Lagrangian ones
have the Kolmogorov scaling, $k^{-2/3}$, in the inertial range.

In section \ref{s:dis} we present discussions, which is followed by 
concluding remarks in section \ref{s:cr}.
To show a possible use of the Novikov-Carini-Quadrio FRR,
we describe one attempt to theoretically estimate the time scales of the response functions
at short times both in the Eulerian and Lagrangian coordinates,
which are in appendices \ref{s:shortNCQE} and \ref{s:shortNCQL}.

\section{\label{s:e}Correlation and linear response functions in the Eulerian coordinates}

\subsection{\label{s:dns}Direct numerical simulation}

We first describe the method of our DNS.
We consider the incompressible Navier-Stokes equations in a periodic cube with the side length $2\pi$:
\begin{equation}
 \partial_t \vec{u} + (\vec{u}\cdot\vec{\nabla}) \vec{u}
 = - \nabla p + \nu \nabla^2 \vec{u} + \vec{F}, 
\quad \nabla \cdot \vec{u} = 0,
\label{ns}
\end{equation}
where $\vec{u}, p$ and $\nu$ denote the velocity, the pressure
and the kinematic viscosity. The fluid density is normalised to unity.
The velocity and the pressure are functions of the spatial 
coordinates $\vec{x}$ and the time $t$.

We add a large-scale forcing, $\vec{F}$, to keep the system in a statistically steady state,
which is expressed in the Fourier space as
\begin{equation}
 \hat{\vec{F}}(\vec{k}, t)
 = 
\begin{cases}
  \frac{\epsilon_{\rm in}}{2E_f} \hat{\vec{u}}(\vec{k}, t) \quad (0 < |\vec{k}| \le k_f), \cr
  \vec{0} \quad ({\rm otherwise}).
\end{cases}
\label{nsf}
\end{equation}
Here $\hat{\vec{F}}(\vec{k}, t)$ and $\hat{\vec{u}}(\vec{k}, t)$ are the Fourier modes
of the forcing and of the velocity, and $\vec{k}$ denotes the wavevector.
The forcing parameters, $\epsilon_{in}$ and $k_f$, are the energy input rate
and the maximum forcing wavenumber, respectively.
By $E_f$, we denote the kinetic energy in the forcing range
\begin{equation}
 E_f = \sum_{\substack{\vec{k}\\ |\vec{k}| \le k_f}}
  \frac{1}{2}|\hat{\vec{u}}(\vec{k}, t)|^2.
\label{nsfc}
\end{equation}
With this setting, the numerically realised energy input rate by the forcing is indeed 
kept constant in time.
This type of forcing is often used in DNSs by various authors including \citet{CQ}. 

Numerically we solve the forced Navier-Stokes equations in the form of the vorticity equations
with the Fourier-spectral method with the $N^3$ grid points in the cube. We set mainly $N = 512$.
The aliasing error is removed by the phase shift and
the isotropic truncation (setting zero to the modes in $|\vec{k}| \ge \sqrt{2}N/3$).
We use the 4th order Runge-Kutta scheme for the time stepping. We set the parameter values
as follows: $\nu = 5.30 \times 10^{-4}$, $\epsilon_{\rm in} = 1.00\times10^{-1}$, $k_f = 2.50$
and the size of the time step $\Delta t = 1.87 \times 10^{-3}$. 
We make ten random initial velocity fields with the energy spectrum 
$E(k) \propto k^4 \exp(-k^2/2)$ by setting identically and independently distributed 
Gaussian random variables to the real and imaginary parts of the incompressible velocity Fourier modes.
The kinetic energy of the initial field is set to $0.50$.
For each initial data, we run the simulation for ten large-scale turnover times
and the statistics are collected since then. 
The resultant velocity fields are regarded as in a statistically
steady state with the Taylor-scale based Reynolds number being $R_\lambda = 210$.
The large-scale eddy turnover time is  $\tau_{to} = \langle L(t)  \rangle / (2 \langle E(t) \rangle/ 3)^{1/2} = 1.80$,
which is calculated with the energy, $E(t) = \sum_{\vec{k}} |\hat{\vec{u}}(\vec{k}, t)|^2/ 2$,
and the integral-length scale, $L(t) = (3\pi)/(4E(t))\times\sum_{\vec{k}} |\hat{\vec{u}}(\vec{k}, t)|^2 / |\vec{k}|$.
Here $\langle \cdot \rangle$ denotes the average over time and the ensemble. The root-mean-square velocity
is $u_{\rm rms} = (2\langle E(t) \rangle /3)^{1/2} = 6.25\times10^{-1}$.
The relation between the truncation wavenumber, $k_{\rm max} = \sqrt{2}N/3$, and the Kolmogorov dissipation
length scale, $\eta = (\nu^3/\langle \epsilon \rangle)^{1/4}$ , is $k_{\rm max} \eta = 1.51$.
Here $\langle \epsilon \rangle$ is the mean energy dissipation rate, which is here indeed 
equal to the prescribed energy input rate $\epsilon_{\rm in}$.

\subsection{Eulerian correlation and response function}

Here we start with a decomposition of the incompressible 
velocity Fourier modes in the Eulerian coordinates, which have
only two independent components. 
Such a decomposition becomes crucially important when we later
consider the FRRs by adding random noise to the Navier-Stokes equations.
We adopt the Craya-Herring decomposition defined with the reference vector
chosen here as $-\vec{e}_z = (0, 0, -1)$,
\citep[see, e.g.,][]{SagautCambon}, which is 
\begin{equation}
 \hat{\vec{u}}(\vec{k}, t) 
= 
  \hat{u}_\varphi(\vec{k}, t) \vec{e}_\varphi
+ \hat{u}_\theta(\vec{k}, t) \vec{e}_\theta.
\label{edeco}
\end{equation}
Here the unit vectors are written in the spherical coordinate system as
$\vec{e}_\varphi = (-\sin\varphi, \cos\varphi, 0)$ and
$\vec{e}_\theta = (\cos\theta \cos\varphi, \cos\theta \sin\varphi, -\sin \theta)$
with
the polar angle $\theta ~(0 \le \theta \le \pi)$
and the azimuthal angle $\varphi ~(0 \le \varphi < 2\pi)$
of the wavevector $\vec{k} = k(\sin\theta \cos\varphi, \sin\theta \sin\varphi, \cos\theta)$,
where $k = |\vec{k}|$.
If $\vec{k}$ is aligned with the $z$-axis ($\theta = 0$ or $\pi$), 
we set $\varphi = 0$.

With this decomposition we define the correlation function
of the velocity Fourier modes in the Eulerian coordinates as
\begin{equation}
 C_{\alpha \beta}(\vec{k}, t | \vec{q}, s)
  =
  \langle 
  \hat{u}_\alpha (\vec{k},  t)
  \hat{u}_\beta  (\vec{q}, s)
  \rangle, 
\label{ce}
\end{equation}
where the indices, $\alpha, \beta$, are either $\varphi$ or $\theta$.

In the numerical simulation, 
we calculate the shell average of the diagonal 
correlation functions
\begin{equation}
 C_{\alpha \alpha}(k, t - s)
 = 
\frac{1}{N(k, k + \Delta k)}
 \sum_{\substack{\vec{k} \\ k \le |\vec{k}| < k + \Delta k}}
 C_{\alpha \alpha}(\vec{k}, t | -\vec{k}, s),
\label{sce}
\end{equation}
where $N(k, k + \Delta k)$ is the number of the Fourier modes
lying in the annulus $k \le |\vec{k}| < k + \Delta k$. We set here  $\Delta k = 1$.
Notice that we assume isotropy in the Fourier space and statistically 
steady state.
We calculate the autocorrelation function of each mode,  $C_{\alpha \alpha}(\vec{k}, t|-\vec{k}, s)$,
by way of the temporal Fourier modes using the Wiener-Khinchin theorem.
In practice, we record the time series of the real 
and imaginary parts of each Fourier mode and calculate mean of the squared modulus 
of the Fourier transform of the time series.

Now we define the mean linear response function 
of the velocity Fourier modes in the Eulerian coordinates as
\begin{equation}
 G_{\alpha \beta}(\vec{k}, t | \vec{q}, s)
  =
  \left\langle 
   \frac{\delta \hat{u}_\alpha (\vec{k},  t)}
        {\delta \hat{u}_\beta  (\vec{q},  s)}
  \right\rangle. 
\label{ge}
\end{equation}

In our numerical calculation of the mean linear response function,
we adopt the method used for the shell model in \citet{bdlv}.
Specifically, we take the numerical solution at time $t_0$ 
in the statistically steady state and consider two solutions: one is starting from $\hat{\vec{u}}_\alpha(\vec{q}, t_0)$
and the other is starting from a perturbed solution, 
$\hat{\vec{u}}_\alpha(\vec{q}, t_0) + \Delta \hat{\vec{u}}_\alpha(\vec{q}, t_0)$.
We then integrate the Navier-Stokes equations starting from the two initial conditions
independently.
At some later time $t ~(> t_0)$, the difference between the two solutions, which is denoted by 
$\Delta \hat{\vec{u}}_\alpha(\vec{k}, t)$, yields one sample of the linear response function 
\begin{equation}
 G_{\alpha \beta}(\vec{k}, t | \vec{q}, t_0)
  \sim 
   \frac{\Delta \hat{u}_\alpha (\vec{k},  t)}
        {\Delta \hat{u}_\beta  (\vec{q}, t_0)},
\label{gsim}
\end{equation}
provided that the difference stays so small that the evolution
is essentially linear.
We then take average of the right hand side of (\ref{gsim}) over time 
$t_0$ and over the ensemble of the several numerical solutions.

As in the correlation function, we calculate the shell average of the diagonal part 
of the response function,
\begin{equation}
 G_{\alpha \alpha}(k, t  - t_0)
  =
\frac{1}{N(k, k + \Delta k)}
 \sum_{\substack{\vec{k} \\  k \le |\vec{k}| < k + \Delta k}}
 G_{\alpha \alpha}(\vec{k}, t | -\vec{k}, t_0).
\label{sge}
\end{equation}
In the calculation of the shell average,
we add the initial perturbation at time $t_0$ (the denominator in (\ref{gsim}))
to all the modes in the shell.
For the initial perturbation, we set only the real part: in other words,
${\rm Im}[\Delta \hat{u}_\alpha (-\vec{k}, t_0)] = 0$.
We set the initial perturbation, ${\rm Re}[\Delta \hat{u}_\alpha (-\vec{k}, t_0)]$, to five percent of 
the standard deviation of $|\hat{u}_\alpha (\vec{k}, t)|$ (the sign of the initial perturbation is always positive).
We check that the shell-averaged response function calculated in this manner
agrees well with the mode-wise response function, $G_{\alpha \alpha}(\vec{k}, t | -\vec{k}, t_0)$,
which is calculated by adding the initial perturbation only to two modes $\hat{u}(\pm \vec{k}, t_0)$ with
$\vec{k}$ and $-\vec{k}$ being within the same shell.

\begin{figure}
\centerline{\includegraphics[scale=0.7]{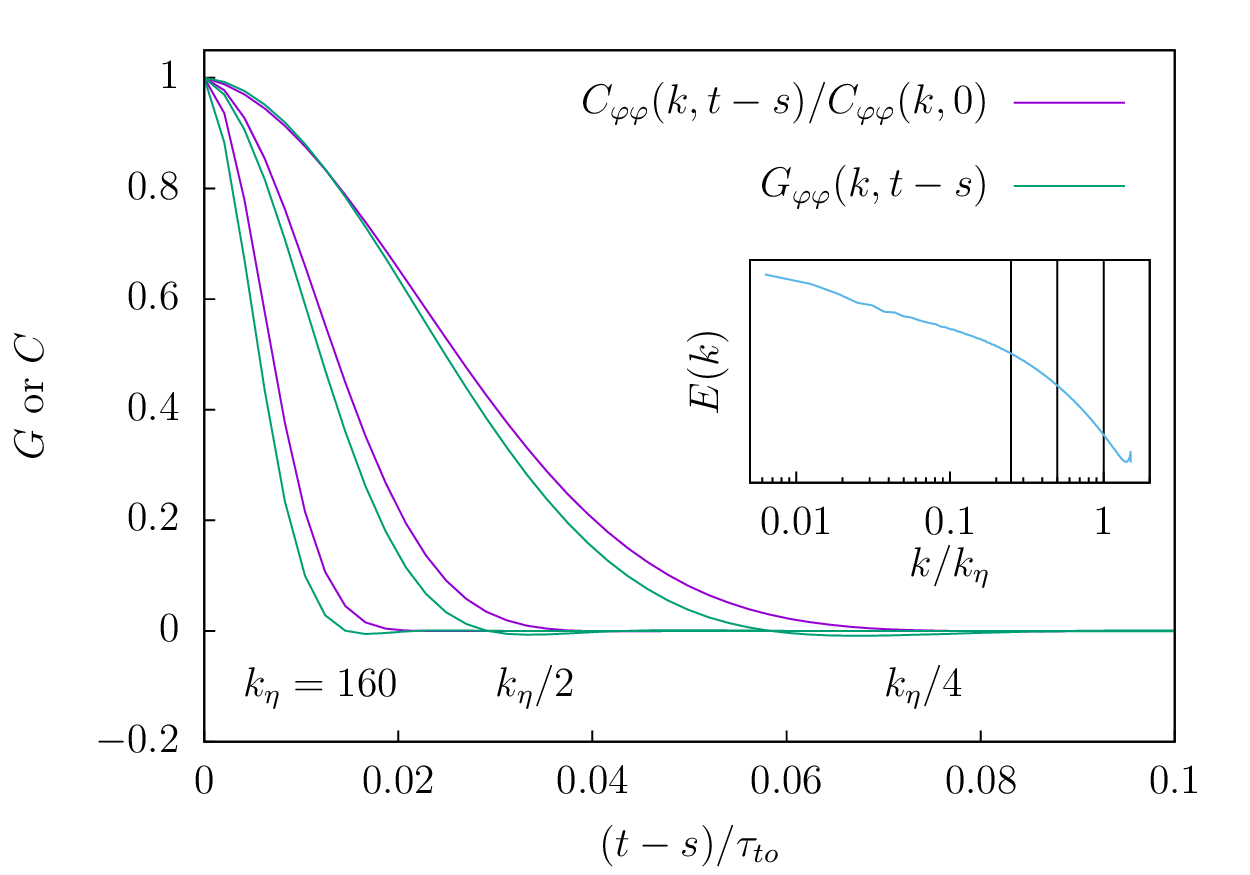}}
\centerline{\includegraphics[scale=0.7]{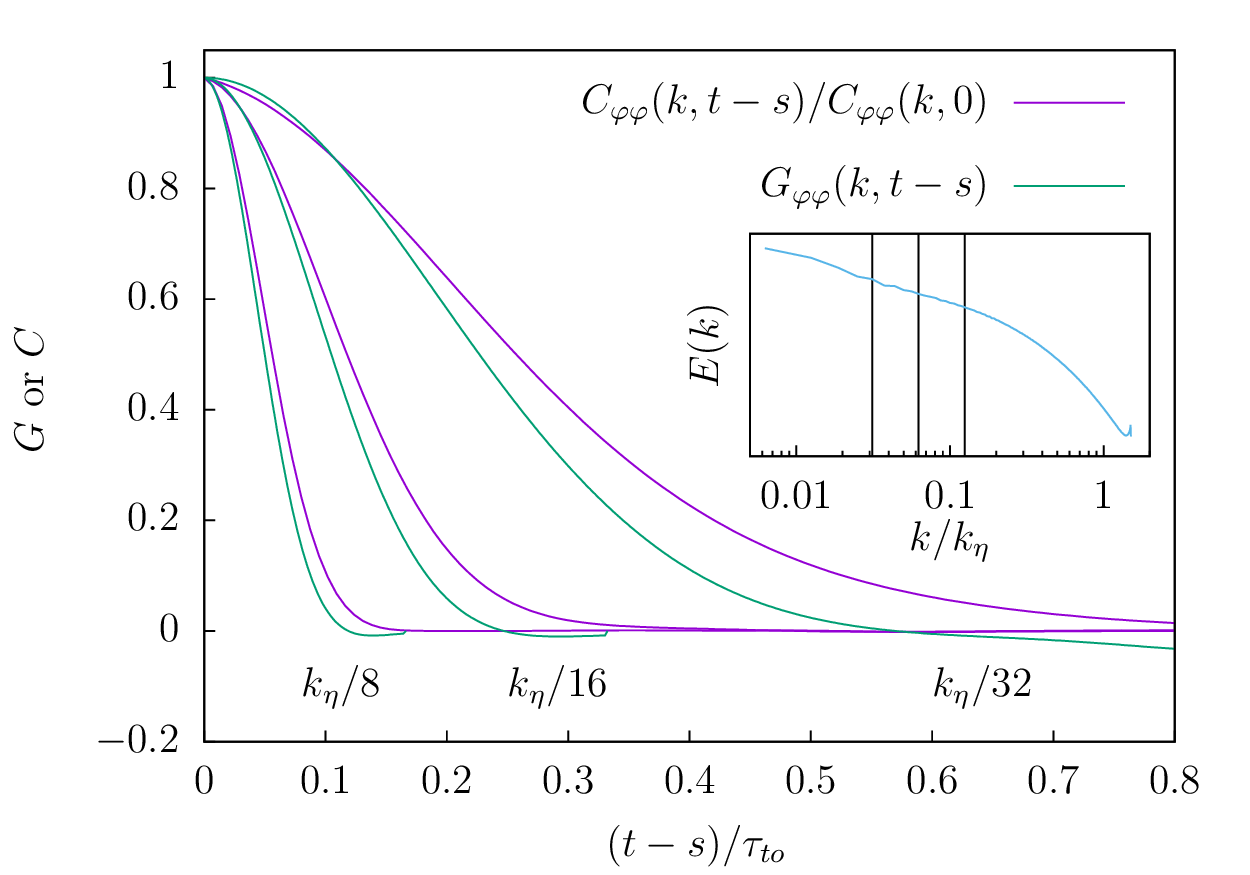}} 
\caption{\label{cge}The shell-averaged correlation function and the shell-averaged mean linear response function of the diagonal $\varphi$-component for $k=k_\eta, ~k_\eta/ 2, ~k_\eta/4$ (Top) and $k = k_\eta/8, k_\eta / 16, k_\eta / 32$ (Bottom). Notice that the correlation function is normalised with the equal-time value $C_{\varphi \varphi}(k, 0)$. Here the large-scale eddy turnover time is $\tau_{to} = 1.80$. Insets: the averaged energy spectrum with the representative wavenumbers depicted by vertical lines.}
\end{figure}

In figure \ref{cge}, 
we show the shell-averaged correlation functions normalised with the equal-time values
and
the shell-averaged linear response functions for six representative wavenumbers.
The wavenumbers are chosen as powers of one half times the Kolmogorov dissipation wavenumber,
$k_\eta = (\langle \epsilon \rangle / \nu^3)^{1/4} = 160$, up to 
the one in the energy-containing range, $k = k_\eta 2^{-5} = 2 k_f = 5$. 
In figure \ref{cge} we show only the real parts of the correlation and response functions since the imaginary
parts are about two orders of magnitude smaller than the real parts.
For the shell-averaged other components, the correlation function $C_{\theta \theta}(k, t - s)$ is nearly identical
to $C_{\varphi \varphi}(k, t - s)$ and the response function $G_{\theta \theta}(k, t - s)$ is nearly identical
to $G_{\varphi \varphi}(k, t - s)$.

We here observe small but measurable difference between the correlation function
and the linear response function. 
In particular, the FDT, $C_{\varphi \varphi} \propto G_{\varphi \varphi}$, is invalid
for all the representative wavenumbers spanning from the inertial range 
to the dissipation range. Here we regard $k_f < k \le k_\eta / 4 = 40$ as the inertial range
and $k > 40$ as the dissipation range based on the shape of the energy spectrum shown in
the inset of figure \ref{cge}. Another observation in figure \ref{cge} is the tendency 
that the response functions are generally smaller than the normalised correlation functions.
We do not have an explanation of this tendency. 

This breakdown of the FDT, which is as expected, 
is a manifestation of the fact that the velocity Fourier modes of turbulence are not
described with the equilibrium statistical mechanics \citep{MPRV08} regardless
of the wavenumber ranges.
Here we point out an apparently contradicting fact: the probability density functions
(PDFs) of the real and imaginary parts of the velocity Fourier modes in all the wavenumber ranges
are known to become closer to the Gaussian distribution as we increase the Reynolds number 
\citep{BrunPumir}. In our simulation, the PDFs are indeed close to Gaussian for all the six 
wavenumbers selected to be shown in figure \ref{cge}. Those PDFs are shown in appendix 
\ref{s:pdf}.
The Gaussian distribution implies the FDT, provided that there is no correlation
among different wavenumber modes.
An example with correlated degrees of freedom, whose marginal PDF is near Gaussian,
is carefully examined by \citet{MPRV08}. Indeed, they showed that the example
does not satisfy the FDT.
The homogeneous isotropic turbulence is another example of having Gaussian (marginal) PDF 
and not showing the FDT. 

Coming back to figure \ref{cge},
despite the difference between the correlation function and the linear response function,
we observe that their characteristic times defined, for example, as 
the integral time scales, seem to be of the same order of magnitude. This point will be studied 
in section \ref{s:timescale} together with the Lagrangian counterparts. 

We end this section by commenting on 
details of the averaging of the correlation and response 
functions.
We set the length of the temporal window for the correlation function
to $1.85\tau_{to}$ for the small wavenumbers, $k = k_\eta /32, k_\eta / 16$ and $k_\eta/8$,
and to $0.265\tau_{to}$ for large wavenumbers, $k = k_\eta / 4, k_\eta / 2$ and $k_\eta$.
The correlation functions shown in figure \ref{cge} are given in one-half of these window lengths.
We take total 15 such windows (5 windows in 3 simulations) in the averaging for the former set of $k$'s 
and total 200 windows (20 windows in 10 simulations) for the latter set of $k$'s.
For the linear response function, the length of the temporal window for each wavenumber
is $0.833\tau_{to},  0.331\tau_{to}, 0.164\tau_{to}$ for the set of the small wavenumbers
and $0.331\tau_{to}$ for the set of the large wavenumbers. The total number of
the windows are 20 (20 windows in 1 simulation), 100 (50 windows in 2 simulations),
and 200 (100 windows in 2 simulations) respectively for the former set of three $k$'s 
and 50 windows (50 windows in 1 simulation) for the latter set of three $k$'s.
Now the question with this sampling is whether the means of the correlation function and the 
linear response function shown in figure \ref{cge}  are converged or not.
To check this, we decrease the number of the samples to $1/3$ and compare the averages
over the full sample with those over the $1/3$ sample.
The difference between the averages is at most a few percent
for both the correlation function and the response function.
This is the case for large wavenumbers $k_\eta /16$
and $k_\eta /32$. For other wavenumbers, the difference between the samples is smaller than a few percent.
We regard the difference as small enough and consider that the average reached convergence.
This difference in the averages is less than the discrepancy between the correlation
function and the response functions shown in figure \ref{cge}.

\subsection{\label{s:NUMFRRE}Fluctuation-response relation with random forcing}
As mentioned in section \ref{intro}, the linear response function cannot 
be written in general with the two-point correlation function.
However, if the uncorrelated Gaussian noise is added to the evolution
equation, we can obtain several expressions of the linear response function
(FRR) in terms of certain correlation functions.
Here we consider two FRRs and compare them to 
the linear response function without the noise shown in the previous subsection.

For the homogeneous and isotropic Navier-Stokes turbulence, 
one of the expressions was derived by \citet{novi} and numerically studied by \citet{CQ}. 
To give the precise expression, we now fix some notation.
We first add the random Gaussian noise $\hat{\xi}_\alpha(\vec{k}, t)$
to the Navier-Stokes equations in the Fourier space in addition to the large-scale forcing as
\begin{align}
 \partial_t \hat{u}_\alpha(\vec{k}, t) 
  = & (\vec{e}_\alpha)_j
  \left(-\frac{{\rm i}}{2}\right)
  P_{jlm}(\vec{k}) 
  \sum_{\substack{\vec{p}, \vec{q} \\ \vec{p} + \vec{q} + \vec{k} = \vec{0}}}
  \hat{u}_l(-\vec{p}, t) \hat{u}_m(-\vec{q}, t) \nonumber \\
&  
  - \nu k^2 \hat{u}_\alpha(\vec{k}, t) 
 + \hat{F}_\alpha(\vec{k}, t)
  + \hat{\xi}_\alpha(\vec{k}, t).
\label{nsr}
\end{align}
Here we take summation over the repeated indices $j, l$ and $m$ and the index $\alpha$ denotes 
the Craya-Herring component $\varphi$ or $\theta$.
The projection operator is $P_{jlm}(\vec{k}) = k_m P_{jl}(\vec{k}) + k_l P_{jm}(\vec{k})$, where
$P_{jl}(\vec{k}) = \delta_{jl} - k_j k_l / k^2$ with $\delta_{jl}$ being the Kronecker delta and $k = |\vec{k}|$.
The real and imaginary parts of the noise, $\hat{\xi}_\alpha(\vec{k}, t)$, are identically and independently distributed
Gaussian random variables with the following mean and covariance
\begin{eqnarray}
 \langle \hat{\xi}_\alpha(\vec{k}, t) \rangle &=& 0,
 \label{nm} \\
 \langle \hat{\xi}_\alpha(\vec{k}, t) \hat{\xi}_\beta(\vec{p}, s) \rangle
  &=& 2 \sigma^2(k) T \delta_{\alpha, \beta} \delta_{\vec{k}, -\vec{p}} \delta(t - s),
\label{nv}
\end{eqnarray}
where $\sigma(k)$ is some function of $k$, $T$ is a parameter which we call ``temperature''
in this paper for convenience and $\delta(t)$ is the Dirac delta function.

The diagonal linear response function with the noise,
denoted by  $G^{(T)}_{\alpha \alpha}(\vec{k}, t|-\vec{k}, s)$,
is expressed as
\begin{equation}
 G^{(T)}_{\alpha \alpha}(\vec{k}, t|-\vec{k}, s)
 = \frac{1}{2\sigma^2(k) T}
   \left\langle
    \hat{u}_\alpha(\vec{k}, t) \hat{\xi}_\alpha(-\vec{k}, s)
   \right\rangle.
\label{J}
\end{equation}
We denote the right hand side of (\ref{J}) as 
$J^{(T)}_{\alpha \alpha}(\vec{k}, t|-\vec{k}, s)$. This is the first FRR which 
we consider.
The value of $J^{(T)}_{\alpha \alpha}(\vec{k}, t|-\vec{k}, s)$ at the equal time,
$t - s = 0$, should be one, which is guaranteed by the variance (\ref{nv}).
In \citet{CQ}, the expression (\ref{J}) was shown numerically
to be equal to the linear response function in the dissipation range without the random noise
if the noise is sufficiently small.

This FRR holds in general for a randomly forced system. 
As the name, FRR, indicates,
it gives the relation between  the fluctuation (the random noise) and the response. 
The FRR (\ref{J}) has been used in statistical mechanics, for example, \citet{CKP94},
and can be obtained also from the statistical field-theoretic formalism on 
the linear response function, see e.g., section 10.4 of \citet{Cardy} 
or Chapter 36 of \citet{ZinnJustin}.
The theoretical basis of \citet{CQ} is \citet{LQZ} in which
the FRR was referred as a well-known result of signal theory.
According to \citet{MPRV08}, this FRR, not only for the Navier-Stokes equations but also for general Langevin equations, 
is ascribed to \citet{novi}. In this paper we call it Novikov-Carini-Quadrio FRR.

Now we move to another expression of the linear response function in terms of
the two-point or multi-point correlation functions of $\hat{u}$, which was 
outlined in \citet{moogn}.
For brevity, we write the nonlinear term and the large-scale forcing as
\begin{eqnarray}
 \Lambda_\alpha(\vec{k}, t) 
=  
 (\vec{e}_\alpha)_j
  \left(-\frac{{\rm i}}{2}\right)
  P_{jlm}(\vec{k}) 
  \sum_{\substack{\vec{p}, \vec{q} \\ \vec{p} + \vec{q} + \vec{k} = \vec{0}}}
  \hat{u}_l(-\vec{p}, t) \hat{u}_m(-\vec{q}, t)     
  + \hat{F}_\alpha(\vec{k}, t).
\label{La}   
\end{eqnarray}
Using this $\Lambda_\alpha(\vec{k}, t)$, we have another expression of the diagonal response function is
\begin{align}
 G^{(T)}_{\alpha \alpha}(\vec{k}, t|-\vec{k}, s)
= &
\frac{1}{2\sigma^2(k) T}
 \bigg[
   2\nu k^2 C_{\alpha \alpha}(\vec{k}, t|-\vec{k}, s) \nonumber \\
 &  -\big\{
       \langle \Lambda_\alpha^*(\vec{k}, t) \hat{u}_\alpha(\vec{k}, s)  \rangle  
     + \langle \Lambda_\alpha^*(\vec{k}, s) \hat{u}_\alpha(\vec{k}, t)  \rangle
    \big\} 
\bigg].
\label{H} 
\end{align}
We denote the right hand side of (\ref{H}) as $H^{(T)}_{\alpha \alpha}(\vec{k}, t | -\vec{k}, s)$.
This form was derived by adapting the Harada-Sasa relation  of nonlinear Langevin equation
in non-equilibrium steady state \citep{hs05, hs06} to the Navier-Stokes equations with Gaussian noise 
(\ref{nsr}).
We call $H^{(T)}_{\alpha \alpha}(\vec{k}, t | -\vec{k}, s)$ Harada-Sasa FRR in this paper.
Heuristically, the Harada-Sasa FRR can be also obtained from (\ref{J}) by re-writing the noise $\hat{\xi}_\alpha(-\vec{k}, s)$ 
with the dissipation term,  $\Lambda_\alpha(\vec{k}, t)$ and the time-derivative term via (\ref{nsr}).
We can next eliminate the time-derivative term by
using the causality of the response function and the symmetry of the auto-correlation function
$C_{\alpha \alpha}(\vec{k}, t|-\vec{k}, s)$. Then we arrive at the Harada-Sasa FRR from the Novikov-Carini-Quadrio FRR.
However, the original derivation of the Harada-Sasa FRR
does not depend on (\ref{J}). A derivation of (\ref{H}) is given in appendix \ref{s:FRRE}. 

Now let us observe structure of the Harada-Sasa FRR (\ref{H}) at the formal level.
It  provides a closed expression of the linear response
function in terms of the second-order correlation function and many third-order correlation functions
(recall that $\Lambda_\alpha(\vec{k}, t)$ involves the nonlinear term as given in (\ref{La})).
In particular, the second and third terms of the FRR (\ref{H}) describe the deviation from 
the FDT, $G_{\alpha \alpha} \propto C_{\alpha \alpha}$, implying that the nonlinearity
is responsible for the deviation. This point will be examined later numerically.
Another observation concerns the value of $H^{(T)}_{\alpha \alpha}(\vec{k}, t | -\vec{k}, s)$ at the equal time,
$t - s = 0$, which should be one.
This is guaranteed by the statistical steadiness of the energy
of the each component of the Fourier mode, i.e.,
$\partial_t \langle |\hat{u}_\alpha(\vec{k}, t)|^2\rangle = 0$.
More precisely, under the steadiness,
the numerator on the right hand side of (\ref{H}) at $t = s$ is equal to the energy
input by the noise, $2\sigma^2 T$. 

The two FRRs, which are basically equivalent expressions, hold
owing to the random noise.
However, the noise's role in this study is not physical but just theoretical
as we mentioned in section \ref{intro}. Let us argue that the two FRRs are consistent
with the FDT in the absolute equilibrium where the velocity
Fourier modes follow the Gaussian distribution and become independent from
each other. For the Harada-Sasa FRR,  the triple correlation vanishes
in the absolute equilibrium and hence it becomes consistent with the FDT.
For the Novikov-Carini-Quadrio FRR, we consider in the following
manner. First, the absolute equilibrium for this case can be realised by the Langevin
noise with a finite $T$ and $\sigma(k)=k^{1}$ as found by \citet{fns}.
Second, let us here ignore the large-scale forcing $\hat{\vec{F}}(\vec{k}, t)$
for the sake of the argument.
In this setting, the Navier-Stokes equations become just an Ornstein-Uhlenbeck process
given by the viscous term and the noise. Then we can see that
the Novikov-Carini-Quadrio FRR expression is consistent  with the FDT.

\begin{figure}
\centerline{\includegraphics[scale=0.7]{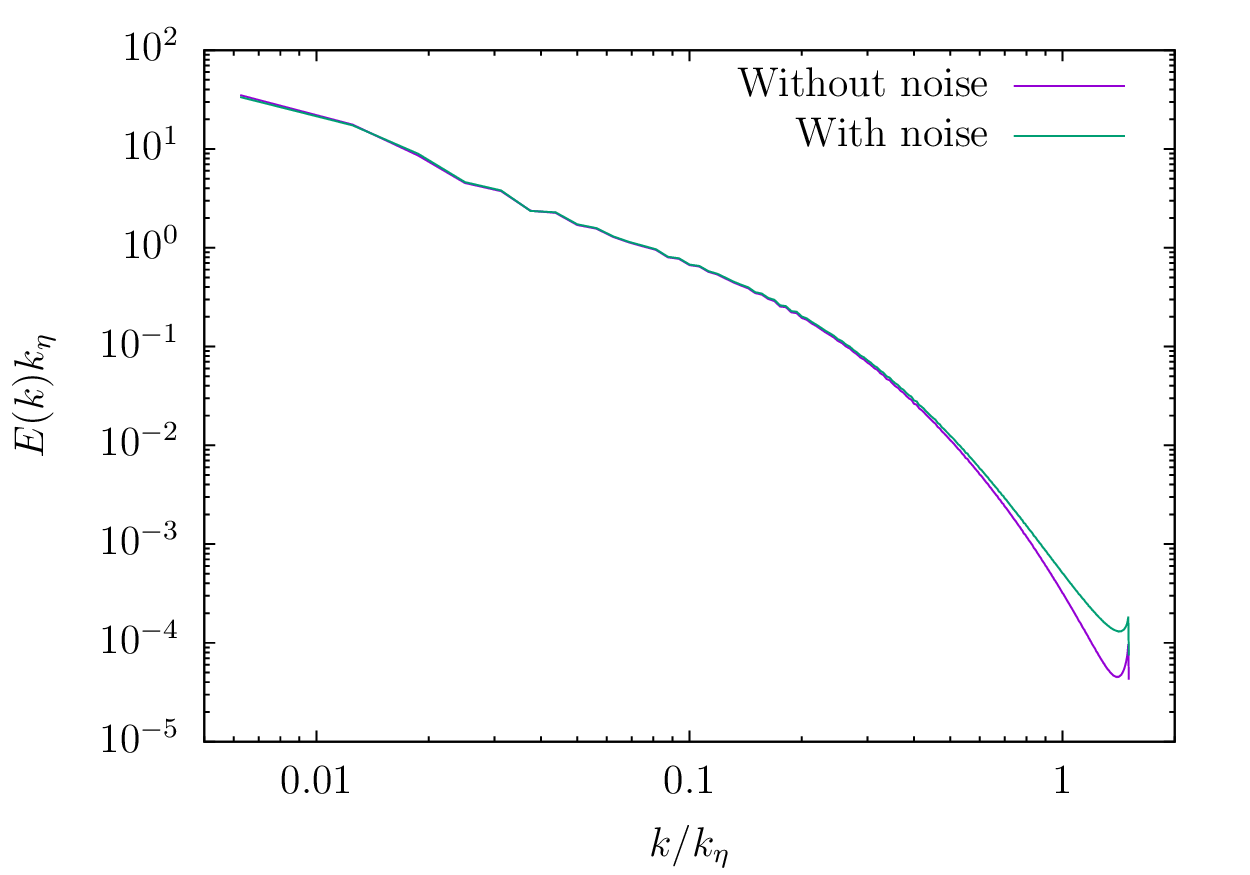}}
 \caption{\label{sc} Comparison between the energy spectra with and without the small random forcing, $\hat{\vec{\xi}}(\vec{k}, t)$. Here the noise variance parameters in (\ref{nv}) are $T = 10^{-6}$ and $\sigma(k) = k^{-1}$.}
\end{figure}

Now a question we numerically address is the same as \citet{CQ}: whether the FRRs with a sufficiently small noise 
amplitude $T$ are good approximations of the response function without the noise.
To answer this question, we compare the shell-averaged Novikov-Carini-Quadrio FRR, $J^{(T)}_{\alpha \alpha}(\vec{k}, t|-\vec{k}, s)$,
and the Harada-Sasa FRR, $H^{(T)}_{\alpha \alpha}(\vec{k}, t|-\vec{k}, s)$, with a small $T$
to the response function without the noise, $G_{\alpha \alpha}(k| t - s)$. 
The shell averages of the FRRs are defined in a similar fashion to (\ref{sge}).
As a small amplitude, we here take the value of the temperature $T = 10^{-6}$ and 
$\sigma(k) = k^{-1}$ (which corresponds to the wavenumber-independent noise spectrum).
With this choice, the energy spectrum is close to that of the noiseless case except for 
the far dissipation range as shown in figure \ref{sc}.
To calculate the FRRs, we solve the stochastic Navier-Stokes equations in terms of the vorticity equations
in the Cartesian $xyz$ components
with the same 4th order Runge-Kutta method as in the deterministic case (we do not use a stochastic scheme). 
The noise is generated 
in the Craya-Herring components, $(\hat{\xi}_\varphi(\vec{k}, t), \hat{\xi}_\theta(\vec{k}, t))$,
and then transformed to the $xyz$ components. 
This noise is added for all the wavenumbers in the computational (Cartesian) Fourier domain.
The time-step size is $\Delta t = 1.87 \times 10^{-3}$, which is the same as in the deterministic case. 
In the Runge-Kutta scheme, we do not generate the random noise at the middle time $t + \Delta t / 2$
but use the same noise generated at the time $t$. 

\begin{figure}
\centerline{\includegraphics[scale=0.7]{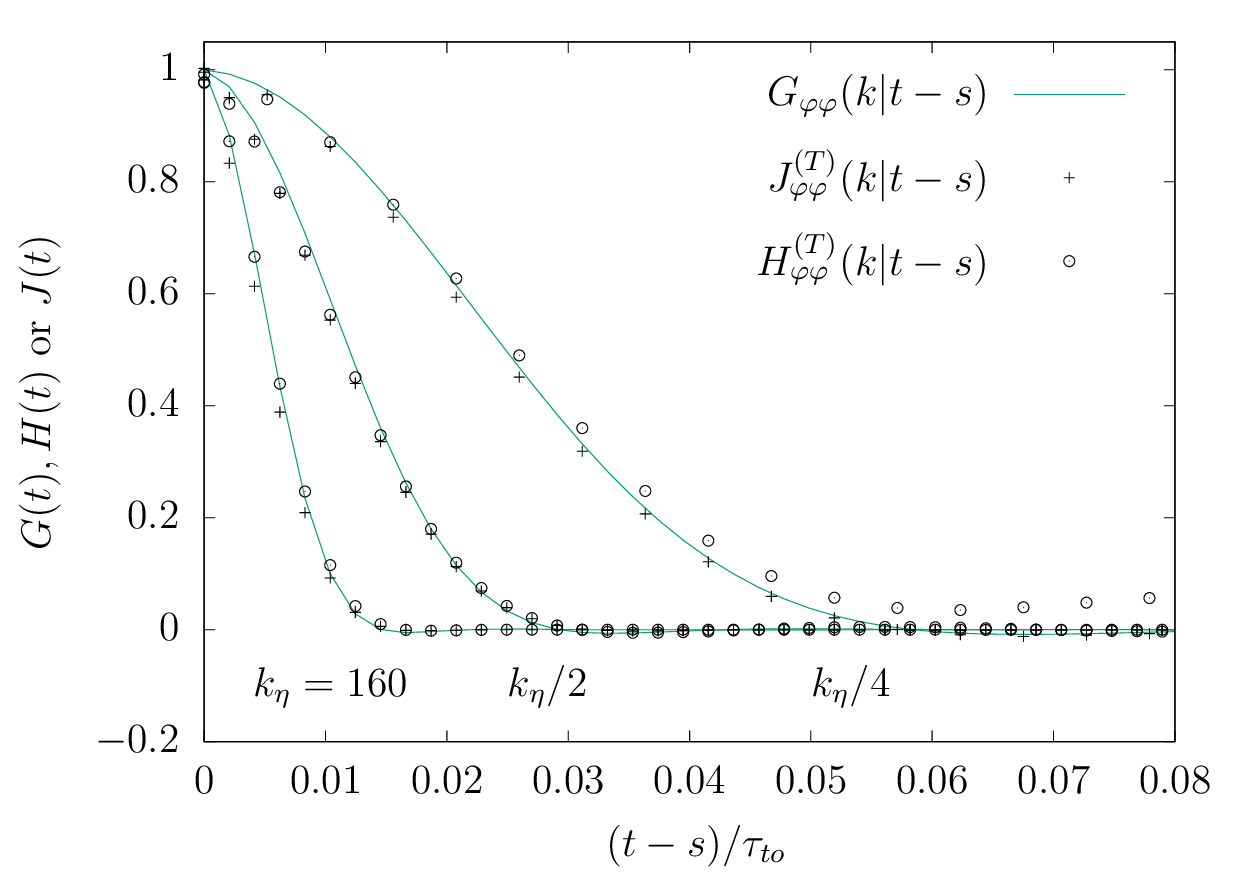}}
\centerline{\includegraphics[scale=0.7]{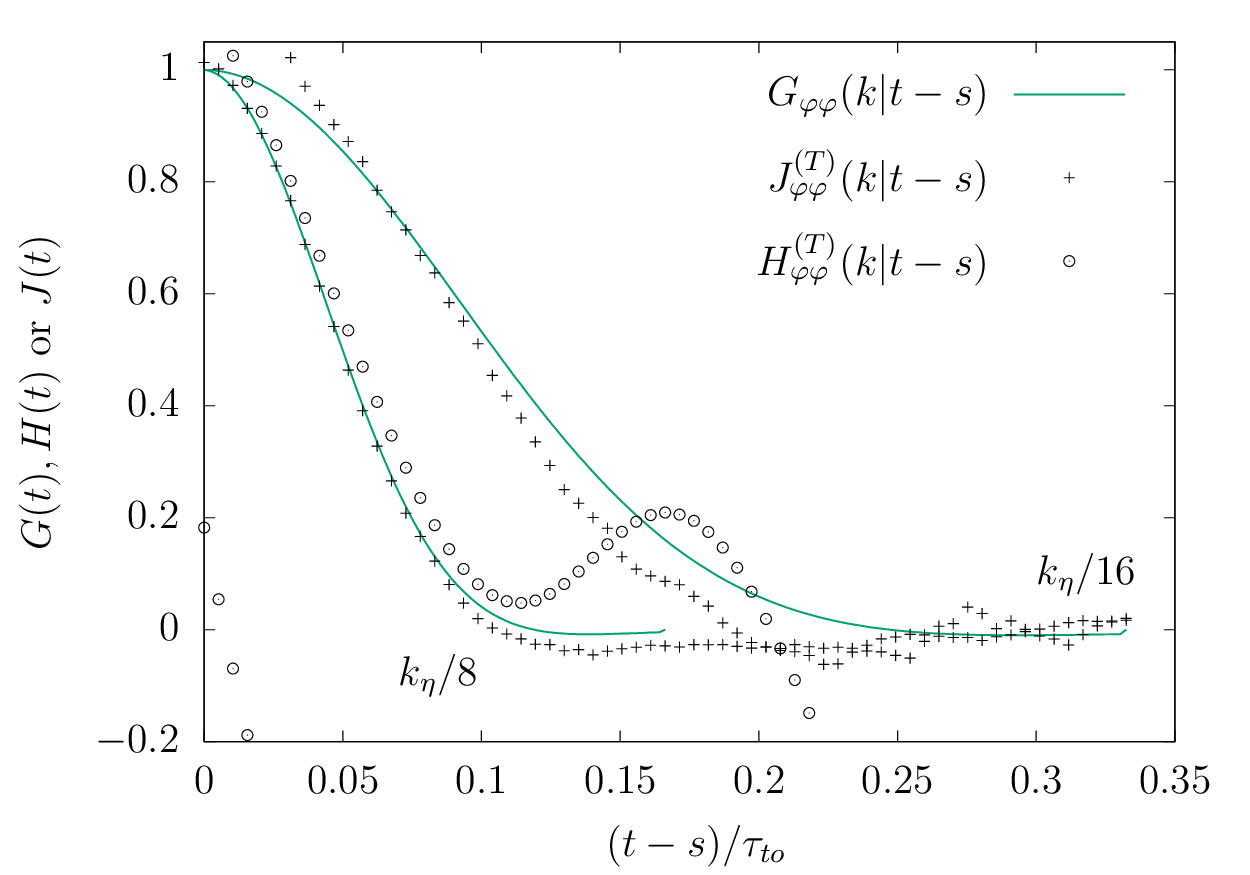}} 
 \caption{\label{hsj} The shell-averaged Novikov-Carini-Quadrio expression of the linear response function, $J^{(T)}(k, t - s)$, the Harada-Sasa expression, $H^{(T)}(k, t - s)$, and the linear response function in the noiseless case, $G(k, t - s)$, which is the same one shown in figure \ref{cge}. Here the noise is specified by $\sigma(k) = k^{-1}$ and $T = 10^{-6}$. Top: for $k = k_\eta, k_\eta/2, k_\eta/4$, Bottom: for $k = k_\eta / 8, k_\eta / 16$. The Harada-Sasa expression $H^{(T)}$ for $k = k_\eta/16$ (plotted with circles) has the numerical value of $0.2$ at the origin and becomes negative for $(t - s)/\tau_{to} > 0.03$, implying that the statistical convergence is not reached.  Here the $k = k_\eta/32 = 5$ case is not shown because of similar but much larger discrepancies.}
\end{figure}

The numerical calculations of the two FRRs are done as follows. We show here only the $\varphi$ component ($\alpha = \varphi$).
For $J^{(T)}_{\varphi \varphi}(\vec{k}, t|-\vec{k}, s)$, we use the same 
method as \citet{CQ}, namely calculate the correlation between $\hat{u}_\varphi(\vec{k}, t)$ and
$\hat{\xi}_\varphi(-\vec{k}, s)$.
The calculation of $H^{(T)}_{\varphi \varphi}(\vec{k}, t|-\vec{k}, s)$ is done by computing
the correlations involved, such as $\Lambda_\varphi^*(\vec{k}, t)$ and $\hat{u}_\varphi(\vec{k}, s)$ 
and so forth.
The results are shown in figure \ref{hsj}. 
We observe that the three response functions agree well for large wavenumbers, more precisely, from the end of 
the inertial range to the dissipation wavenumber $k_\eta$. 
For smaller wavenumbers, the two expressions start to deviate from each other. 
While the Novikov-Carini-Quadrio 
expression $J^{(T)}$ keeps a better agreement with $G$, the Harada-Sasa expression $H^{(T)}$ shows 
sizeable deviations. 
By increasing the number of samples, the deviations becomes smaller, though.
The worse agreement of $H^{(T)}$ has been anticipated from our previous study of the shell model \citep{moogn}
since the summations in the shell-model equivalent of (\ref{H}) caused loss of significant digits,
in particular, in the inertial range. This is also the case for the Navier-Stokes case as we will show now.
The Novikov-Carini-Quadrio FRR, $J^{(T)}$, does not have such a cancellation and hence exhibits better agreement.

\begin{figure}
\centerline{\includegraphics[scale=0.7]{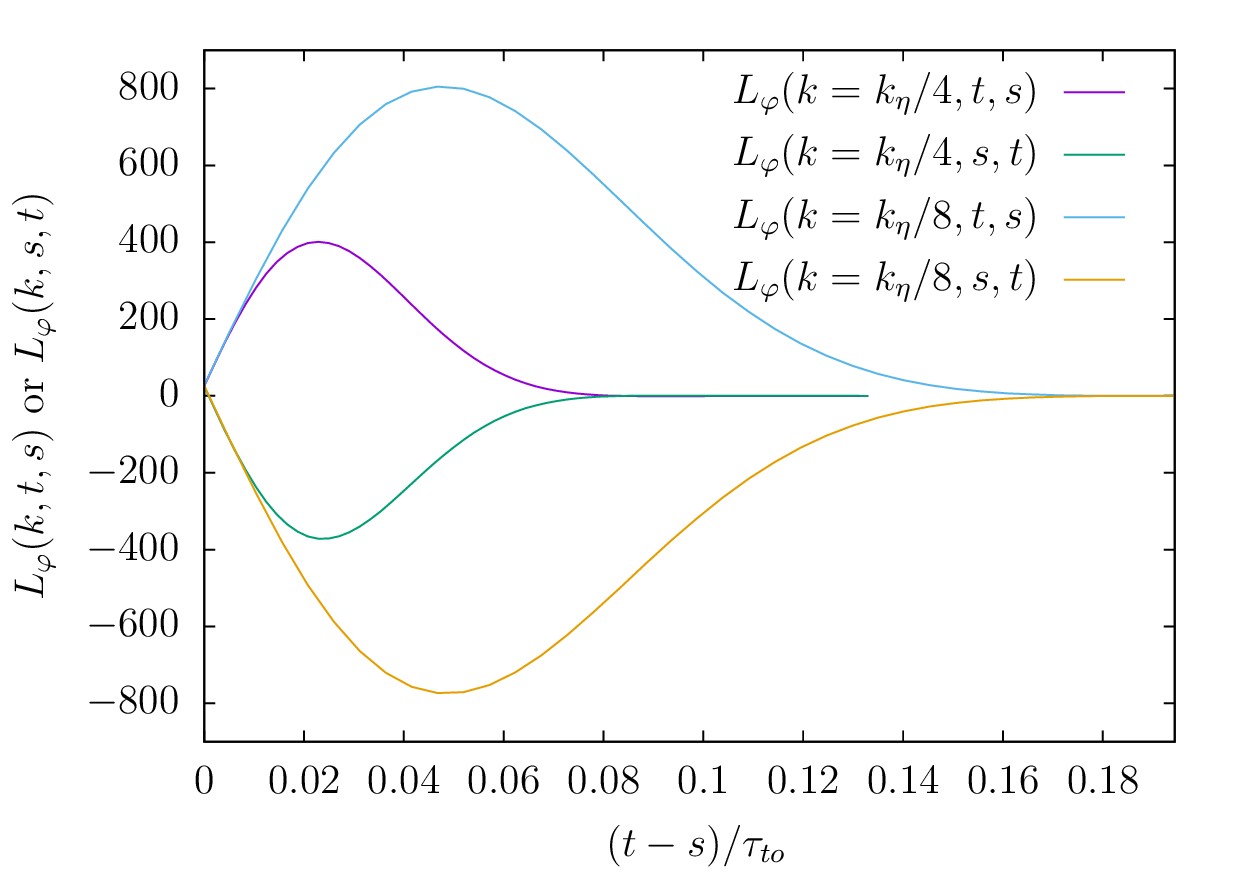}}
\centerline{\includegraphics[scale=0.7]{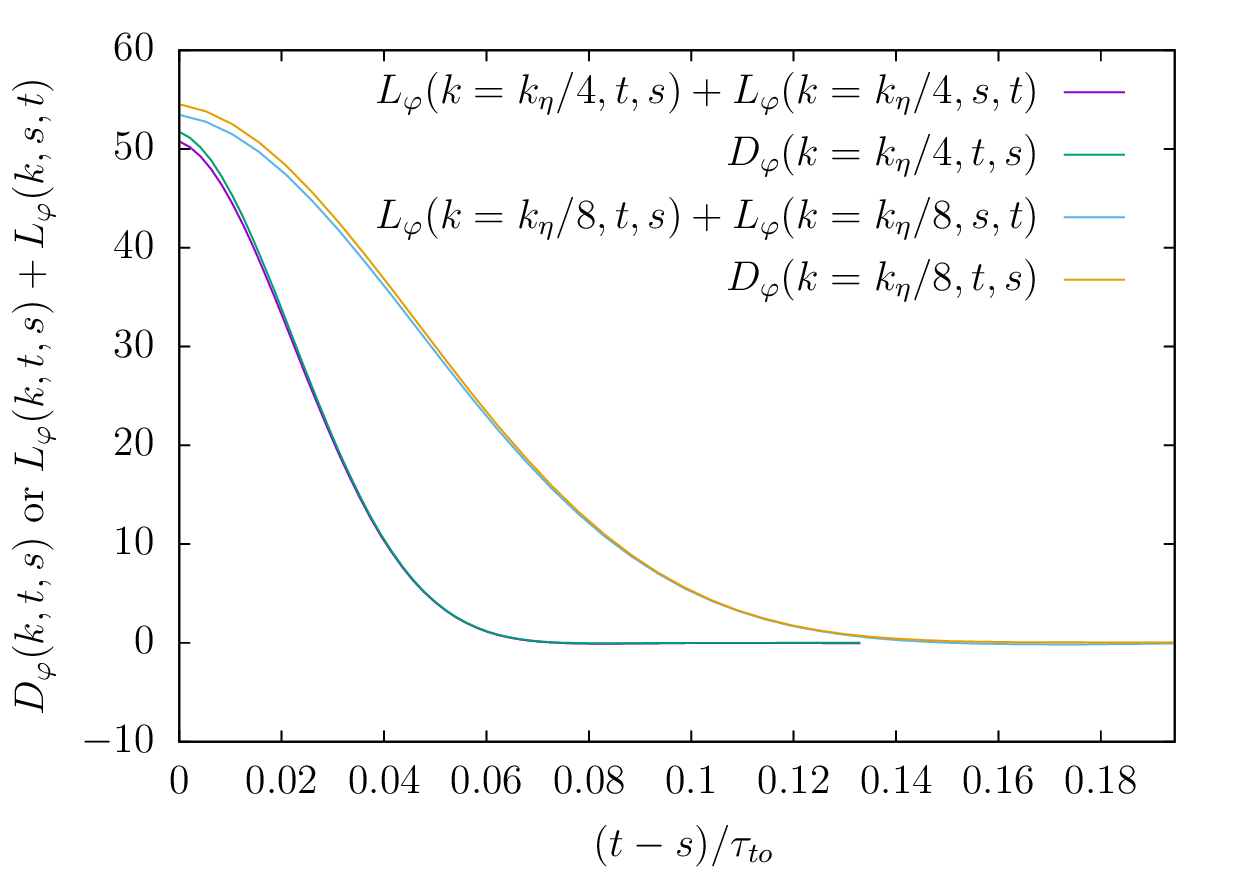}} 
 \caption{\label{hsc}Twofold cancellations involved in evaluation of the Harada-Sasa FRR, $H_{\varphi \varphi}^{(T)}(k, t - s) = D_\varphi(k, t - s) - [L_\varphi(k, t, s) + L_\varphi(k, s, t)]$, with $R_\lambda = 210 ~(k_\eta = 160)$ .}
\end{figure}

To discuss the cancellation of significant digits in (\ref{H}),  we write separately the shell-averages 
of the nonlinear and linear parts of the Harada-Sasa FRR as 
\begin{align}
 L_\alpha(k, t, s) =
 \frac{1}{N(k, k + \Delta k)}
 \sum_{\substack{\vec{k} \\  k \le |\vec{k}| < k + \Delta k}}
 \frac{1}{2\sigma^2(k) T}
 \langle \Lambda^*_\alpha(\vec{k}, t) 
 \hat{u}_\alpha(\vec{k}, s) \rangle, \\
 D_\alpha(k, t, s) = 
 \frac{1}{N(k, k + \Delta k)} 
 \sum_{\substack{\vec{k} \\ k \le |\vec{k}| < k + \Delta k}}
 \frac{1}{\sigma^2(k) T} 
\nu k^2 C_{\alpha \alpha}(\vec{k}, t| - \vec{k}, s).
\end{align}
Here notice that the wavenumber factors, $\sigma(k)$ and $k^2$, are 
inside the summation. This is necessary for our limited range of
the wavenumbers, $k_\eta / 16 = 10 \le k \le 160 = k_\eta$ with $\Delta k = 1$.  
The shell-averaged Harada-Sasa FRR is given as
$H^{(T)}_{\alpha \alpha}(k, t -  s) = D_\alpha(k, t, s) - [L_\alpha(k, t, s) + L_\alpha(k, s, t)]$
(which is shown in figure \ref{hsj}).
These shell-averaged parts are plotted in figure \ref{hsc} for $k = k_\eta / 4$ and $k_\eta / 8$.
Although the overall shapes of the triple correlations $L_\varphi(k, t, s)$ and $L_\varphi(k, s, t)$ ($t > s$)
are nearly symmetrical with respect to the horizontal axis, the former is slightly larger 
than the latter in magnitude. The positive sign of $L_\varphi(k, t, s)$
can be a reflection of the direct energy cascade. 
We can now see that the cancellation is twofold: the first is in the sum 
$L_\varphi(k, t, s) + L_\varphi(k, s, t)$ and the second is in the subtraction of the sum
from the viscous term. Roughly one significant digit is lost in each cancellation.
This implies that, in order to calculate $H^{(T)}$ with the right order of magnitude, 
the correlations involved should be calculated with more than 3-digit accuracy. This is 
a demanding numerical requirement in particular for those in small wavenumbers since 
a very long integration time is required to make fluctuation of the average small.

\begin{figure}
\centerline{\includegraphics[scale=0.7]{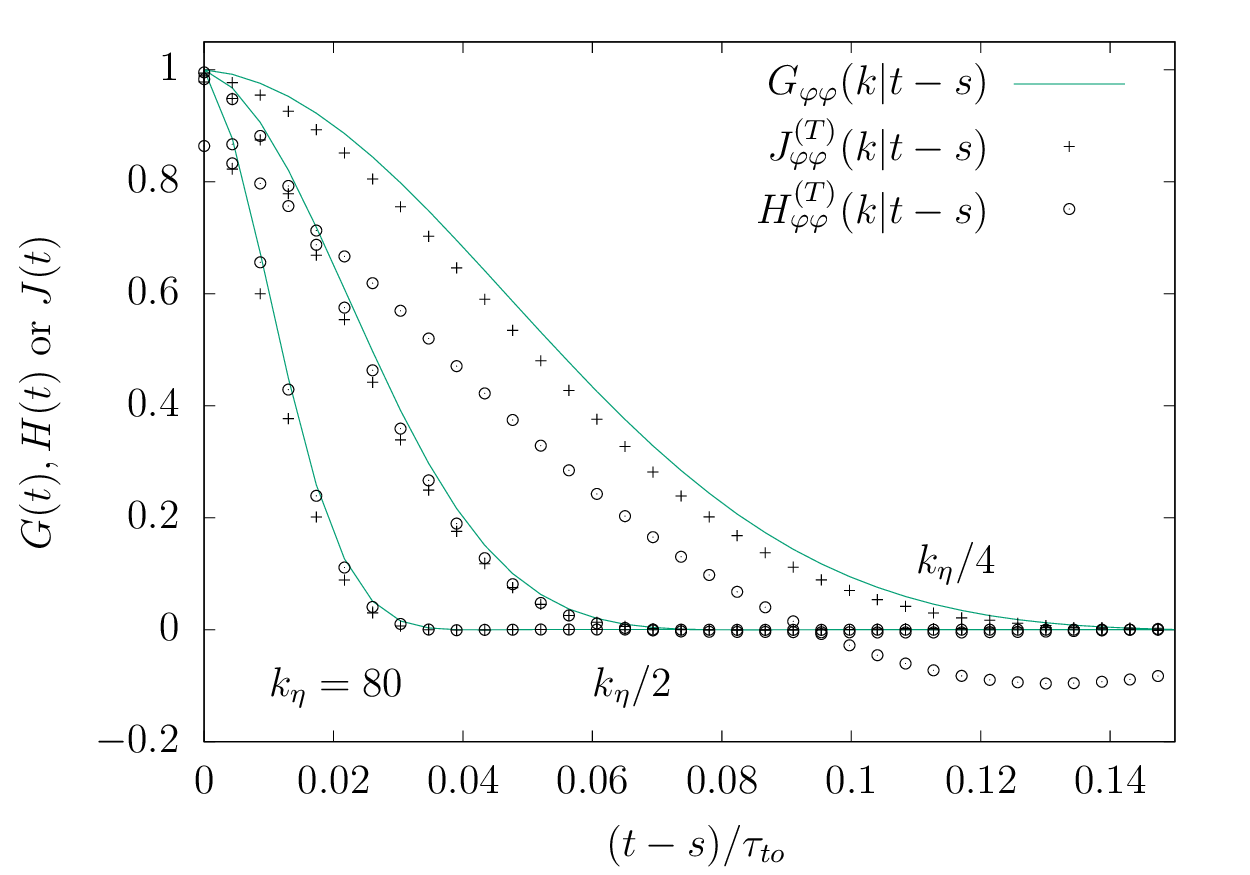}}
 \caption{\label{hsjlow}Same as figure \ref{hsj} but with a lower Reynolds number $R_\lambda = 130 ~(k_\eta = 80)$.}
\end{figure}

\begin{figure}
\centerline{\includegraphics[scale=0.7]{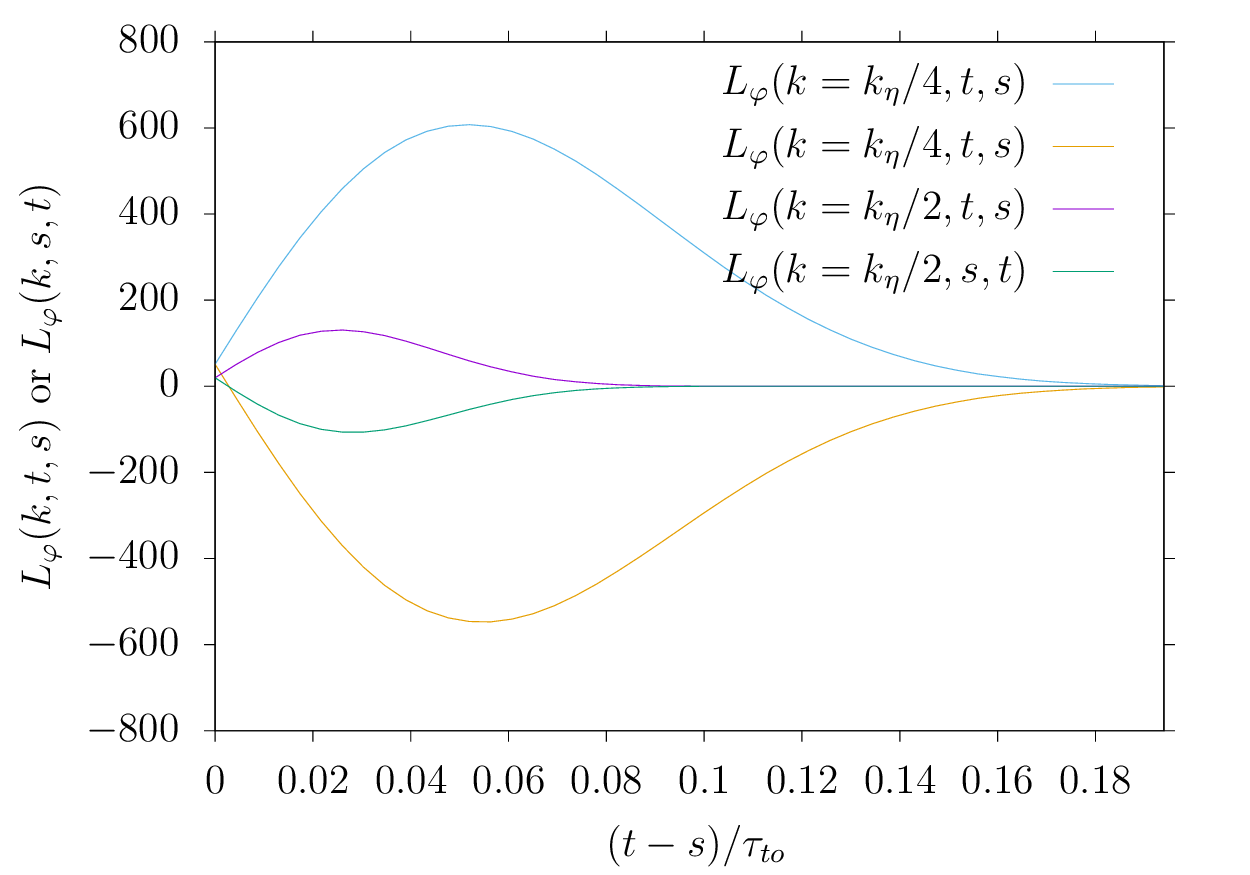}}
\centerline{\includegraphics[scale=0.7]{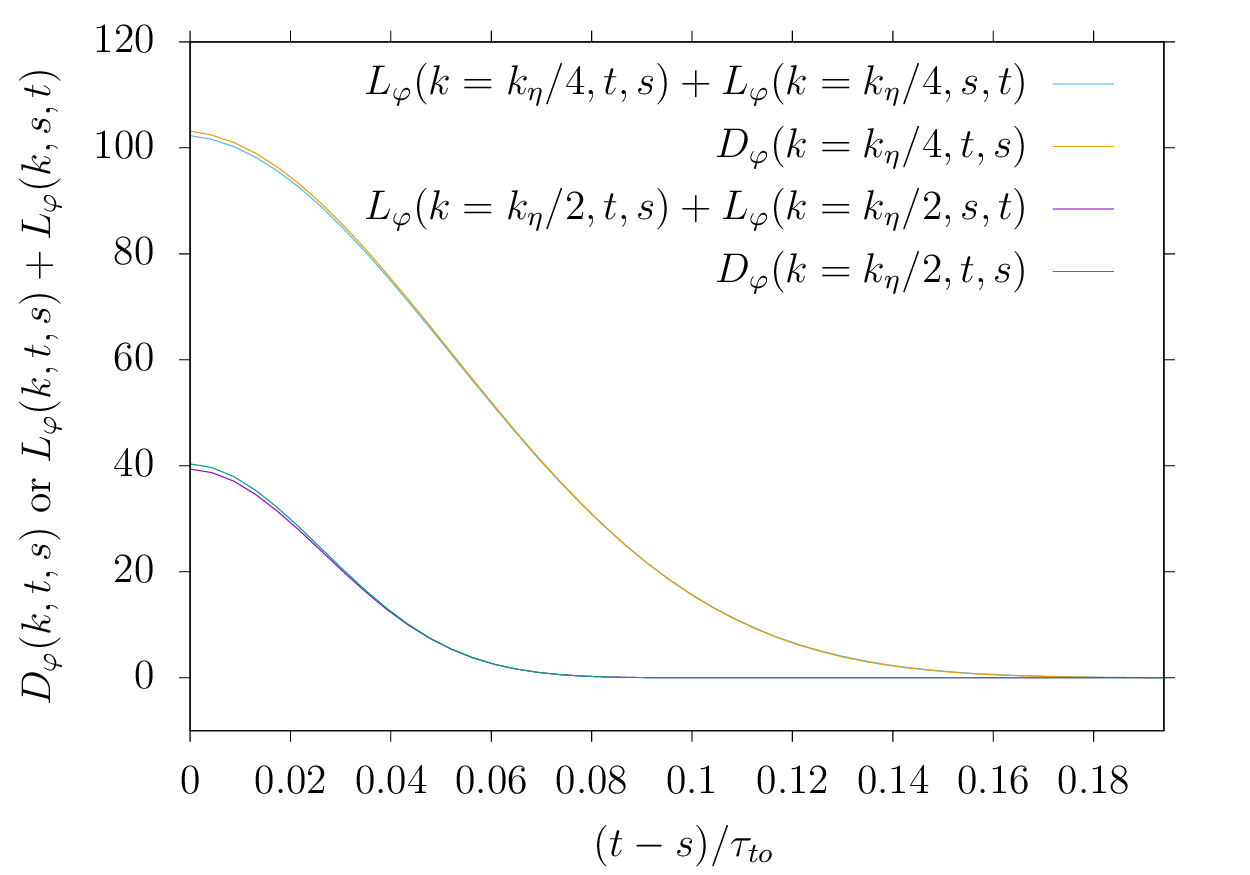}} 
 \caption{\label{hsclow}Same as figure \ref{hsc} but with a lower Reynolds number $R_\lambda = 130 ~(k_\eta = 80)$.}
\end{figure}

Next we consider Reynolds-number effect on the Harada-Sasa FRR. 
With a smaller Reynolds number, $R_\lambda = 130$ with $\nu = 1.34 \times 10^{-3}$, 
let us show the FRR in figure \ref{hsjlow} and the cancellations in the Harada-Sasa FRR 
in figure \ref{hsclow}.
In these figures, the noise is specified by $\sigma(k) = k^{-1}$ and $T = 10^{-6}$. 
Comparing figure \ref{hsjlow} with the top panel of figure \ref{hsj} for the higher Reynolds number, 
we find that the FRR's behaviour is similar, although
the agreement for the largest $k$  becomes poor for the lower Reynolds-number case.
We now argue that this poor agreement is due to the cancellations which become severer as we decrease the Reynolds number.
As shown in figure \ref{hsclow}, the twofold cancellations occur also for the lower Reynolds-number case. 
Here we notice in the bottom panel of figure \ref{hsclow} that the values of $D_\varphi$ and the sum of $L_\varphi$
for $k = k_\eta /4 = 20$ around the origin ($t - s = 0$) with $R_\lambda = 130$ is about $100$. In contrast,
the corresponding value is $55$ for $k = k_\eta / 8 = 20$ with $R_\lambda = 210$,
as shown in the bottom panel of figure \ref{hsc}. 
If this value at the origin is smaller, then the second cancellation, namely the loss of significant digits, 
becomes less severe. This gives
rise to the poor agreement for $k = k_\eta / 4$ shown in figure \ref{hsjlow}.
These observation suggest that, as we increase the Reynolds number, the Harada-Sasa FRR agrees better  
with the linear response function for small wavenumbers.

Now let us come back to the formal observation of the Harada-Sasa FRR (\ref{H}).
We previously noted that the sum of the second and third terms on the right hand side of (\ref{H}) is formally
responsible for the deviation from the FDT.
This formal observation assumes that the functional form of the sum as the time difference, $t - s$,
is very much different from that of the first term in (\ref{H}). 
However, this assumption is not valid as indicated by the second cancellation.
As shown in the bottom panel of figure \ref{hsc},
the numerical data demonstrates that the functional form of the sum is quite close to that of 
the viscous contribution.
Contrary to the formal observation of (\ref{H}), in reality
the deviation from the FDT arises equally both from the viscous contribution
and the nonlinear contributions. 
The viscous contribution to the deviation is not at all negligible for all the time range.
By contrast, in the shell-model study of the Harada-Sasa FRR \citep{moogn},
the second cancellation was not observed. This is probably owing to the extremely small kinematic viscosity 
of the shell model. It is also consistent with our observation that the second cancellation for the Navier-Stokes
case becomes less severe as we decrease the Reynolds number.
It is then suggested that the second cancellation does not occur for the Navier-Stokes 
case if the Reynolds number is sufficiently large.
There is one technical remark, however:
when we increase the Reynolds number, we may need to adjust the noise temperature $T$
to have the same energy spectra with $T = 0$ as we illustrated in figure \ref{sc}.
Another implication of the second cancellation shown in the bottom panel of figure \ref{hsc}
is that the sum of the triple correlations, $L_\varphi(k,t,s) + L_\varphi(k, s, t)$,
is very close to $D_\varphi(k, t - s)$ which is the correlation function multiplied by $\nu k^2$
in the whole $t - s$ domain. This suggests that this combination of the triple correlations
can be well approximated with the pair correlation with a suitable constant depending on $k$, 
which is considered to be a kind of eddy viscosity.

Regarding the wavenumber-dependent noise amplitude $\sigma(k)$, we have considered numerically
so far only one case, $\sigma(k) = k^{-1}$ with $T = 10^{-6}$.
In principle, the Novikov-Carini-Quadrio FRR and the Harada-Sasa FRR hold for any $\sigma(k)$ 
and $T$. We now briefly mention that the FRRs hold numerically for other choices of $\sigma(k)$
and $T$. Indeed, with an arbitrary choice of $\sigma(k)$ and $T$, the corresponding numerical 
solution can be quite different from what we wish to simulate as turbulent flow in nature. 
For example, the energy spectrum may be different from the Kolmogorov spectrum $E(k) \propto k^{-5/3}$.
But our focus here is to use those FRRs with the random noise to approximate the linear 
response function without the noise. For this purpose, given $\sigma(k)$, we expect that
sufficiently small $T$ enables us to have the Eulerian velocity with the noise being
statistically close to the one without the noise. It should be noticed here that, in addition to the noise,
we have the large-scale forcing $\hat{\vec{F}}(\vec{k}, t)$. 
As one variation of $k$-dependence of $\sigma(k)$, we set $\sigma(k) = k^{-2}$ with $R_\lambda = 130$.
In this case we find that $T=10^{-4}$ is small enough to have the same $E(k)$ as in the noiseless case. 
The FRRs for this noise behave similarly to those as depicted in figure \ref{hsj}.
To test the FRRs for the velocity field with $E(k) \not\propto k^{-5/3}$, an easy way is
to use a suitable $\sigma(k)$ and to increase the temperature $T$. 
In one numerical experiment with $R_\lambda = 130$, we consider 
$\sigma(k) = k^{-1}$ and $T = 10^{-4}$, where the energy spectrum
becomes $E(k) \propto k^{-1}$ for all the wavenumbers
(if $T$ is much smaller than that value, 
$E(k)$ follows $k^{-5/3}$ because of the large-scale forcing $\hat{\vec{F}}(\vec{k}, t)$).
This spectrum $E(k) \propto k^{-1}$ is
consistent with the renormalisation group analysis for $\sigma(k) = k^{-1}$, see, e.g., \citet{UF} and \citet{smp}.
In this non-Kolmogorov case, the directly measured
linear response function $G_{\varphi\varphi}^{(T)}$ also differ from the noiseless one $G_{\varphi\varphi}$.
We here also find that the Novikov-Carini-Quadrio FRR and the Harada-Sasa FRR 
agree well with the directly measured response function $G_{\varphi\varphi}^{(T)}$.
From these examples it is now clear that those FRRs are not
limited to the case characterised with the Kolmogorov spectrum.

To summarise, we conclude that the two FRRs with the sufficiently small random noise 
agree well with the linear response function in the deterministic setting (without 
the random noise) at least from the dissipation range up to the middle of the inertial 
range. In practice, the Novikov-Carini-Quadrio FRR, $J^{(T)}$, is numerically easier 
to calculate accurately than the Harada-Sasa FRR, $H^{(T)}$.
The direct evaluation of the linear response 
function \citep{bdlv} yields the least fluctuating result among the three methods, although
its computational cost is high. It requires to solve simultaneously two solutions of the
Navier-Stokes equations.
In principle, one solution of the Navier-Stokes equations and one solution
of the linearised Navier-Stokes equations are sufficient for the direct evaluation.
However, its cost is not very different from solving the two fully nonlinear equations.
We show a possible theoretical use of the Novikov-Carini-Quadrio FRR in appendix \ref{s:shortNCQE}, 
which is related to the subject of section \ref{s:timescale}.

The details of the averaging of the FRRs  in this section are as follows.
With $R_\lambda = 210$, the length of the temporal window of calculating $H^{(T)}$ and $J^{(T)}$ is 
$0.267\tau_{to}$ for the large wavenumbers $k = k_\eta, k_\eta / 2$ and $k_\eta / 4$. 
The total number of the windows is 200.
In one simulation we take 20 windows consecutively. Furthermore we repeat this for
the ensemble of the 10 simulations.
For the small wavenumbers, $k = k_\eta / 8$ and $k_\eta / 16$, the window
length is $0.667\tau_{to}$. The number of total windows is 100, which consists
of 10 consecutive windows in each simulation of the ensemble.
In the low Reynolds-number ($R_\lambda = 130$) case shown in figures \ref{hsjlow} and \ref{hsclow},
the time window to calculate the FRRs is $0.554\tau_{to}$ and the total number of the windows
is 200. We take 100 windows consecutively in one simulation. We repeat this for the ensemble
of the two simulations.

\section{\label{s:l}Correlation and linear response function in the Lagrangian  coordinates}
In this section we numerically study the correlation function and the linear
response function of the velocity Fourier coefficients in the Lagrangian
coordinates. To calculate them numerically with spectral accuracy, we employ
the passive vector method proposed by \citet{kpv}. This leads to the linear
response function with respect to the labelling time, not the measuring time 
of the Lagrangian velocity. Hence the linear response function studied here
is the one used in Kraichnan's ALHDIA \citep{K66}. 
Our goal here is to measure the two functions in the deterministic setting reliably.
This result will be used to extract time scales in the next section.
Contrary to the previous section, we study only theoretically 
FRR expressions of the Lagrangian response function
in appendices \ref{s:FRRL} and \ref{s:shortNCQL}.

\subsection{\label{s:pv}Passive vector method for the Lagrangian velocity}

We use Kraichnan's notation of the Lagrangian velocity,
$\vec{v}(\vec{a}, t_\ell| t_m)$, also known as the generalised velocity \citep{K65}. 
This is the velocity of a fluid particle measured at time $t = t_m$ 
which is called the measuring time.
This particular fluid particle passes the point whose coordinate 
is $\vec{a}$ at time $t = t_\ell$. The coordinate $\vec{a}$ and the time $t_\ell$ are called 
the Lagrangian label and the labelling time, respectively.
An intuitively natural choice is $t_m \ge t_\ell$ as made in the LRA and the LDIA.
In the ALHDIA the opposite choice $t_\ell \ge t_m$ was made:
one lets the labelling time vary by keeping the measuring time constant. 
In this case, the velocity measured at the fixed time, $t_m$, is
a Lagrangian invariant later in the labelling time. Therefore we have
the following passive vector equations describing the labelling-time evolution of the Lagrangian velocity,
\begin{eqnarray}
 \partial_{t_\ell} \vec{v}(\vec{x}, t_\ell| t_m)
 + 
 (\vec{u}(\vec{x}, t_\ell) \cdot \nabla) \vec{v}(\vec{x}, t_\ell| t_m)
 = \vec{0}.
\label{pv}
\end{eqnarray}
The initial condition of the Lagrangian velocity is given by the Eulerian velocity 
at the measuring time.

In our numerical simulation of the statistically steady state, we set the initial Lagrangian velocity
from the Eulerian velocity by $\vec{v}(\vec{x}, t_m| t_m) = \vec{u}(\vec{x}, t_m)$,
we then solve the passive vector equation (\ref{pv}) and the Navier-Stokes equations 
(\ref{ns}) simultaneously with the same numerical method as described 
in section \ref{s:dns}. 
After some long time, we reset the Lagrangian velocity to the current Eulerian velocity. 
Repeating this procedure, we obtain an ensemble of the Lagrangian velocity evolving in
the labelling time.

It is known that this passive vector method for the Lagrangian 
velocity has a serious numerical difficulty due to the lack of any dissipation in (\ref{pv}), 
see e.g.,\citet{grhk}. The difficulty is that the energy spectrum of the Lagrangian velocity, $E_v(k, t_\ell)$,
starting with the same spectrum of the Eulerian one at $t_\ell = t_m$,
increases quickly, especially in high wavenumbers. 
Hence the truncation error of the Lagrangian velocity becomes large
in a finite time.
The question is how long we can trust the Lagrangian velocity calculated 
with the passive vector method for a given spatial resolution. We will examine
this point in the next subsection.

\subsection{Lagrangian correlation and linear response function}

Having obtained the Fourier coefficient of the Lagrangian velocity, $\hat{\vec{v}}(\vec{k}, t_\ell|t_m)$,
we consider the following Lagrangian correlation function 
\begin{eqnarray}
 C^{(L)}_{jn}(\vec{k}, \vec{q}, t_\ell| t_\ell, t_m)
  = \langle 
    \hat{v}_j(\vec{k}, t_\ell|t_\ell)
    \hat{v}_n(\vec{q}, t_\ell|t_m)
    \rangle
 =  \langle 
    \hat{u}_j(\vec{k}, t_\ell)
    \hat{v}_n(\vec{q}, t_\ell|t_m)
    \rangle 
\label{cl}
\end{eqnarray}
and the linear response function 
\begin{eqnarray}
 G^{(L)}_{jn}(\vec{k}, \vec{q}, t_\ell| t_\ell, t_m)
  = \left\langle 
    \frac{\delta \hat{v}_j(\vec{k}, t_\ell|t_\ell)}
         {\delta \hat{v}_n(\vec{q}, t_\ell|t_m)}
    \right\rangle
  = \left\langle 
    \frac{\delta \hat{u}_j(\vec{k}, t_\ell)}
         {\delta \hat{v}_n(\vec{q}, t_\ell|t_m)}
    \right\rangle
  \label{gl}
\end{eqnarray}
for $t_\ell \ge t_m$.
Notice that the two functions in this time ordering are the same 
as used in the ALHDIA.
The indices, $j$ and $n$,
take values $1, 2$ or $3$ which correspond to the $x$, $y$ and $z$ 
components, respectively. The Lagrangian velocity ceases to be solenoidal
for $t_\ell > t_m$. To define the linear response function in (\ref{gl}), 
we adopt the notation used by \citet{LDIA}.
Hence the variation is with respect to
the velocity, not the infinitesimal probe force or the source term.

Let us compare the Lagrangian response function (\ref{gl}) with
those Lagrangian response functions used in the DIAs. 
The one used in the LHDIA  \citep{K65} corresponds to 
\begin{eqnarray}
 G^{(L)}_{jn}(\vec{k}, t_\ell| t_m; \vec{q}, s_\ell| s_m)
  = \left\langle 
    \frac{\delta \hat{v}_j(\vec{k}, t_\ell|t_m)}
         {\delta \hat{f}_n(\vec{q}, s_\ell|s_m)}
    \right\rangle
\label{gllhdia}  
\end{eqnarray}
with $t_\ell \ge t_m$ and $s_\ell \ge s_m$. Hence four time variables are involved 
in the LHDIA.
Here $\hat{\vec{f}}(\vec{q}, s_\ell|s_m)$ is the Fourier mode of
the infinitesimal probe force added to the right hand side of (\ref{pv}). 
Another one used in the ALHDIA \citep{K65, K66} is an abridged version of (\ref{gllhdia}),
\begin{eqnarray}
 G^{(L)}_{jn}(\vec{k}, t_\ell| t_\ell; \vec{q}, t_\ell| t_m)
  = \left\langle 
    \frac{\delta \hat{v}_j(\vec{k}, t_\ell|t_\ell)}
         {\delta \hat{f}_n(\vec{q}, t_\ell|t_m)}
    \right\rangle,
  \label{galhdia}
\end{eqnarray}
where only two time variables are involved and the numerator is an equal-time velocity.
This (\ref{galhdia}) coincides with (\ref{gl}) which we will study numerically.
Yet another one used in the LRA and the LDIA is
\begin{eqnarray}
 G^{(L)}_{jn}(\vec{k}, t_\ell| t_m; \vec{q}, t_\ell| t_\ell)
  = \left\langle 
    \frac{\delta \hat{v}_j(\vec{k}, t_\ell|t_m)}
         {\delta \hat{g}_n(\vec{q}, t_\ell|t_\ell)}
    \right\rangle
\label{gllra}  
\end{eqnarray}
with $t_\ell \le t_m$ \citep{LRA, LDIA}. This time ordering is different from
that in the ALHDIA and our DNS study.
Here $\hat{\vec{g}}(\vec{q}, t_\ell|t_m)$ is the infinitesimal probe force added to 
the right hand side of the measuring-time evolution equation of the Lagrangian velocity (\ref{eqv}):
see also appendices \ref{s:FRRL} and \ref{s:shortNCQL}.

\begin{figure}
\centerline{\includegraphics[scale=0.7]{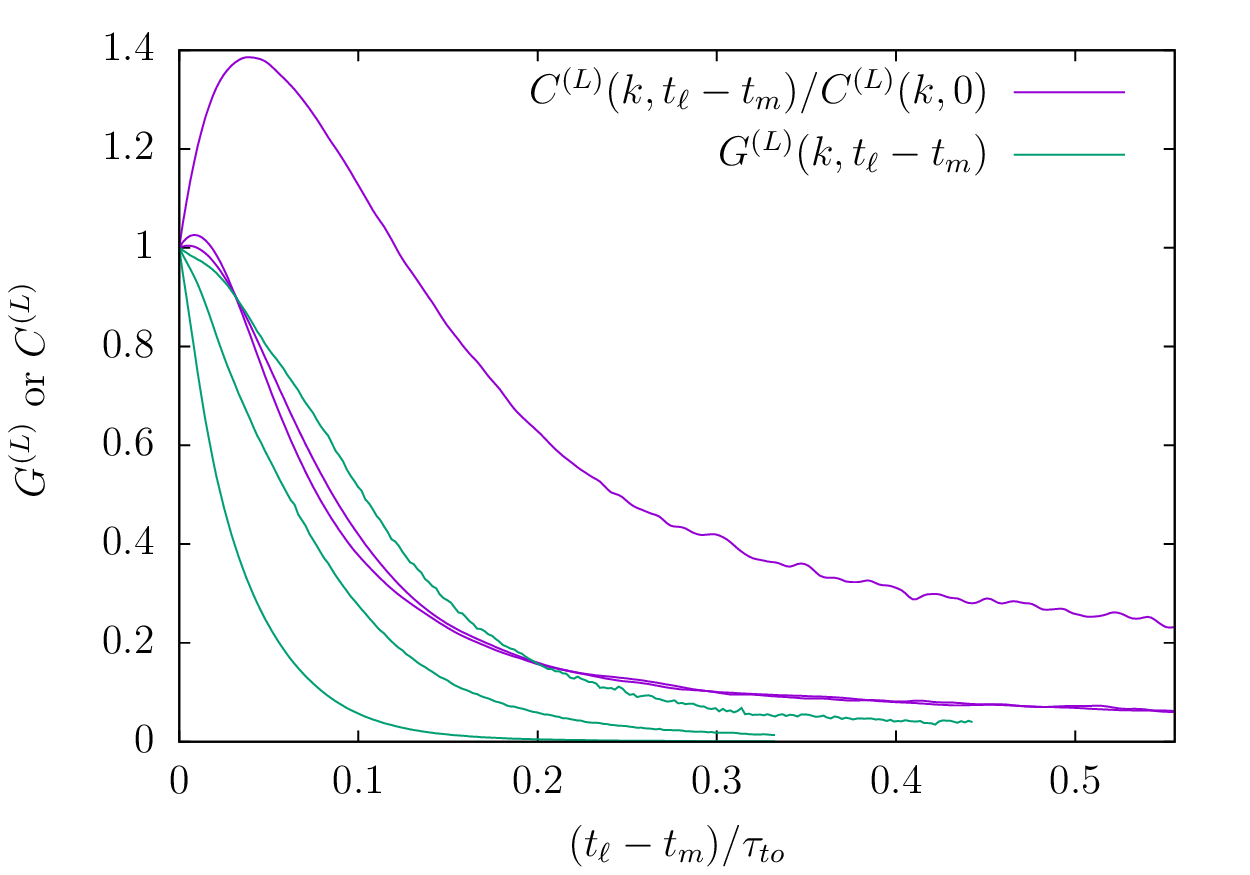}}
\centerline{\includegraphics[scale=0.7]{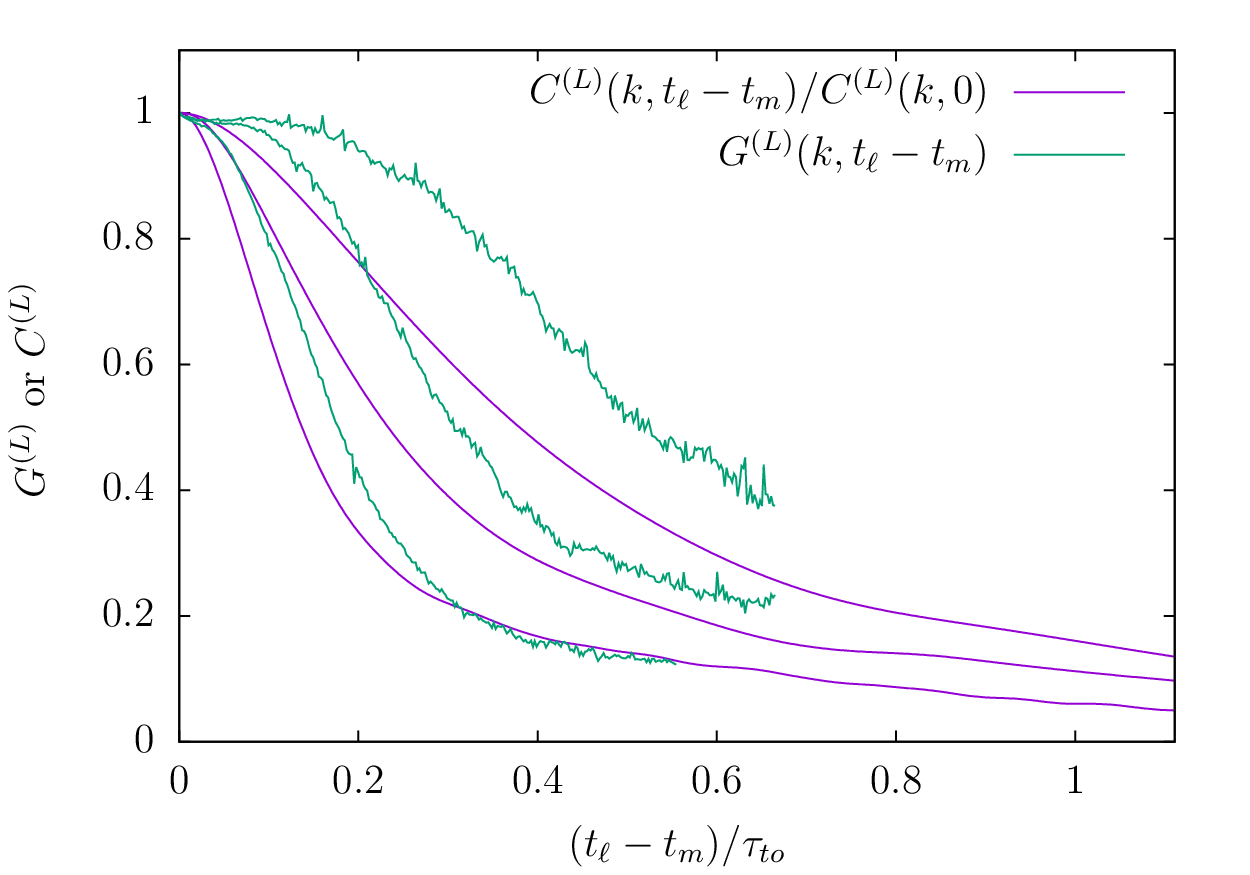}}
\caption{\label{lcg} Shell-averaged correlation functions and linear response functions in the Lagrangian coordinates
for $k = k_\eta, k_\eta / 2, k_\eta / 4$ from left to right curves (Top) and $k = k_\eta / 8, k_\eta / 16, k_\eta / 32$ from left to right curves (Bottom). We show only the real parts since the imaginary parts are orders-of-magnitude smaller. The correlations of $k = k_\eta / 2$ and $k_\eta / 4$ are very close to each other. The increase of the correlation functions at short time is discussed in the text. The noisy behaviour of the response functions is due to lack of the statistical samples. In the bottom panel, the response functions are calculated up to around $t_\ell - t_m = 0.55\tau_{to}$.} 
\end{figure}
We now move to DNS of (\ref{cl}) and (\ref{gl}).
We calculate the correlation function (\ref{cl}) from 
the Lagrangian velocity in the same way as the Eulerian one.
Let us here write the time of the Eulerian simulation as $t$.
For the direct calculation of the response function (\ref{gl}), we 
add small initial perturbation $\Delta \hat{\vec{u}}(\vec{k}, t_0)$ at $t = t_0$, to both the Eulerian and Lagrangian
velocity fields. This ``initial'' time $t_0$ for the perturbation becomes the measuring time for the Lagrangian 
velocity, namely, $t_0 = t_m$.
We then consider evolution in the labelling time $t = t_\ell ~(\ge t_m)$ of the two pairs of the velocity fields:
the unperturbed pair,   $\big( \hat{\vec{u}}(\vec{k}, t),         ~\hat{\vec{v}}(\vec{k}, t|t_m)\big)$,
and the perturbed pair, $\big( \hat{\vec{u}}^\prime (\vec{k}, t), ~\hat{\vec{v}}^{\prime}(\vec{k}, t|t_m)\big)$.
For the two pairs, we solve the Navier-Stokes equations (\ref{ns}) and the passive vector equations (\ref{pv}) 
simultaneously.
More specifically, the starting condition of the perturbed pair is
$\hat{\vec{u}}^\prime (\vec{k}, t_m) = \hat{\vec{u}}(\vec{k}, t_m) + \Delta \hat{\vec{u}}(\vec{k}, t_m)$
and $\hat{\vec{v}}^\prime (\vec{k}, t_m|t_m) = \hat{\vec{u}}^\prime (\vec{k}, t_m)$. For the unperturbed
Lagrangian velocity, $\hat{\vec{v}}(\vec{k}, t_m|t_m) = \hat{\vec{u}}(\vec{k}, t_m)$.
The response function at a labelling time $t$ can be obtained via 
$[\hat{\vec{u}}^\prime (\vec{k}, t) - \hat{\vec{u}}(\vec{k}, t)]/ [\hat{\vec{v}}^\prime (\vec{k}, t|t_m) - \hat{\vec{v}}(\vec{k}, t|t_m)]$. We repeat this procedure to calculate the Lagrangian response function. 

In figure \ref{lcg}, we show the shell-averaged correlation and response functions
\begin{align}
 C^{(L)}(k, t_\ell - t_m)
 = \frac{1}{3N(k, k + \Delta k)}
 \sum_{\substack{\vec{k} \\ k \le |\vec{k}| < k + \Delta k}}
 C^{(L)}_{jj}(\vec{k}, -\vec{k}, t_\ell| t_\ell, t_m),
 \label{scl}\\
 G^{(L)}(k, t_\ell -  t_m) 
=
\frac{1}{3N(k, k + \Delta k)}
 \sum_{\substack{\vec{k} \\ k \le |\vec{k}| < k + \Delta k}}
 G^{(L)}_{jj}(\vec{k}, -\vec{k}, t_\ell| t_\ell, t_m).
 \label{sgl}
\end{align}
Here we take summation over the index $j$. 
We add the initial perturbation $\Delta \hat{u}_j(\vec{k}, t_m)$ for all the modes 
within the shell $k \le |\vec{k}| < k + \Delta k$. 
The initial perturbation for each mode is set as
$\Delta \hat{\vec{u}}(\vec{k}, t_m) = \Delta u_0 \vec{e}_\varphi + \Delta u_0 \vec{e}_\theta$.
Here  $\Delta u_0$ is a real positive number whose magnitude is five percent of the standard deviation
of $|\hat{u}_\varphi(\vec{k}, t)|$.

\begin{figure}
\centerline{\includegraphics[scale=0.7]{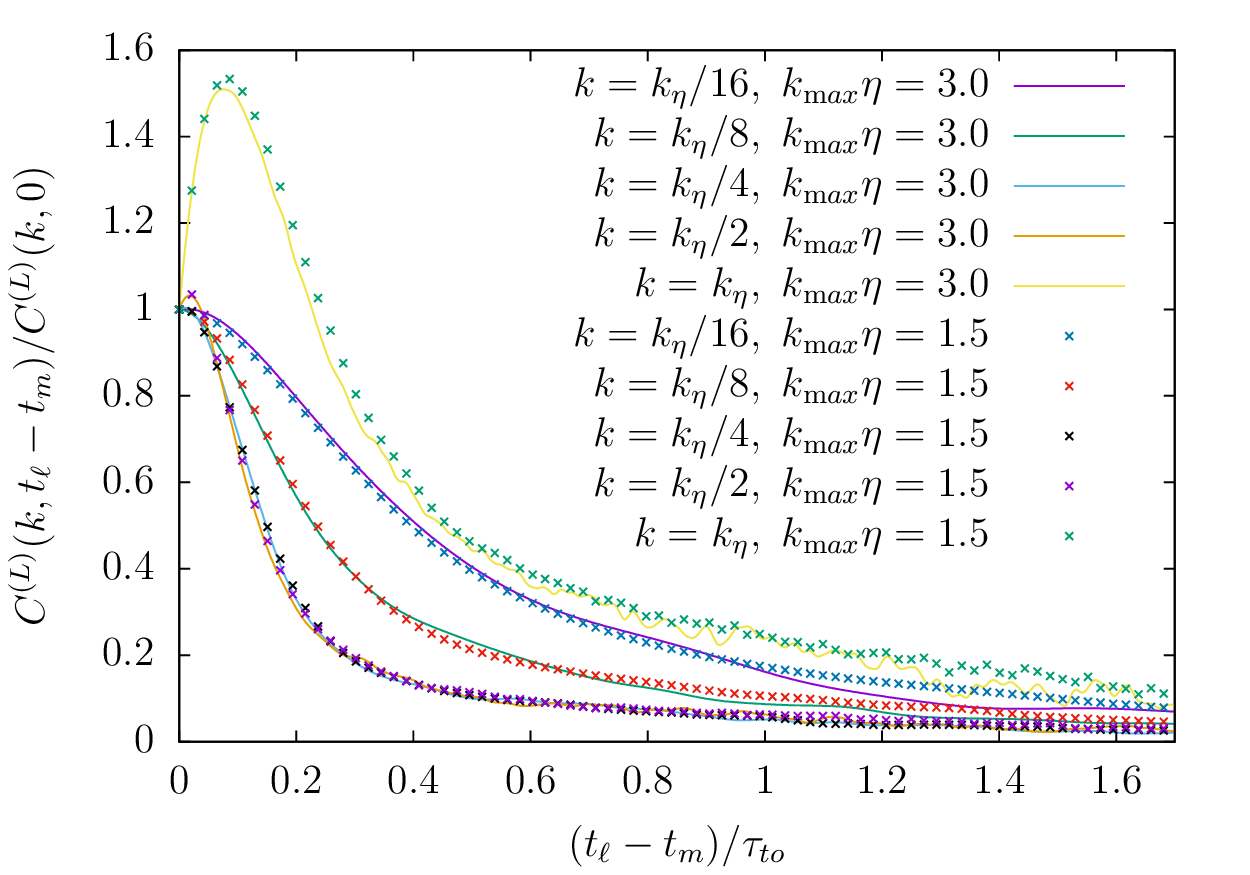}}
 \caption{\label{rslc}Resolution dependence of the Lagrangian correlation function
 for $R_\lambda = 130$. The Kolmogorov dissipation wavenumber is $k_\eta = 80$.
 The truncation wavenumbers of the two resolutions are $k_{\rm max} = 120$ and $241$.
 The correlation functions of $k = k_\eta / 4$ and $k_\eta / 2$ are again very close as in figure \ref{lcg}.}
\end{figure}

\begin{figure}
\centerline{\includegraphics[scale=0.7]{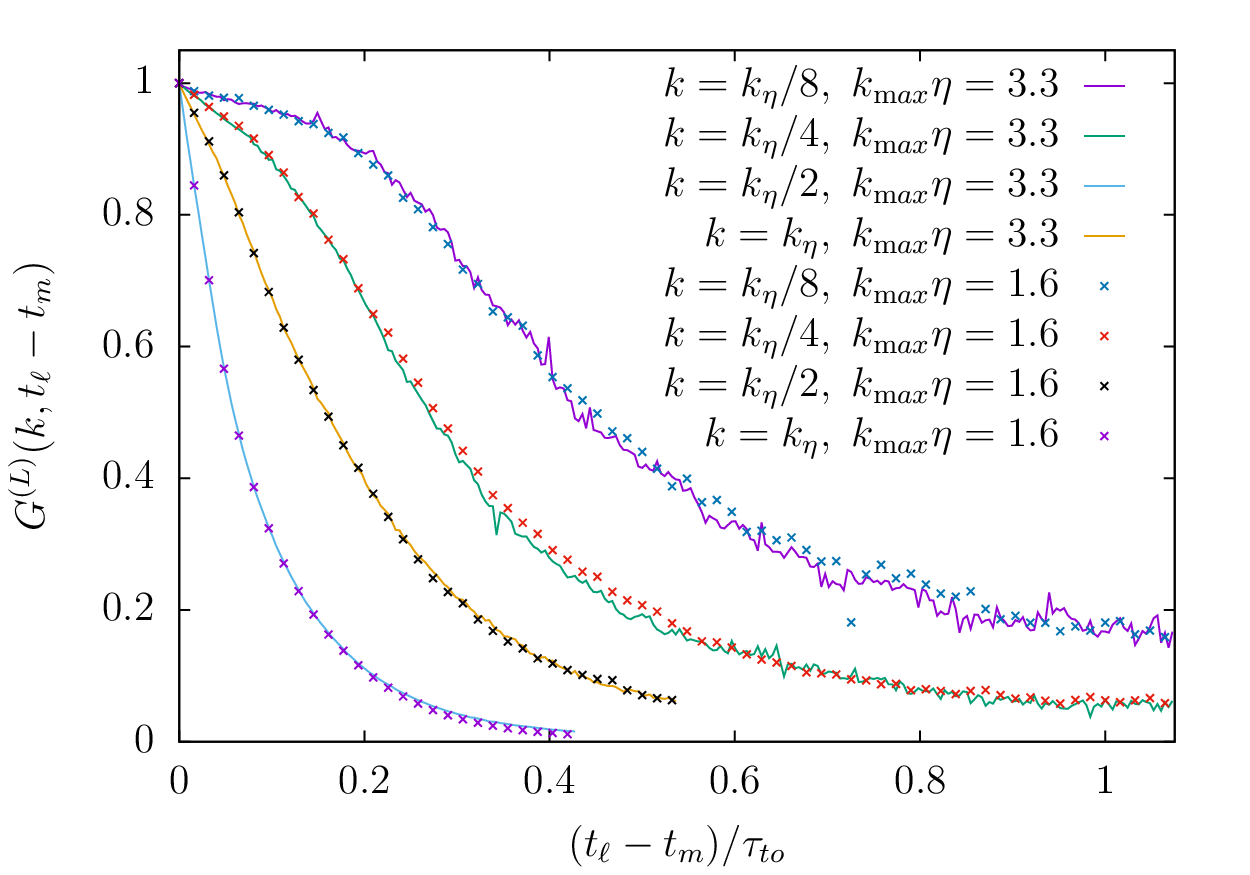}}
 \caption{\label{rslg}Resolution dependence of the Lagrangian linear response function
 for $R_\lambda = 70$. The Kolmogorov dissipation wavenumber is here $k_\eta = 37$
 For simplicity, we take $k_\eta / 2, k_\eta / 4$ and $k_\eta / 8$ as $k = 18, 9$ and $5$, respectively.
 The truncation wavenumbers of the two resolutions are $k_{\rm max} = 60$ and $120$.}
\end{figure}

Before discussing results shown in figure \ref{lcg}, 
let us first consider effect of the truncation error of the passive vector method.
As shown in figure \ref{lcg}, it takes about one large-scale turnover time, which is about $\tau_{to} = 1.80$,
for the correlation function to decrease to $20$\% of the value at the origin $t_\ell - t_m = 0$ 
for the small wavenumber $k = k_\eta / 16$.
The question is then, with the resolution $k_{\rm max} \eta = 1.50$, whether we can trust 
the computed correlation function and response function up to this time lag, $t_\ell - t_m \simeq \tau_{to}$.
To answer this question, we do the following resolution study. 
We decrease the Reynolds number to $R_\lambda = 130$ by setting
the kinematic viscosity to $\nu = 1.34\times 10^{-3}$. We compute this case
with two resolutions, $256^3$ and $512^3$ grid points.
The two simulations yield $k_{\rm max}\eta = 1.50$ and $3.00$, respectively. 
The large-scale turnover time is about $1.74$.
We compare the correlation functions  calculated with 
the two resolutions in figure \ref{rslc}. We observe that the correlation functions 
agree well up to about one large-scale turnover time. 
Regarding the Lagrangian response function, whose numerical calculation is costly,
we further decrease the Reynolds number to $R_\lambda = 70$ ($\nu = 3.75 \times 10^{-3}$)
and calculate the solutions with two resolutions, $128^3$ and $256^3$ grid points. The two simulations
have $k_{\max} \eta = 1.63$ and $3.26$. The large-scale turnover time is now $1.86$.
The comparison of the response function is shown in figure \ref{rslg}.
We observe that the response functions for the small wavenumbers agree well up to one turnover time.
Therefore we infer that $k_{\rm max}\eta = 1.50$ is sufficient to study the correlation function 
and the response function up to one large-scale turnover time in our study. 
In \cite{grhk}, $k_{\rm max}\eta = 2.0$ is recommended though.

\begin{figure}
\centerline{\includegraphics[scale=0.7]{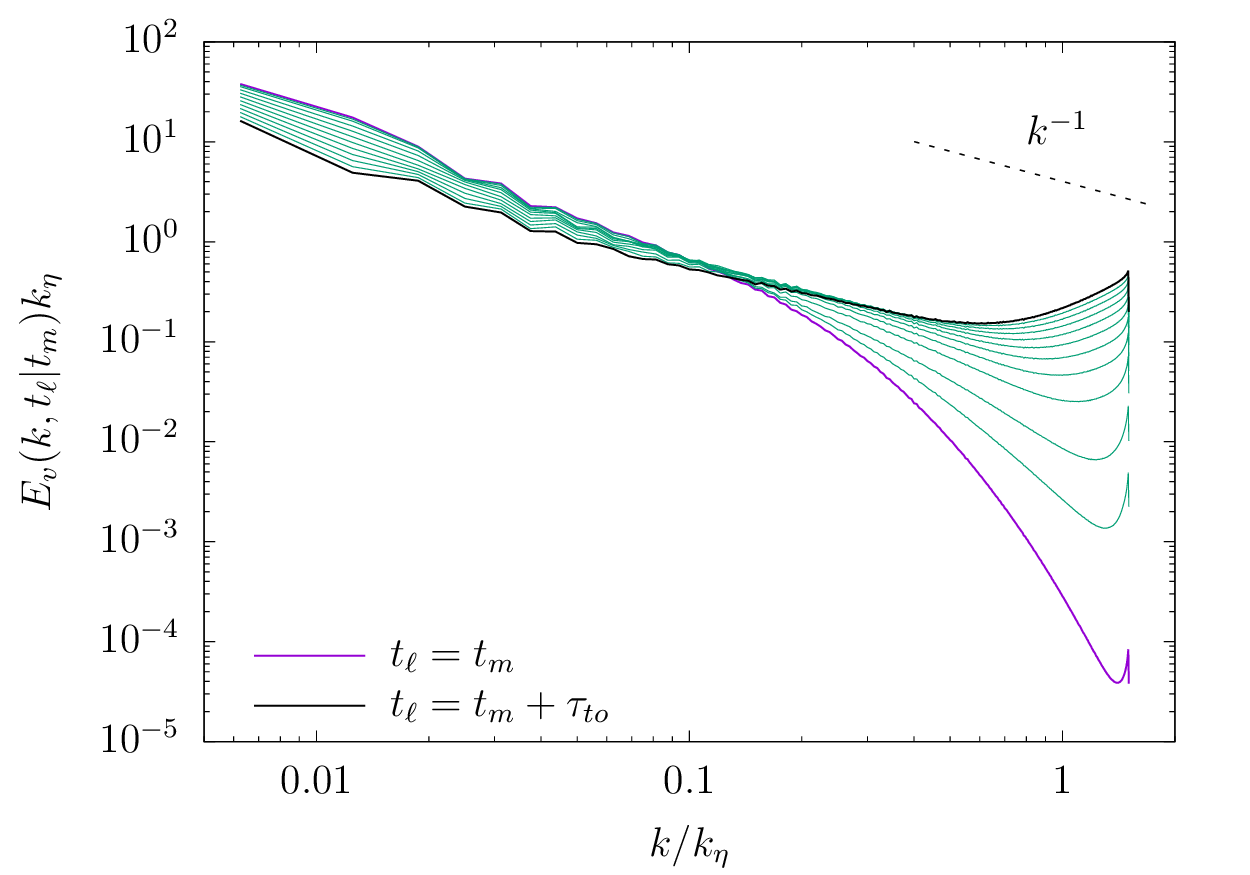}}
 \caption{\label{lagspc}Labelling time evolution of the energy spectrum of the Lagrangian velocity with $R_\lambda = 210$.
 The plotted curves' labelling times correspond to $t_\ell = t_m, ~t_m + 0.05\tau_{to}, ~t_m + 0.1 \tau_{to}, ~t_m + 0.2 \tau_{to}, \ldots, ~t_m + 1.0\tau_{to}$. We take neither temporal nor ensemble averages for each curve.}
\end{figure}

Having checked that the correlation function and the response function are
reliable up to $t_\ell - t_m \simeq \tau_{to}$, we now list observations from figure \ref{lcg}.
First, the Lagrangian correlation functions decrease more slowly than the Eulerian ones, which 
will be analysed quantitatively in the next section. 
Second, the correlation functions at the equal time $t_\ell - t_m = 0$ have positive slopes (time-derivatives)
and hence
their peaks are shifted from the origin for all the six wavenumbers shown in figure \ref{lcg}.
For small wavenumbers, the slopes at the origin in figure \ref{lcg} are hardly seen as positive, but 
we verify that they are positive by magnifying the figure.
This positive slope at the origin is peculiar. It is caused
by the asymmetry of the Lagrangian correlation function with respect to 
swapping the labelling and measuring times, $t_\ell$ and $t_m$,
as discussed in \citet{K66} and \citet{grhk} for the physical space.
It can be shown that the slope at the origin, 
$\partial_{t_\ell} C_{jj}^{(L)}(\vec{k}, -\vec{k}, t_\ell|t_\ell, t_m)$ as $t_\ell \to t_m$ (from above),
is equal to $\partial_{t_\ell} \langle |\hat{\vec{v}}(\vec{k}, t_\ell|t_m)|^2 \rangle/2$ in the same limit.
In order to observe how the Lagrangian modes change in time,
we show labelling-time evolution of the energy spectrum of the Lagrangian
velocity in figure \ref{lagspc}, which is defined as
\begin{equation}
 E_v(k, t_\ell|t_m) 
  =  
  \sum_{\substack{\vec{k}\\ k \le |\vec{k}| < k + \Delta k}} 
  \frac{1}{2}  |\hat{\vec{v}}(\vec{k}, t_\ell |t_m)|^2,
\end{equation}
where $\Delta k = 1$.
As shown in figure \ref{lagspc}, the Lagrangian spectrum $E_v(k, t_\ell|t_m)$ for large $k ~(\sim k_\eta)$ 
becomes much larger than the initial spectrum that is identical to the Eulerian energy spectrum.
As explained in \citet{K66} and \citet{grhk}, the spectrum $E_v(k, t_\ell|t_m)$ for $k \ge k_\eta$, if it is 
fully numerically resolved, 
grows toward $k^{-1}$ spectrum by the same mechanism as the viscous-convective-range spectrum for the passive scalar.
This growth corresponds to 
$\partial_{t_\ell} \langle |\hat{\vec{v}}(\vec{k}, t_\ell|t_m)|^2 \rangle > 0$  at short times
for large $k$ and this implies $\partial_{t_\ell} C_{jj}^{(L)}(\vec{k}, -\vec{k}, t_\ell|t_\ell, t_m) > 0$ for large $k$.
Therefore
the Lagrangian velocity correlation of large $k$ grows at short times as seen in the top panels of figures \ref{lcg}
and \ref{rslc}.
At the same time, the total kinetic energy of the Lagrangian velocity is conserved.
To respect this conservation, the spectrum $E_v(k, t_\ell|t_m)$ for small $k$ decreases
as seen in figure \ref{lagspc}.
This implies
that $\partial_{t_\ell} C_{jj}^{(L)}(\vec{k}, -\vec{k}, t_\ell|t_\ell, t_m) < 0$ for small $k$ at short times.
Indeed, in our simulation, for the two smallest wavenumbers, $k = 1$ and $2$, 
the slopes of the Lagrangian velocity correlation at the origin are negative (figure not shown).
Lastly, we observe that the FDT, $C^{(L)} \propto G^{(L)}$, is violated also in the Lagrangian coordinates
for all the wavenumbers shown in figure \ref{lcg}. 
In the Lagrangian coordinates, the response functions are larger than the correlation functions
for small wavenumbers. In the Eulerian coordinates, the opposite is true as shown in figure \ref{cge}.
The same tendency as in the Lagrangian coordinates, namely $G > C$, was observed in the shell model \citep{moogn}.
We do not have an explanation for this tendency in the Lagrangian coordinates.
Comparing the present result to that of the ALHDIA, 
we note that the difference between the Lagrangian correlation function and the response function obtained here 
is much larger than that of the ALHDIA for the inertial range, 
which was shown in figure 2 of \citet{K66}. The ALHDIA's two functions  are nearly
identical in the range of the vertical axis, $0.6 \le G^{(L)} \le 1.0$ and then they deviate 
from each other in $G^{(L)} \le 0.6$.
A possible explanation of 
this discrepancy between ours and the ALHDIA's is finite Reynolds number effect since the ALHDIA treated
the inertial-range quantities by setting $\nu = 0$.

Let us here comment on the FRR in the Lagrangian coordinates.
An FRR expression of the Lagrangian response function (\ref{gl}) analogous to the Eulerian FRRs
cannot be obtained by adding the Gaussian random forcing to the right hand side of 
the passive vector equation (\ref{pv}) in the Fourier space. 
This is because
the Lagrangian velocity measured at $t_\ell$, which is in the numerator of (\ref{gl}), is 
not a solution to the passive vector equation.
Instead, to obtain an FRR, we should add the Gaussian noise to the equation of
the Lagrangian velocity $\vec{v}(\vec{a}, t_\ell|s_m)$,
which describes the evolution in the measuring time $s_m$ ranging from $t_m$ to $t_\ell$  
(see (\ref{eqv}) in appendix \ref{s:FRRL}).
Notice that this equation is different from the passive vector equation (\ref{pv}) describing the evolution
in the labelling time. 
Under this setting, the Novikov-Carini-Quadrio FRR and the Harada-Sasa FRR
for the Lagrangian response function are obtained on a formal level as we describe
in appendix \ref{s:FRRL}. 
Contrary to the Eulerian case, numerical computation of these Lagrangian FRRs
as they are is nearly impossible since it requires evaluation of the  position function \citep{LRA}.
This is beyond the scope of our present study. 
Therefore we do not pursue numerical study of the Lagrangian FRRs here.

We end this section by describing the number of samples used in the calculation
of the Lagrangian correlation functions and the response functions.
The correlation functions shown in figure \ref{lcg} are averaged in the following
way. The length of the temporal window for calculating the correlation function 
is $1.67\tau_{to}$ and we take five consecutive windows
from two simulations. Thus the total number of samples is 10.
The response functions shown in figure \ref{lcg} are averaged as follows.
For the large wavenumbers, $k=k_\eta, k_\eta / 2$ and $k_\eta / 4$, the total
number of samples is 20. Here the length of the time windows are 
$0.333\tau_{to}, 0.333\tau_{to}$ and $0.444\tau_{to}$, respectively.
We take the 20 windows consecutively in one simulation. 
For the small wavenumbers, $k=k_\eta/  8 , k_\eta / 16$ and $k_\eta / 32$,
the total number of samples is 20, 32 and 32, respectively. The window lengths
are $0.556\tau_{to}$, $0.667\tau_{to}$ and $0.667 \tau_{to}$, respectively. We place the windows consecutively 
in one simulation. As in the Eulerian case, we check the convergence of the
correlation function and the response function by comparing the average over
the full samples and some smaller samples. Here we decrease the number of the samples
by $1/2$. The difference between the averages between the two sample data is 
within a few percent for both the correlation function and the response function.
In this sense, we regard the correlation function and the response function 
have reached convergence.

In the resolution study, the correlation functions shown in figure \ref{rslc}
are averaged as follows. For the $k_{\rm max}\eta = 1.5$ case with $256^3$ grid points, 
we take
$29$ consecutive windows with the length $1.69\tau_{to}$ from one simulation. Hence the 
total number of samples is $29$. For the $k_{\rm max}\eta = 3.0$ case with $512^3$ grid points, 
we take 5 consecutive windows with the length $3.0$ from one simulation. Hence the 
total number of samples is 5. Although the sample sizes differ by about a factor 6, 
good agreement is observed.
In the resolution study of the response function shown in figure \ref{rslg},
we use $128^3$ and $256^3$ grid points. We take the windows of the following sizes, 
$0.430\tau_{to}, 0.538\tau_{to}, 1.08\tau_{to}$ and $1.08\tau_{to}$, for $k = k_\eta, k_\eta/2, k_\eta / 4$ and $k_\eta / 8$,
respectively. The number of the windows are $30$ for $k = k_\eta, k_\eta/2$ and
$60$ for $k_\eta / 4, k_\eta / 8$.
The windows are placed consecutively in one simulation. This setting is the same for 
the two resolutions.

\section{\label{s:timescale}Scaling of characteristic times associated with the correlation and response functions in the Eulerian and Lagrangian coordinates}

Having measured the correlation functions and the linear response functions
in both Eulerian and Lagrangian coordinates, we evaluate characteristic times associated
with them. The purpose is to identify how they vary as a function of wavenumber.

Here, as a characteristic time, we consider the halving time, at which the function
becomes one-half of the value at the time origin.
There are two reasons for this choice.
One is that the halving time 
of the linear response function of the shell model 
was found to be statistically stable in quantifying its decrease by \citet{bdlv}. 
The other reason concerns the similarity assumption made in solving the integro-differential 
equations obtained in the DIA type closures as we briefly discussed in the Introduction.
For this, we intend to focus on characteristic times representing the inertial range. 
They are intermediate in the sense that they are smaller than the large-scale turnover time 
and larger than the Kolmogorov dissipation time.
Notice that obtaining the two functions up to the large-scale turnover time accurately 
is quite demanding since it requires a huge number of samples. 
The halving time is a good numerical compromise.

\begin{figure}
\centerline{\includegraphics[scale=0.7]{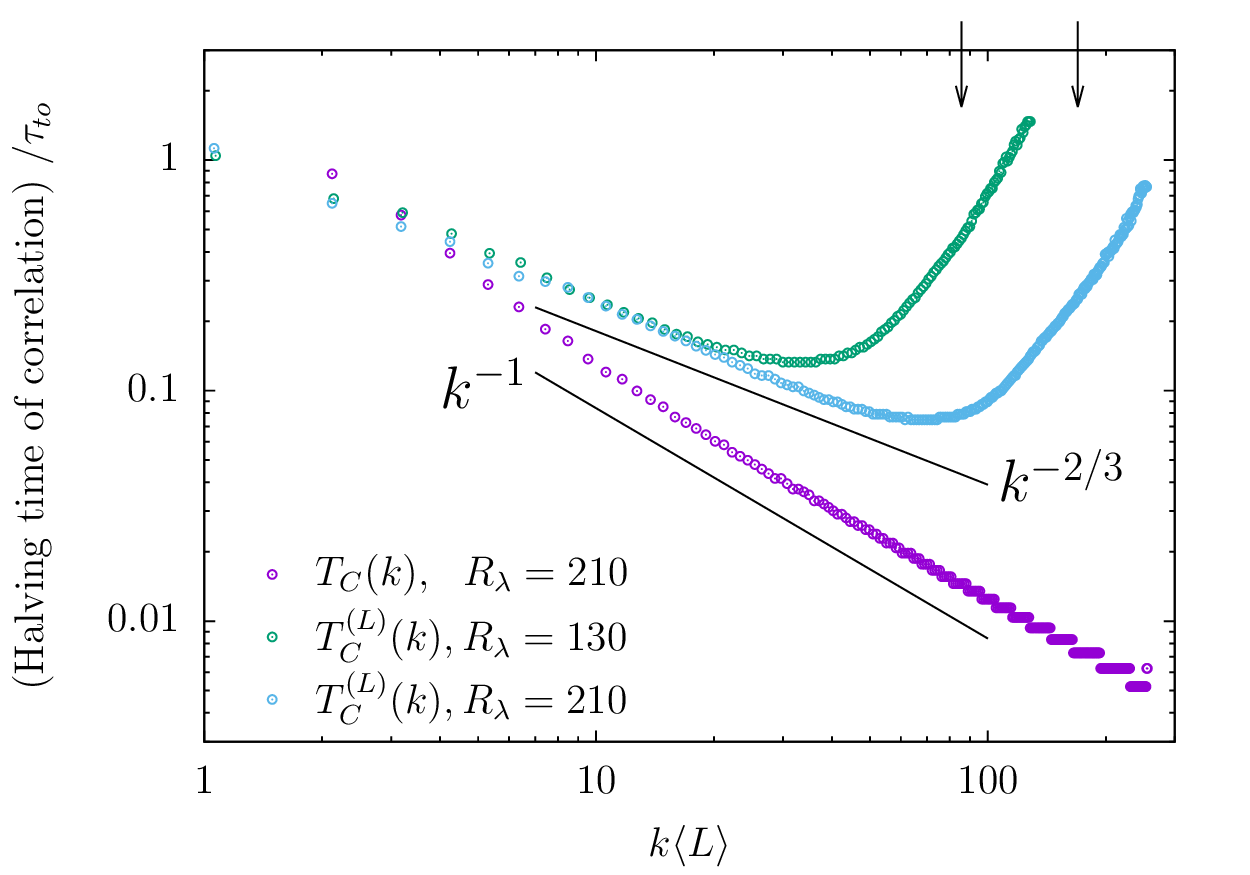}} 
 \caption{\label{htc} Halving time of the correlation function as a function of wavenumber $k$. The halving time of the Eulerian correlation function, $C_{\varphi\varphi}(k, t - s)$, and that of Lagrangian one, $C^{(L)}(k, t_\ell - t_m)$ are denoted by $T_C(k)$ and $T_C^{(L)}(k)$, respectively. For the Lagrangian halving time, we plot also the small Reynolds number case calculated with $k_{\rm max}\eta = 1.5$, whose correlation functions were shown in figure \ref{rslc}. The wavenumber is normalised with the integral scale, $\langle L \rangle \simeq 1.05$, for both Reynolds number cases. The arrows in the top right corner indicate the Kolmogorov dissipation wavenumbers, $k_\eta$, of both cases. The halving time is normalised with the large-scale eddy turnover time, $\tau_{to}$.  The staircase-like behaviour of the Eulerian halving time, $T_C(k)$, for large $k$'s is caused by the fact that the sampling of the correlation function becomes too coarse to resolve the halving time accurately. The Eulerian halving time at the smallest wavenumber, $T_C(k=1)$, is not measurable within the size of the temporal window used here since the Eulerian correlation at the smallest wavenumber, $k=1$, does not decrease by one half.}
\end{figure}

We calculate the halving time of the Eulerian correlation function $C_{\varphi\varphi}(k, t - s)$,
which was already shown in figure \ref{cge}, 
and of the Lagrangian correlation function $C^{(L)}(k, t_\ell - t_m)$, which was shown in figures \ref{lcg} and \ref{rslc}.
We denote the Eulerian and Lagrangian halving times of the correlations by $T_C(k)$ and $T^{(L)}_C(k)$,
respectively. The results are plotted in figure \ref{htc}.
Some cautions are needed to interpret the Lagrangian characteristic time, $T^{(L)}_C(k)$.
We show two data sets with different Reynolds numbers for the Lagrangian halving time.
Both exhibit the decreasing part and the increasing part.
As we raise the Reynolds number, the range of the decreasing part in figure \ref{htc}, which behaves as $k^{-2/3}$,
becomes larger. 
Hence we conclude that the Lagrangian characteristic time 
follows the Kolmogorov scaling, $T^{(L)}_C(k) \propto k^{-2/3}$, in the inertial range.

In \citet{kig}, the characteristic time of the Lagrangian velocity correlation at short times were
studied with the Taylor expansion of the correlation function. Although their Lagrangian velocity, 
which evolves in the measuring time, and their definition of the characteristic time
are different from ours, they observed $k^{-2/3}$ behaviour of their characteristic time 
at short times if $k$ is in the inertial range.

We consider  the increasing part of $T^{(L)}_C(k)$ (which is close to $k^2$) as follows.
First, it is in the large wavenumber region, $k\langle L \rangle > 0.47 k_\eta $. Let us look back
at the graphs of the Lagrangian correlation functions shown in the top panel of figure \ref{lcg}.
We observed that the correlation functions for the corresponding wavenumbers, $k = k_\eta$ and $k_\eta / 2$, 
grow at short time $t_\ell - t_m$, as we discussed in the previous section \ref{s:l}.
As a result, we have much larger $T^{(L)}_C(k)$ at large $k$'s than those at small $k$'s
as inferred from the top panel of figure \ref{lcg}.
Because of this steep growth of the correlations, meaning of the halving times for large $k$'s
is likely to be different from that of the halving times for the smaller wavenumbers $k < k_\eta / 2$.
In \citet{kig}, they observed a similar rapid growth in the large $k$ region of their short-time characteristic 
time of the Lagrangian measuring-time evolved velocity correlation. 
They showed that the high-$k$ growth is due to the viscosity from the equations for the Taylor coefficients.
In our case, the rapid-growth of $T^{(L)}_C(k)$ is considered to occur in the large wavenumber range $k \sim k_\eta$,
where the Lagrangian correlation grows at short times.
This wavenumber range corresponds to the ``viscous-convective'' range 
for the Lagrangian history velocity.
Hence, we infer that the high-$k$ growth of $T^{(L)}_C(k)$ is due to lack of dissipation in the passive vector 
equations (\ref{pv}).

In contrast to the behaviour of the Lagrangian halving time $T^{(L)}_C(k)$, 
we observe that the Eulerian halving time follows $T_C(k) \propto k^{-1}$.
Moreover, this $k^{-1}$ behaviour covers not only the inertial range, but also the dissipation range. 
These power laws for the correlation functions 
are as expected and already obtained numerically, see e.g., \citet{K64} and \citet{grhk}.
The Eulerian time scale $T_C(k) \propto k^{-1}$ is known as the sweeping time scaling and the Lagrangian 
time scale $T^{(L)}_C(k) \propto k^{-2/3}$ is the time scale of the Kolmogorov dimensional analysis 
in the inertial range, namely $\epsilon^{-1/3}k^{-2/3}$. 
From figure \ref{htc}, 
we cannot rule out deviations of $T_C(k)$ from $k^{-1}$ and of $T_C^{(L)}(k)$ from $k^{-2/3}$
in the inertial range, which may be due to some subdominant effect or due to
the intermittency effect. The latter effect can be weak, if it exists, since we are treating the second
order moments. The nondimensional constants involved in the characteristic times are estimated 
in the scaling range with the naked eye as $T_C(k) = 1.3 / (k u_{\rm rms})$ and 
$T^{(L)}_C(k) = 0.90 \epsilon^{-1/3} k^{-2/3}$. 
Here $u_{\rm rms}$ is the root-mean square Eulerian velocity.
A similar $k^{-1}$ behaviour was obtained for the characteristic time of the Eulerian velocity correlation
at short times in \citet{kig}.

\begin{figure}
\centerline{\includegraphics[scale=0.7]{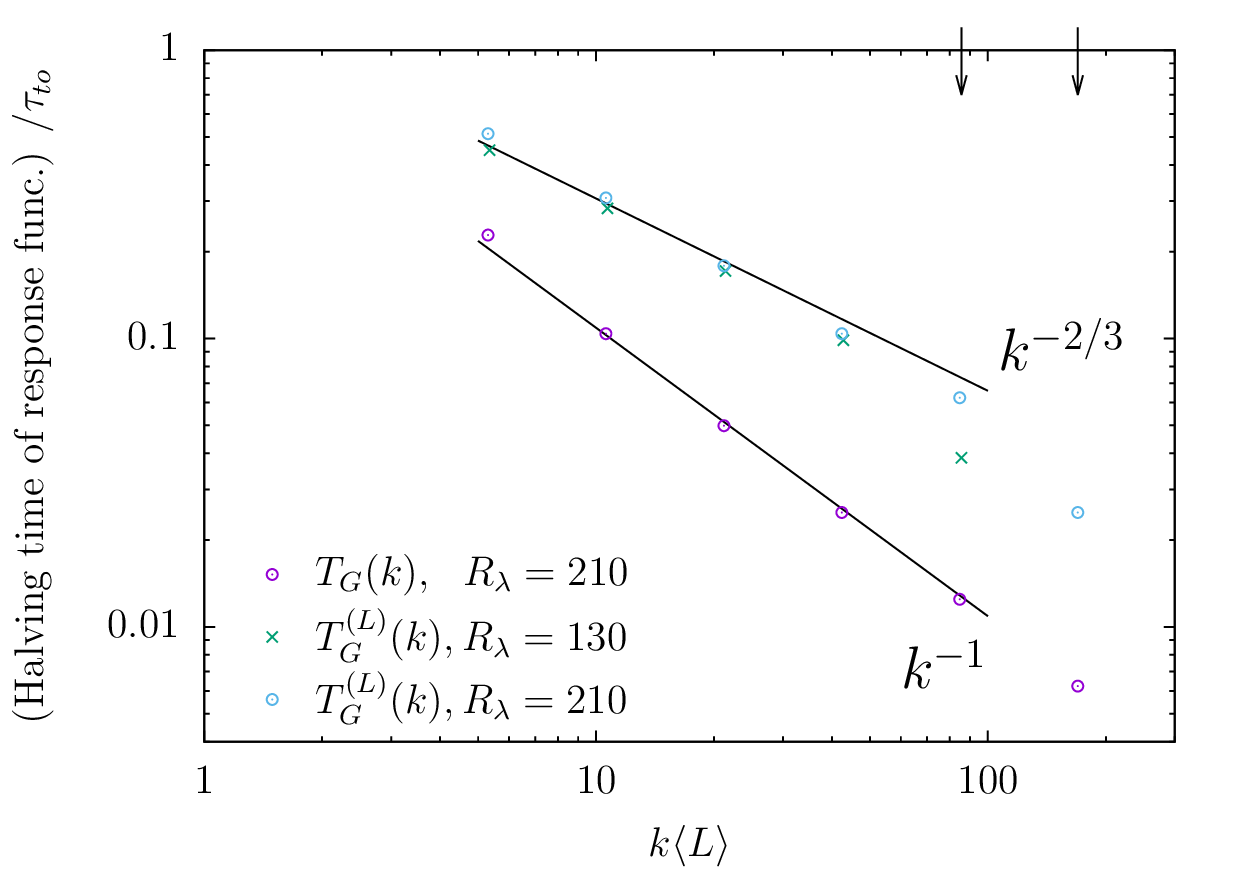}}
 \caption{\label{htr} Halving time of the linear response function as a function of $k$. The halving time of the Eulerian response function, $G_{\varphi\varphi}(k, t - s)$, and that of the Lagrangian one, $G^{(L)}(k, t_\ell - t_m)$, are denoted by $T_G(k)$ and $T_G^{(L)}(k)$, respectively.  As in figure \ref{htc}, we plot the Lagrangian halving time of the small Reynolds number case calculated with $k_{\rm max} \eta = 1.5$.}  
\end{figure}

In figure \ref{htr}, we plot the halving time of the Eulerian response function $G_{\varphi \varphi}(k, t - s)$,
which was already shown in figure \ref{cge} and of the Lagrangian response function $G^{(L)}(k, t_\ell - t_m)$,
which was shown in figure \ref{lcg}.
We denote their time scales as $T_G(k)$ and $T^{(L)}_G(k)$, respectively. 
The number of the sampling wavenumbers are only six since the numerical calculation of the response functions 
are very costly.
The variation of the halving time of the Eulerian
response functions is $T_G(k) \propto k^{-1}$ up to the dissipation range.
For the Lagrangian response functions, we observe that the variation is
close to $T_G^{(L)}(k) \propto k^{-2/3}$ except for the rightmost data for each $R_\lambda$, 
which are probably affected by the viscous dissipation.
We also notice for $T_G^{(L)}(k)$ that small deviations from $k^{-2/3}$ is present.
The nondimensional constants in the scaling range are estimated with the naked eye
as $T_G(k) = 1.1 / (k u_{\rm rms})$ and $T^{(L)}_G(k) = 1.1 \epsilon^{-1/3} k^{-2/3}$.

We conclude that  the time scale of the response function obeys
the same scaling laws as that of the correlation function of the corresponding coordinates.
While we have seen that the FDT, $C\propto G$, holds in neither coordinates, the characteristic 
times of the two functions follow the same scaling laws in $k$.
This numerical result supports the self-similar assumptions made
upon solving the integro-differential equations of the Eulerian DIA and the ALHDIA.
To check robustness of these scaling laws, instead of the halving time,
we calculate the $3/4$-time at which the correlation function
and the response function decrease to $3/4$ of the value at the time origin. Using the $3/4$ time, 
we observe the same scaling laws as shown in figures \ref{htc} and \ref{htr}.

In the numerical calculation of the Harada-Sasa FRR in the Eulerian coordinates,
we pointed out the twofold cancellation. In the description of the cancellation, 
we presented figure \ref{hsc} where  the triple correlation
$L_\varphi(k, t, s)$ has a maximum and that $L_\varphi(k, s, t)$ has a minimum for $t > s$.
A characteristic time of $L_\varphi(k, t, s)$ can be obtained
as the time at which $L_\varphi(k, t, s)$ becomes maximum or as the time in which
$L_\varphi(k, s, t)$ becomes minimum. We plot these times as a function of $k$ and found that 
they vary as $k^{-1}$ (figure not shown). It indicates that this characteristic time of the energy transfer 
correlation $L_\varphi(k, t, s)$ has the sweeping scaling as the Eulerian correlation and response 
functions.
We also plot the maximum values of $L_\varphi(k, t, s)$ and the absolute
values of the minimum of $L_\varphi(k, s, t)$ as a function of $k$ and find that they
do not follow a power law of $k$. 

\section{\label{s:dis}Discussion}
 
In the Eulerian coordinates, we have calculated the linear response function in the deterministic case
with the method used in \cite{bdlv}. We have called it the direct method. 
In contrast, for the two FRRs, we have needed to add the Gaussian random forcing.
If the random forcing is sufficiently small, we have showed that the two FRRs agree
with the response function in the deterministic setting calculated
with the direct method, at least for large wavenumbers.

The following question then arises: which way of calculating the response function in
the deterministic setting is the numerically best for the Navier-Stokes turbulence?  
In terms of yielding the least fluctuating results, the direct method is the best
and the Novikov-Carini-Quadrio FRR is the second. Due to the twofold cancellations, 
the Harada-Sasa FRR is the worst. 
However there is a price to pay for each approach.
The FRRs allow one to compute the shell-averaged response function
for all of the wavenumber shells at one time in principle (apart from the statistical convergence).
The drawback is that we need to identify how small the noise should be
in order for the response function with the noise to agree with the response function without the noise.
The direct method requires to simultaneously 
follow two neighbouring solutions of the Navier-Stokes equations to compute one shell-averaged
response function for a given wavenumber. To obtain the response function
for a different wavenumber shell, we have to repeat this calculation by changing the shell
to which we add the initial perturbation.
This is more numerically expensive than following one solution of the randomly 
forced Navier-Stokes equations. 
However, our evaluation that the Harada-Sasa FRR underperforms in calculation of 
the linear response function of the Navier-Stokes turbulence ignores its physical 
significance in the microscopic systems, which connects
the hard-to-measure heat dissipation to other easy-to-measure statistical quantities 
and identifies the dissipation as the deviation from the FDT 
(for this, see, e.g., \citet{PSV17}).

Apart from the numerical convergence problem of the FRRs, 
the next question we address is the following: 
can the FRRs yield a better understanding of turbulence otherwise unavailable?
At present, we are not able to answer yes.
Our numerical results suggested that the FRRs contain more information than the FDT does.
In particular, the twofold cancellations of the Harada-Sasa
FRR indicated that the deviation from the FDT is caused by both the nonlinearity and 
the dissipation. Whether or not the effect of the dissipation diminishes as we increase
the Reynolds number, as suggested by the shell model study \citep{moogn},
remains to be seen.
If the nonlinear part of the Harada-Sasa FRR, which corresponds to the energy transfer 
function at equal times, becomes dominant at high Reynolds numbers, it is tempting to connect
the breakdown of the FDT with the energy cascade.

Granted that our current use of the FRRs does not improve closure approximations,
the FRRs' closed expressions 
of the response function, such as (\ref{J}) and (\ref{H}), in which no closure approximation
is made, stimulate further analysis or comparison concerning assumption or consequence of 
a closure theory. 
As one example of such attempts, in appendix \ref{s:shortNCQL} we study
a short-time behaviour of the response function with the Novikov-Carini-Quadrio FRR
by applying the field-theoretical method as used, for example, in \cite{reichman}.
Such studies can be done not only in the Eulerian coordinates but also in the Lagrangian
coordinates  with the both time orderings $t_\ell \ge t_m$ and $t_\ell \le  t_m$. 
As shown in appendix \ref{s:shortNCQL}, we express the temporal Taylor expansion
of the response functions, more precisely the Novikov-Carini-Quadrio FRRs,
up to the second order in terms of the instantaneous two-point correlation 
functions. These expressions become independent of the temperature of the noise at short times, 
suggesting that such an analysis is meaningful also for studying the response function 
in the deterministic setting.
Therefore, we consider that this line of research can yield an important insight
only obtainable with the FRR.
With the short-time expansion of the response function, we can study the dominant 
time scale of the response function in the inertial range. 
The results suggest that the sweeping scaling is dominant at short times
for the Eulerian response function (appendix \ref{s:shortNCQE}) and 
that the Kolmogorov scaling is dominant for Lagrangian response function 
in $t_m > t_\ell$ (appendix \ref{s:shortNCQL}). However for the Lagrangian 
history response function $t_\ell > t_m$,
the time scale at short times depends on the ultra-violet cut-off wavenumber (appendix \ref{s:shortNCQL}).
How the Kolmogorov scaling becomes dominant in the half-life of the Lagrangian history response function
as observed in section \ref{s:timescale}, is an interesting theoretical problem.

Another way of going further with the FRRs can be to relax the constraint of
the sufficiently small noise. With a moderately large noise, the statistical quantities
of the velocity field like the energy spectrum differ measurably
from the deterministic case. If we tolerate this discrepancy, 
we can explore non-equilibrium characters using the FRRs (and recent fluctuation relations)
numerically with various techniques of the statistical mechanics.
It should be noticed that, without the random forcing,
obtaining a FRR-like exact expression of the linear response function
in terms of the correlation functions is a challenge.

We have calculated the Lagrangian correlation function and the linear
response function with the passive
vector method. Despite the large truncation error of the Lagrangian velocity, we have
obtained reliable numerical results from small to moderate wavenumbers. The characteristic times
of the Lagrangian functions measured as the halving times obey the Kolmogorov scaling,
$k^{-2/3}$, in the inertial range. This supports the assumption made in solving
the integro-differential equations of the ALHDIA.
In contrast, the characteristic times of the Eulerian correlation and response functions
have the sweeping scaling $k^{-1}$. This Eulerian result 
is also consistent with the analysis made upon studying the failure of the Eulerian
DIA, see, e.g., \cite{K64}.
There was an attempt to circumvent the failure within the Eulerian 
framework by assuming that the Eulerian response function has 
the Kolmogorov scaling $k^{-2/3}$ as the characteristic time, see section 6.4 of \citet{Leslie}. 
This assumption is not valid according to the numerical result obtained here.

We have observed that the FDT, the proportionality between the correlation 
function and the linear response function, does not hold in the Eulerian coordinates
and the ``abridged Lagrangian history coordinates'' where
the labelling time is the present and the measuring time is the past, $t_\ell \ge t_m$.
This is a manifestation of non-Gaussian, non-equilibrium
statistical mechanical character of turbulence \citep{MPRV08}.
These breakdowns of the FDT are consistent with the original Eulerian DIA and the ALHDIA,
although the two DIAs lead to the FDT in the absolute equilibrium case or the fully 
thermalized Galerkin-truncated Euler case.
On the other hand, the characteristic times of the correlation and linear response functions 
obey the same power-law scaling. 
In this sense, the discrepancy may not be taken so seriously.

In fact, our original motivation of this study was to numerically examine the FDT that holds
as a consequence of the LRA and the LDIA.  However, the Lagrangian quantities used in the LRA
and the LDIA are hard to calculate with the spectral accuracy. 
To aim at the accuracy we hence employ the passive vector method.
Then the price to pay for the accuracy is the time ordering, $t_\ell \ge t_m$.   
In the LRA and the LDIA, the ordering is opposite, i.e., $t_m \ge t_\ell$.
Numerical study of the same ordering of the LRA and the LDIA can be done by adopting the Lagrangian particle tracking method 
as pioneered by \citet{YP}. We speculate that a similar breakdown of the FDT occurs in $t_m \ge t_\ell$
and that the characteristic times of the two functions follow the same scaling $k^{-2/3}$.
As a further remark on the FDT in the LRA and the LDIA, the solenoidal projection of
the measuring-time evolving Lagrangian velocity and the linear response function 
leads to the simplified closure equations as described in, e.g., \citet{kaneda2007}. 
For the compressible components, the FDT may be broken even within the LRA and the LDIA.

For the Eulerian correlation function and the linear response function, we have estimated
their characteristic times as $T_C(k) = 1.3 /(k u_{\rm rms})$ and $T_G(k) = 1.1/(k u_{\rm rms})$,
which are defined as halving time.
Here one of the referees pointed out that the Gaussian function $\exp[-k^2 u_{\rm rms}^2(t - s)^2/2]$ 
has the halving time
$\sqrt{2\log 2}/(k u_{\rm rms}) \simeq 1.18/(k u_{\rm rms})$, which is close to the estimated
characteristic times.
Certainly, this  Gaussian function with the sweeping scaling yields a fair approximation
both to the Eulerian correlation functions and the linear response functions shown in figure \ref{cge},
provided that we ignore the breakdown of the FDT.

\section{\label{s:cr}Concluding remarks}
We have numerically studied the correlation function and the linear response function
both in the Eulerian and Lagrangian coordinates in homogeneous isotropic
turbulence with moderate Reynolds numbers.

In the Eulerian coordinates,
the directly measured response function was compared numerically 
with the two FRRs.
With the sufficiently small amplitude of the Gaussian random forcing, 
the two FRRs agreed with the the directly measured response function in the deterministic
setting for the large and moderate wavenumbers.
For the small wavenumbers we expect that they will agree better by increasing
the number of statistical samples.
In the Lagrangian coordinates,
the correlation function and the linear response function were numerically obtained
with the passive vector method for the time ordering $t_\ell \ge t_m$.
In particular, the Lagrangian response function was calculated with DNS for
the first time. The Lagrangian FRRs were considered only theoretically in appendices
since they involve the position function that is beyond the scope of the present
DNS study.

Having calculated the two functions in both coordinates, we studied the characteristic
times of them as a function of the wavenumber. The Eulerian times
obey the sweeping scaling, $k^{-1}$. 
The Lagrangian times follow the Kolmogorov scaling, $k^{-2/3}$, in the inertial range,
which is consistent with the assumption of the ALHDIA. All these results in both coordinates
are as expected. However, these scaling laws of the characteristic times in the Lagrangian coordinates
were verified numerically for the first time.
To illustrate a possible use of the FRRs, in appendices we calculate theoretically
the time scales of the FRR expression of the response function at short times and discuss
their dominant scaling in the inertial range. 

We have considered the Eulerian and Lagrangian velocity statistics separately
and have not addressed how they are related each other. The problem is
a substantial challenge as pointed out by \citet{hrev} among many issues 
reviewed therein about two-point and two-time velocity correlations.
An exact relation between the Eulerian and Lagrangian correlation functions of 
the velocity Fourier modes can be obtained by using the passive vector equation or 
the position function.
The exact one can then be reduced to a closed relation between, for example, 
the Eulerian and Lagrangian characteristic times. 
To do this, we need to apply a closure approximation to the exact relation which 
involves third-order correlations, which is beyond the scope of this paper.

The linear response function has been employed mostly in DIA-type closure
approximations. Our present study here is not directly relevant to developing
a new closure approximation that is manageable for inhomogeneous
and anisotropic turbulence. We recall that the classical role of the linear response function
with the FDT is to describe how a system in the thermally equilibrium state responds to a small
perturbation and how it comes back to the equilibrium state.
Much beyond the classical role, the FRRs have been developed to describe non-equilibrium systems
which include Navier-Stokes turbulence, as we studied.
We hope that these non-equilibrium  FRRs will reveal unknown non-equilibrium character of
turbulent flow and lead to its better understanding.

\backsection[Acknowledgements]{We acknowledge stimulating discussions with So Kitsunezaki and Yukio Kaneda. This work is also supported by the Research Institute for Mathematical Sciences (RIMS) in Kyoto University. We thank the anonymous referees for critical reading, suggestions, and thoughtful comments.}

\backsection[Funding]{This research is funded by Grants-in-Aid for Scientific Research 
KAKENHI (C) No.~24540404 and (C) No.~18K03459 from JSPS.}

\backsection[Declaration of interests]{The authors report no conflict of interest.}

\appendix

\section{\label{s:FRRE}Derivation of the Harada-Sasa FRR in the Eulerian coordinates}
We derive the FRR (\ref{H}) by applying the formalism proposed by \citet{hs05, hs06}
to the Navier-Stokes equations (\ref{nsr}).
The starting point of this formalism is the transition probability 
of the Gaussian random forcing, $(\hat{\xi}_\varphi(\vec{k}, t), \hat{\xi}_\theta(\vec{k}, t))$,
from one state at time $t_0$ to another at time $t$. 
It can be written as a path-integral form of the corresponding Brownian paths,
\begin{equation}
 P(\vec{\Xi}_0, t_0|\vec{\Xi}, t)
 = 
\int_{(\vec{\Xi}_0, t_0)}^{(\vec{\Xi}, t)}
 D[\vec{\Xi}]
\exp
\left(
-\frac{1}{2} 
 \sum_{\vec{k}} 
 \int_{t_0}^{t} \d s 
  \frac{
    |\hat{\xi}_\varphi(\vec{k}, s)|^2
    + |\hat{\xi}_\theta (\vec{k}, s)|^2}
  {\sigma^2(k) T}
\right).
\label{pixi}
\end{equation}
Here $\vec{\Xi}$ denotes one instantaneous realisation of the random forcing for all $\vec{k}$'s and
$D[\vec{\Xi}]$ represents a measure associated with a Brownian path.
To consider the linear response, we next 
add a probe force $(f^{(p)}_\varphi(\vec{k}, t), f^{(p)}_\theta(\vec{k}, t))$
to the right hand side of the Navier-Stokes equations (\ref{nsr}).
This probe force is infinitesimally small.
Now we change variables from the random forcing $\vec{\Xi}$ to the velocity $\vec{U}$, which 
denotes all the velocity Fourier modes. This results in
\begin{align}
 P(\vec{U}_0, t_0|\vec{U}, t)
 = &
\int_{(\vec{U}_0, t_0)}^{(\vec{U}, t)}
 D[\vec{U}]
\exp
\bigg[
-
 \sum_{\vec{k}}
\frac{1}{2\sigma^2(k)T}  
 \int_{t_0}^{t} \d s 
  (
 |\dot{\hat{u}}_\varphi(\vec{k}, s) - \Lambda_\varphi(\vec{k}, s) \nonumber \\
& + \nu k^2 \hat{u}_\varphi(\vec{k}, s)
 - f^{(p)}_\varphi(\vec{k}, s)|^2 
 + |\dot{\hat{u}}_\theta(\vec{k}, s) - \Lambda_\theta(\vec{k}, s) + \nu k^2 \hat{u}_\theta(\vec{k}, s) \nonumber \\
& - f^{(p)}_\theta(\vec{k}, s)|^2
  )
\bigg] {\mathcal J},
\label{piu}
\end{align}
where  $\dot{\hat{u}}$ denotes the time derivative of $\hat{u}$
and ${\mathcal J}$ is the Jacobian due to the change of variables.
With this probability, we formally write the mean of $\hat{u}_\varphi(\vec{k}, t)$ in the presence of the probe force, which
are denoted by $\langle \hat{u}_\varphi(\vec{k}, t) \rangle_p$, as
\begin{align}
\langle \hat{u}_\varphi(\vec{k}, t) \rangle_p
= &
\int_{(\vec{U}_0, t_0)}^{(\vec{U}, t)}
 D[\vec{U}]
 \hat{u}_\varphi(\vec{k}, t)
\exp
\bigg[
-
 \sum_{\vec{k}}
\frac{1}{2\sigma^2(k)T}  
 \int_{t_0}^{t} \d s~
  \big(
 |\dot{\hat{u}}_\varphi(\vec{k}, s) - \Lambda_\varphi(\vec{k}, s) \nonumber \\
& + \nu k^2 \hat{u}_\varphi(\vec{k}, s)
 - f^{(p)}_\varphi(\vec{k}, s)|^2 
 + |\dot{\hat{u}}_\theta(\vec{k}, s) - \Lambda_\theta(\vec{k}, s) + \nu k^2 \hat{u}_\theta(\vec{k}, s) \nonumber \\
& - f^{(p)}_\theta(\vec{k}, s)|^2
  \big)
\bigg] {\mathcal J}
\label{pimean}
\end{align}
(if needed, one can further take the average over the initial velocity $\vec{U}_0$ at time $t_0$ by specifying
the probability distribution of $\vec{U}_0$).
We then expand this mean velocity up to the first order of the probe force as
\begin{align}
\langle \hat{u}_\varphi(\vec{k}, t) \rangle_p
\simeq &
\int_{(\vec{U}_0, t_0)}^{(\vec{U}, t)}
 D[\vec{U}]
\exp
\bigg[
-
 \sum_{\vec{q}}
\frac{1}{2\sigma^2(q)T}  
 \int_{t_0}^{t} \d s 
  (
   |\dot{\hat{u}}_\varphi(\vec{q}, s) - \Lambda_\varphi(\vec{q}, s) \nonumber \\ 
& + \nu q^2 \hat{u}_\varphi(\vec{q}, s)|^2  
 + |\dot{\hat{u}}_\theta(\vec{q}, s) - \Lambda_\theta(\vec{q}, s) + \nu q^2 \hat{u}_\theta(\vec{q}, s)|^2
  )
\bigg] {\mathcal J} 
 \times
  \hat{u}_\varphi(\vec{k}, t) \nonumber \\
& 
\times \bigg\{
 1 
+ 
 \sum_{\vec{q}}
 \frac{1}{\sigma^2(q)T}
 \int_{t_0}^{t}
 \d s
  [\dot{\hat{u}}_\varphi(\vec{q}, s) - \Lambda_\varphi(\vec{q}, s) 
 + \nu q^2 \hat{u}_\varphi(\vec{q}, s)]^* f^{(p)}_\varphi(\vec{q}, s)  \nonumber \\
& + [\dot{\hat{u}}_\theta(\vec{q}, s) - \Lambda_\theta(\vec{q}, s) + \nu q^2 \hat{u}_\theta(\vec{q}, s)]^* f^{(p)}_\theta(\vec{q}, s)
\bigg\}.
\label{piexpand}
\end{align}
Here we assume that ${\mathcal J}$ does not contribute to this form and
ignore formally the first-order term of the complex conjugate of the probe force.
Notice that the first term (i.e., the zeroth order term) in the integrand of (\ref{piexpand}) is 
\begin{align}
\langle \hat{u}_\varphi(\vec{k}, t) \rangle
=
\int_{(\vec{U}_0, t_0)}^{(\vec{U}, t)}
 D[\vec{U}]
\exp
\bigg[
-\frac{1}{2\sigma^2(k)T} 
 \sum_{\vec{q}} 
 \int_{t_0}^{t} \d s 
  (
   |\dot{\hat{u}}_\varphi(\vec{q}, s) - \Lambda_\varphi(\vec{q}, s) + \nu q^2 \hat{u}_\varphi(\vec{q}, s)|^2  \nonumber \\
 + |\dot{\hat{u}}_\theta(\vec{q}, s) - \Lambda_\theta(\vec{q}, s) + \nu q^2 \hat{u}_\theta(\vec{q}, s)|^2
  )
\bigg] {\mathcal J} 
 \times
  \hat{u}_\varphi(\vec{k}, t)
\label{piexpand2}
\end{align}
where $\langle \cdot \rangle$ denotes the average under the setting without the probe force.
Since $\delta \hat{u}_\varphi(\vec{k}, t) = 
\langle \hat{u}_\varphi(\vec{k}, t) \rangle_p - \langle \hat{u}_\varphi(\vec{k}, t) \rangle$,
the expansion (\ref{piexpand}) and (\ref{piexpand2}) yield
the expression of the mean linear response function
\begin{equation}
\frac{\delta \hat{u}_\varphi(\vec{k}, t)} 
     {\delta f^{(p)}_\varphi(\vec{q}', s')} 
= \frac{1}{2\sigma^2(q)T} 
  \langle \hat{u}_\varphi(\vec{k}, t) 
[\dot{\hat{u}}_\varphi(\vec{q}', s') - \Lambda_\varphi(\vec{q}', s') \nonumber \\ + \nu q^2 \hat{u}_\varphi(\vec{q}', s')]^* 
  \rangle.
\end{equation}
We can eliminate the time-derivative term for the diagonal component, $\vec{q}' = \vec{k}$,
in the following manner. For $t \ge s$, the diagonal component can be written as
\begin{equation}
\frac{\delta \hat{u}_\varphi(\vec{k}, t)} 
     {\delta f^{(p)}_\varphi(\vec{k}, s)} 
= \frac{1}{2\sigma^2(k)T} 
[ \partial_s   \langle \hat{u}_\varphi(\vec{k}, t) \hat{u}^*_\varphi(\vec{k}, s)  \rangle
 -    \langle \Lambda^*_\varphi(\vec{k}, s) \hat{u}_\varphi(\vec{k}, t) \rangle
 + \nu k^2 \langle \hat{u}_\varphi(\vec{k}, t) \hat{u}^*_\varphi(\vec{k}, s)  \rangle].
\label{diag}
\end{equation}
Now we interchange $t$ and $s$ in (\ref{diag}). 
Because of the causality, the left hand side becomes zero:
\begin{equation}
  0
= \frac{1}{2\sigma^2(k)T} 
[ \partial_t   \langle \hat{u}_\varphi(\vec{k}, s) \hat{u}^*_\varphi(\vec{k}, t)  \rangle
 -    \langle \Lambda^*_\varphi(\vec{k}, t) \hat{u}_\varphi(\vec{k}, s) \rangle
 + \nu k^2 \langle \hat{u}_\varphi(\vec{k}, s) \hat{u}^*_\varphi(\vec{k}, t)  \rangle].
\label{diag2}
\end{equation}
Here $\partial_t   \langle \hat{u}_\varphi(\vec{k}, s) \hat{u}^*_\varphi(\vec{k}, t)  \rangle = - \partial_s   \langle \hat{u}_\varphi(\vec{k}, t) \hat{u}^*_\varphi(\vec{k}, s)  \rangle$
since the autocorrelation function is a function of $t - s$ due to
the statistical steadiness.
Adding the two equations together, the diagonal component becomes
\begin{align}
\frac{\delta \hat{u}_\varphi(\vec{k}, t)} 
     {\delta f^{(p)}_\varphi(\vec{k}, s)} 
= \frac{1}{2\sigma^2(k)T} 
[2  \nu k^2 \langle \hat{u}_\varphi(\vec{k}, t) \hat{u}^*_\varphi(\vec{k}, s)  \rangle
 -  \langle \Lambda^*_\varphi(\vec{k}, s) \hat{u}_\varphi(\vec{k}, t) \rangle
 -  \langle \Lambda^*_\varphi(\vec{k}, t) \hat{u}_\varphi(\vec{k}, s) \rangle],
\label{hderived}
\end{align}
which is the Harada-Sasa FRR in the Eulerian coordinates given in (\ref{H}).

\section{\label{s:FRRL}Derivation of the Novikov-Carini-Quadrio and Harada-Sasa FRRs in the Lagrangian coordinates}
We first recall the measuring-time evolution equations of the Lagrangian velocity, which involve
the position function. The position function introduced by \citet{LRA} is 
\begin{equation}
 \psi(\vec{x}, t_m|\vec{a}, t_\ell) = \delta(\vec{x} - \vec{X}(\vec{a}, t_\ell|t_m)),
\end{equation}
with which the Lagrangian velocity is written through the Eulerian velocity as
\begin{equation}
 \vec{v}(\vec{a}, t_\ell|t_m) = 
\int \d\vec{x}~ \vec{u}(\vec{x}, t_m)  
\psi(\vec{x}, t_m|\vec{a}, t_\ell).
\end{equation}
Let us write the Fourier series of the position function as
\begin{equation}
\psi(\vec{x}, t_m|\vec{a}, t_\ell)
 = \sum_{\vec{k}, \vec{p}}
   \hat{\psi}(\vec{k}, t_m|\vec{p}, t_\ell) {\e}^{\i (\vec{k}\cdot\vec{x} + \vec{p}\cdot\vec{a})}.
\label{fpsi}   
\end{equation}

The measuring-time evolution equation of the Lagrangian velocity in the 
Fourier space can be obtained from the Eulerian Navier-Stokes equations (\ref{ns})
as 
\begin{eqnarray}
 \partial_{t_m} \hat{v}_j(\vec{k}, t_\ell|t_m)
 &=& -(2\pi)^3 \nu \sum_{\vec{p}} p^2 \hat{u}_j(\vec{p}, t_m) \hat{\psi}(-\vec{p}, t_m|\vec{k}, t_\ell) \nonumber \\
&& -\i (2\pi)^3
 \sum_{\vec{p}}\sum_{\substack{\vec{q}, \vec{r} \\ \vec{q} + \vec{r}+ \vec{p} = \vec{0}}}
 \frac{p_j p_a p_b}{p^2}
 \hat{u}_a(\vec{q}, t_m)
 \hat{u}_b(\vec{r}, t_m)
 \hat{\psi}(-\vec{p}, t_m|\vec{k}, t_\ell) \nonumber \\
&&
+ (2\pi)^3 
  \sum_{\vec{p}}
  \hat{F}_j(\vec{p}, t_m) \hat{\psi}(-\vec{p}, t_m |\vec{k}, t_\ell),
\label{eqv}
\end{eqnarray}
where $p = |\vec{p}|$.
The measuring-time evolution of the position function and
the relation between the Eulerian and Lagrangian velocity modes are given by
\begin{eqnarray}
 \partial_{t_m} \hat{\psi}(\vec{k}, t_m|\vec{p}, t_\ell)
 &=& -\i k_j 
 \sum_{\substack{\vec{p}, \vec{q} \\ \vec{k} + \vec{p} + \vec{q} = \vec{0}}}
  \hat{u}_j(-\vec{p}, t_m)
  \hat{\psi}(-\vec{q}, t_m|\vec{p}, t_\ell),
\label{eqpsi}\\
\hat{u}_j(\vec{k}, t_m) 
&=& (2\pi)^3 \sum_{\vec{p}} 
   \hat{v}_j(\vec{p}, t_\ell|t_m) 
   \hat{\psi}(\vec{k}, t_m|-\vec{p}, t_\ell).
\label{v2u}
\end{eqnarray}
This set of the equations (\ref{eqv})--(\ref{v2u}) is the same as derived by, for example, 
\citet{LDIA}. 
Notice that direct numerical simulation of these equations is nearly impossible because
the degrees of freedom of the position function is prohibitively large.
To obtain the FRRs, we consider the Lagrangian linear response function by adding the probe force $g^{(p)}_j(\vec{k}, t_\ell|t_m)$ 
and the Gaussian random force $\hat{\zeta}_j(\vec{k}, t_\ell|t_m)$ to the right hand side of (\ref{eqv}).
This addition of the random force is not the same as adding the Gaussian noise to the velocity
of the Lagrangian particles.

As in the Eulerian case, let us set the mean and the covariance of the random 
forcing to
\begin{eqnarray}
 \langle \hat{\zeta}_j(\vec{k}, t_\ell|t_m) \rangle &=& 0,\\
 \langle \hat{\zeta}_j(\vec{k}, t_\ell|t_m) \hat{\zeta}_n(\vec{p}, t_\ell|t'_m) \rangle
  &=& 2\tilde{\sigma}^2(k) T \delta_{j,n} \delta_{\vec{k}, -\vec{p}}\delta(t_m - t'_m), \label{zetavar}
\end{eqnarray}
where $\tilde{\sigma}^2(k)$ is some function of $k$.
The Novikov-Carini-Quadrio FRR in the Lagrangian coordinates for $s \ge s'$ can be written as
\begin{equation}
 \frac{\delta \hat{v}_j(\vec{k}, t_\ell|s)}
      {\delta \hat{g}^{(p)}_n(\vec{q}, t_\ell|s')}
      = \frac{1}{2\tilde{\sigma}^2(k)T}
      \langle \hat{v}_j(\vec{k}, t_\ell|s) \hat{\zeta}^*_n(\vec{q}, t_\ell|s')  \rangle,
\end{equation}
by applying the method of \citet{novi}. 
The Harada-Sasa  FRR in the Lagrangian coordinates 
can also be obtained in the same way as Appendix \ref{s:FRRE},
which reads
\begin{align}
 \frac{\delta \hat{v}_j(\vec{k}, t_\ell|s)}
 {\delta g^{(p)}_n(\vec{q}, t_\ell|s')}
 = 
 \frac{1}{2\tilde{\sigma}^2(k) T}
 \left[
 \langle \hat{v}_j(\vec{k}, t_\ell|s) \{\partial_{s'} \hat{v}^*_n(\vec{p}, t_\ell|s') \} \rangle
 - \langle \tilde{\Lambda}_n^*(\vec{k}, t_\ell|s') \hat{v}_j(\vec{k}, t_\ell|s) \rangle 
\right].
\end{align}
Here $\tilde{\Lambda}_j(\vec{k}, t_\ell|t_m)$ denotes the right hand side of (\ref{eqv}).

Let us narrow down these expressions to those of the diagonal component ($j = n$ and $\vec{p} = \vec{k}$)
with $s = t_\ell$ and $s' = t_m$ so that they become consistent with the response function (\ref{sgl}).
The Novikov-Carini-Quadrio FRR becomes
\begin{eqnarray}
 \frac{\delta \hat{v}_j(\vec{k}, t_\ell|t_\ell)}
      {\delta \hat{g}^{(p)}_j(\vec{k}, t_\ell|t_m)}
      &=& \frac{1}{2\tilde{\sigma}^2(k)T}
      \langle \hat{u}_j(\vec{k}, t_\ell) \hat{\zeta}^*_j(\vec{k}, t_\ell|t_m) \rangle.
\label{dlj}
\end{eqnarray}
Here we do not take summation over the index $j$.
Although this looks simple,
numerical calculation of the right hand side of (\ref{dlj}) requires solving (\ref{eqv}) with the random forcing
(more precisely, equation (\ref{eqvz}) or (\ref{nszeta})). This is not an easy task.
The Harada-Sasa FRR for the diagonal component becomes
\begin{align}
 \frac{\delta \hat{v}_j(\vec{k}, t_\ell|t_\ell)}
      {\delta \hat{g}^{(p)}_j(\vec{k}, t_\ell|t_m)}
 =  \frac{1}{2\tilde{\sigma}^2(k) T}
 \left[
 \partial_{t_m} \langle \hat{u}_j(\vec{k}, t_\ell)  \hat{v}^*_j(\vec{k}, t_\ell|t_m) \rangle
 - \langle \tilde{\Lambda}_j^*(\vec{k}, t_\ell|t_m) \hat{u}_j(\vec{k}, t_\ell) \rangle 
\right].
\label{dlh}
\end{align}
We cannot eliminate the time derivative term in (\ref{dlh}) by using the same causality and symmetry argument made
in the Eulerian case. Notice that the second term in (\ref{dlh}) involves correlations between the position function
and the Eulerian velocity.

\section{\label{s:shortNCQE}Short-time expansion of the Novikov-Carini-Quadrio FRR in the Eulerian coordinates}

We consider the Taylor expansion of the Novikov-Carini-Quadrio FRR in the Eulerian coordinates
at short time difference $\varepsilon \ge 0$,
\begin{equation}
 G^{(T)}_{\alpha \alpha}(\vec{k}, t + \varepsilon| -\vec{k}, t)
 = \frac{1}{2\sigma^2(k)T} \langle \hat{u}_\alpha(\vec{k}, t + \varepsilon) \hat{\xi}_\alpha(-\vec{k}, t) \rangle
 = a_0 + a_1 \varepsilon + a_2 \varepsilon^2 + O(\varepsilon^3),
\label{Tege}
\end{equation}
where we do not take summation over the index $\alpha$ which is either $\varphi$ or $\theta$.
We use the standard field-theoretical technique, although we do not employ a renormalisation procedure.
In particular, we write the coefficients $a_0, a_1$ and $a_2$ in terms of the equal-time correlation functions
of the Eulerian velocity Fourier modes. Then we argue that the dominant scaling behaviour 
is $a_2 \propto k^2$ in the inertial range. This shows that the dominant time scale at short time
is given by the sweeping scaling $k^{-1}$, as expected. 
In \citet{kshortc} and \citet{kig}, the Taylor expansion of the Eulerian and Lagrangian velocity correlations 
with the LRA and DNS data has been studied to identify the characteristic times from the second-order coefficients.
In the current and next appendices, we consider the Taylor expansion of the Eulerian and Lagrangian 
linear response functions without using DNS data.

Now let us consider the Navier-Stokes equations in the following form
\begin{align}
 \partial_t \hat{u}_\alpha(\vec{k}, t) 
  =&  (\vec{e}_\alpha(\vec{k}))_j
  \left(-\frac{{\rm i}}{2}\right)
  P_{jlm}(\vec{k}) 
  \sum_{\substack{\vec{p}, \vec{q} \\ \vec{p} + \vec{q} + \vec{k} = \vec{0}}}
  \hat{u}_l(-\vec{p}, t) \hat{u}_m(-\vec{q}, t)     
  - \nu k^2 \hat{u}_\alpha(\vec{k}, t)  \nonumber \\
 & + \hat{F}_\alpha(\vec{k}, t)
   + \hat{\xi}_\alpha(\vec{k}, t),
\label{nsrm}
\end{align}
where $\hat{F}_\alpha(\vec{k}, t)$ is the large-scale forcing term.
The Duhamel-type formal solution of (\ref{nsrm}) is
\begin{equation}
 \hat{u}_\alpha(\vec{k}, t + \varepsilon)
  = \e^{- \nu k^2 \varepsilon} 
  \hat{u}_\alpha(\vec{k}, t)
  + \int_{t}^{t + \varepsilon} 
    \e^{-\nu k^2 (t + \varepsilon - s)} 
    [\Lambda_\alpha(\vec{k}, s) + \hat{\xi}_\alpha(\vec{k}, s)] \d s,
\label{duhamel-e}
\end{equation}
where $\Lambda_\alpha(\vec{k}, s)$ denotes the nonlinear and
the large-scale forcing terms in (\ref{nsrm}).
Using this form, the correlation between  
the velocity Fourier mode and the random forcing 
in the Novikov-Carini-Quadrio FRR can be written as
\begin{align}
 \langle \hat{u}_\alpha(\vec{k}, t + \varepsilon) \hat{\xi}_\alpha(-\vec{k}, t) \rangle
 &= 
 \e^{-\nu k^2 \varepsilon}
 \langle \hat{u}_\alpha(\vec{k}, t) \hat{\xi}_\alpha(-\vec{k}, t) \rangle \nonumber \\
&+ \int_{t}^{t + \varepsilon} \d s~
    \e^{-\nu k^2(t + \varepsilon - s)} 
 [ \langle \Lambda_\alpha(\vec{k}, s) \hat{\xi}_\alpha(-\vec{k}, t) \rangle 
 + \langle \hat{\xi}_\alpha(\vec{k}, s) \hat{\xi}_\alpha(-\vec{k}, t) \rangle ]. 
\label{ncqcor}
\end{align}
In this calculation, we interpret the stochastic differential equation (\ref{nsrm}) in the sense of It\^{o}.
So that we have
\begin{align}
 \langle \hat{u}_\alpha(\vec{p}, t) \hat{\xi}_\alpha(-\vec{k}, t) \rangle = 0,
\label{insta}
\end{align}
(which means that the noise generated at time $t$ is independent of the velocity at the same time)
and 
\begin{align}
 \int_{t}^{t + \varepsilon} 
   \e^{-\nu k^2 (t + \varepsilon - s)}
 \langle \hat{\xi}_\alpha(\vec{k}, s) \hat{\xi}_\alpha(-\vec{k}, t) \rangle ~\d s
 = \e^{-\nu k^2 \varepsilon} 2 \sigma^2(k) T
\end{align}
from the covariance (\ref{nv}).
Therefore the velocity-noise correlation (\ref{ncqcor}) can be calculated as
\begin{align}
 \langle \hat{u}_\alpha(\vec{k}, t + \varepsilon) \hat{\xi}_\alpha(-\vec{k}, t) \rangle
 = \e^{-\nu k^2 \varepsilon}
 \left[2 \sigma^2(k) T 
  + \int_{t}^{t + \varepsilon} 
    \e^{-\nu k^2 (t - s)} 
 \langle \Lambda_\alpha(\vec{k}, s) \hat{\xi}_\alpha(-\vec{k}, t) \rangle \d s 
\right].
 \label{eulerd}
\end{align}
In what follows, we ignore the viscous term and the large-scale forcing term in order
to discuss the inertial-range behaviour.
The correlation in the integrand of (\ref{eulerd}) is then
\begin{align}
  \langle \Lambda_\alpha(\vec{k}, s) \hat{\xi}_\alpha(-\vec{k}, t) \rangle
   = 
   (\vec{e}_\alpha(\vec{k}))_j 
   \tilde{P}_{jlm}(\vec{k})
   \sum_{\substack{\vec{p}, \vec{q} \\ \vec{p} + \vec{q} + \vec{k} = \vec{0}}}
   \langle
   \hat{u}_l(-\vec{p}, s)
   \hat{u}_m(-\vec{q}, s)
   \hat{\xi}^*_\alpha(\vec{k}, t)   
   \rangle,
 \label{elambda}
\end{align}
where we substitute the corresponding Duhamel solution for $\hat{u}_l(-\vec{p}, s)$ 
and $\hat{u}_m(-\vec{q}, s)$. Here $\tilde{P}_{jlm}(\vec{k}) = (-\i/2) P_{jlm}(\vec{k})$.
By repeating this procedure, we arrive at the expansion of
the correlation (\ref{ncqcor}),
\begin{align}
\langle \hat{u}_\alpha(\vec{k}, t + \varepsilon) \hat{\xi}_\alpha(-\vec{k}, t) \rangle
& = 2\sigma^2 T \nonumber \\
&
 + 4 \varepsilon^2 \sigma^2 T
    (\vec{e}_\alpha(\vec{k}))_j
     (\vec{e}_\alpha(\vec{k}))_c
  \tilde{P}_{jlm}(\vec{k})
  \sum_{\vec{p}}
  \tilde{P}_{mcd}(\vec{p} + \vec{k})
 \langle \hat{u}^*_l(\vec{p}, t)
         \hat{u}_d(\vec{p}, t)  \rangle \nonumber \\
&  + O(\varepsilon^3)
 \nonumber \\
& = 2\sigma^2 T 
  - \varepsilon^2
 \bigg[
  2 k_l (\vec{e}_\alpha(\vec{k}))_d 
  \sum_{\vec{p}} 
  (\vec{p}\cdot\vec{e}_\alpha(\vec{k})) 
  \langle 
   \hat{u}^*_l(\vec{p}, t) 
   \hat{u}_d(\vec{p}, t) 
  \rangle \nonumber \\
&\quad
 + k_l k_d
  \sum_{\vec{p}}
  \left(1 - \frac{2(\vec{p}\cdot\vec{e}_\alpha(\vec{k}))^2}{|\vec{p} + \vec{k}|^2} \right)
  \langle 
   \hat{u}^*_l(\vec{p}, t) 
   \hat{u}_d(\vec{p}, t) 
  \rangle \nonumber \\
&\quad
 - 2 k_d (\vec{e}_\alpha(\vec{k}))_l
  \sum_{\vec{p}}
   (\vec{p}\cdot\vec{e}_\alpha(\vec{k}))
 \frac{\vec{k} (\vec{k} + \vec{p})}{|\vec{k} + \vec{p}|^2}
  \langle 
   \hat{u}^*_l(\vec{p}, t) 
   \hat{u}_d(\vec{p}, t) 
  \rangle
 \bigg]
+ O(\varepsilon^3).
\end{align}
To make it simpler, we further assume that the diagonal parts ($l = d$) are dominant. 
This leads to an expression of the Eulerian response function as
\begin{align}
 G^{(T)}_{\alpha \alpha}(\vec{k}, t + \varepsilon|-\vec{k}, t)
  \simeq 1 -
  \frac{1}{2} 
 \varepsilon^2 k^2 
  \sum_{\vec{p}}
 \left(1 - \frac{2(\vec{p}\cdot\vec{e}_\alpha(\vec{k}))^2}{|\vec{p} + \vec{k}|^2}\right) 
  \langle |\hat{\vec{u}}(\vec{p}, t)|^2 \rangle + O(\varepsilon^3).
 \label{ed1}
\end{align}
Clearly, the first term of the summand in (\ref{ed1}) 
represents the sweeping scaling.
The scaling behaviour of the second term in the summand can be estimated 
in the inertial range by changing the summation to an integral and 
assuming the Kolmogorov scaling law,
$\langle |\hat{\vec{u}}(\vec{p}, t)|^2 \rangle = C_K/(2\pi) \epsilon^{2/3} p^{-11/3}$
in the wavenumber range $(0 < )~k_0 \le p \le k_1$.
Here $C_K$ is the Kolmogorov universal constant and $\epsilon$ is the energy dissipation rate.
The $k_0$ and $k_1$ are the wavenumber cut-off's. 
The second term in the summation can be estimated as follows
\begin{align}
 k^2 \sum_{\vec{p}}
  \frac{(\vec{p}\cdot\vec{e}_\alpha(\vec{k}))^2}{|\vec{k} + \vec{p}|^2}
  \langle |\hat{\vec{u}}(\vec{p}, t)|^2 \rangle
&\simeq
k^2 
\left(\frac{L}{2\pi}\right)^3 \frac{C_K}{2\pi} \nonumber \\
& \qquad \times \int_{k_0}^{k_1} \d p \int_0^{\pi} \d \theta \int_0^{2\pi} \d \varphi
  \frac{p^4 \sin^3\theta \cos^2\varphi}
  {k^2 + p^2 + 2 k p \cos\theta}
  \epsilon^{2/3} p^{-11/3} \nonumber \\
&= 
\frac{1}{2}C_K \epsilon^{2/3}
\int_{k_0}^{k_1} \d p~
 p^{1/3}
 \left[
     \frac{k^2 + p^2}{2 p^2}
  -  \frac{(p^2 - k^2)^2}{4 k p^3}
  \log\frac{p + k}{|p - k|}\right] \nonumber \\
 &=
 - 
\frac{9\sqrt{3}}{35} 
\pi
 C_K
\epsilon^{2/3} k^{4/3} 
+ \mbox{(subleading terms)}.
\label{ed2}
\end{align}
Here we set $\vec{p}\cdot\vec{e}_\alpha(\vec{k}) = p \sin\theta \cos\varphi$,
where we regard $\vec{k} = (0, 0, 1)$ and $\vec{e}_\alpha = (1, 0, 0)$ and the
box size is $L = 2\pi$.
The integral in the second line of (\ref{ed2}) can be calculated analytically.
From that result, we calculate the leading behaviour of (\ref{ed2}) by assuming $k_0 \ll k \ll k_1$.

Therefore, the Novikov-Carini-Quadrio FRR leads to the following 
expansion of the Eulerian response function in the inertial range as
\begin{eqnarray}
 G^{(T)}_{\alpha \alpha}(\vec{k}, t + \varepsilon|-\vec{k}, t)
 &\simeq&
1 
 -
\varepsilon^2
\left[
\langle E \rangle
k^2 
-
\frac{9\sqrt{3}}{35}\pi C_K \epsilon^{2/3} k^{4/3}
\right]
 + O(\varepsilon^3) \nonumber \\
 &\simeq&
1 
 -
\frac{\varepsilon^2}{2}
\left[
 3 u_{\rm rms}^2 k^2 
-
\frac{18\sqrt{3}}{35}\pi C_K \epsilon^{2/3} k^{4/3}
\right]
 + O(\varepsilon^3) \nonumber \\  
\label{eexp}
\end{eqnarray}
where the energy is $\langle E \rangle = \sum_{\vec{p}} \langle |\hat{\vec{u}}(\vec{p}, t)|^2\rangle /2$
and  $u_{\rm rms}$ is the root-mean-square of the velocity defined as
$u_{\rm rms} = (2\langle E \rangle /3)^{1/2}$.  
Hence, the short-time characteristic time of the Eulerian response function
is indeed the sweeping time scaling, $(k u_{\rm rms})^{-1}$.
Here the Kolmogorov time scaling $\epsilon^{-1/3} k^{-2/3}$ is present 
but subdominant. The result (\ref{eexp}) can be compared to a theoretical result in \citet{kshortc},
$G^{(T = 0)} = 1 - (\varepsilon^2/2)[k^2 u^2_{\rm rms} - 1.66 \epsilon^{2/3}k^{4/3}] + O(\varepsilon^3)$,
obtained as the time expansion for the Eulerian velocity correlation function.
For comparison, we need to assume the FDT contrary to what we find in section \ref{s:e}.
The numerical constants in (\ref{eexp}) are larger than those of Kaneda's result.
The prefactor in (\ref{eexp}) of the Kolmogorov time scaling can be estimated as
$(18\sqrt{3}/35) \pi C_K \simeq 2.80 C_K = 4.76$ using an estimate of the Kolmogorov 
constant $C_K = 1.70$.

The viscous term brings a linear term of $\varepsilon$
in the short time expansion of the response function as seen in (\ref{eulerd}).
There is also a viscous correction in the second-order term of $\varepsilon$, which 
are ignored here.

As we have seen, up to the second order, the coefficients of the expansion do not
depend on the noise covariance.
If the expansion (\ref{Tege}) is continued to higher orders of $\varepsilon$, it is expected that
the coefficients involve positive powers of $\sigma^2(k)T$ and consequently that 
the limit $T \to 0$ is not singular.
This may be consistent to the fact that the Novikov-Carini-Quadrio FRR with the small random forcing agrees 
well with the linear response function under the deterministic setting.

\section{\label{s:shortNCQL}Short-time expansion of the Novikov-Carini-Quadrio FRR in the Lagrangian coordinates}
We consider the Taylor expansion of the Novikov-Carini-Quadrio FRR in the Lagrangian coordinates at short times
as in the previous appendix \ref{s:shortNCQE}.
In our DNS study we dealt with the Lagrangian response function only for the ordering, $t_\ell \ge t_m$.
Here we study theoretically both orderings, $t_\ell \ge t_m$ and $t_m \ge t_\ell$.
In this appendix \ref{s:shortNCQL} we first consider the latter ordering and then 
switch to the former which is more complicated.

To be specific, we first consider the expansion for $t_\ell \le t_m = t_\ell + \varepsilon$ ($\varepsilon \ge 0$)
\begin{align}
 G^{(L, T)}_{jj}(\vec{k}, -\vec{k}, t_\ell |t_\ell + \varepsilon, t_\ell)
 &= \frac{1}{3\times 2\tilde{\sigma}^2(k)T}
   \langle \hat{v}_j(\vec{k}, t_\ell|t_\ell + \varepsilon) \hat{\zeta}_j(-\vec{k}, t_\ell|t_\ell) \rangle \nonumber \\
 &= \tilde{a}_0 + \tilde{a}_1 \varepsilon + \tilde{a}_2 \varepsilon^2 + \ldots.
 \label{lgm}
\end{align}
Here we take summation over the index $j$.
In what follows, we express the coefficients $\tilde{a}_0, \tilde{a}_1$ and $\tilde{a}_2$  
in terms of the equal-time (at $t_\ell$) correlation functions of the Eulerian velocity modes
and discuss their dominant scaling behaviour as a function of $k$ in the inertial range. 
We will show the dominant scaling is the Kolmogorov scaling, $\epsilon^{-1/3} k^{-2/3}$,
in the inertial range.

The expansion procedure is the same as in the Eulerian case. 
However, one should be careful of where to add the noise.
As discussed in appendix \ref{s:FRRL}, 
the noise $\hat{\zeta}_j(\vec{k}, t_\ell| t_m)$ is added to the right hand side of 
the measuring-time evolution equation of the Lagrangian velocity (\ref{eqv}), namely
\begin{eqnarray}
 \partial_{t_m} \hat{v}_j(\vec{k}, t_\ell|t_m)
 &=& -(2\pi)^3 \nu \sum_{\vec{p}} p^2 \hat{u}_j(\vec{p}, t_m) \hat{\psi}(-\vec{p}, t_m|\vec{k}, t_\ell) \nonumber \\
&& -\i (2\pi)^3
 \sum_{\vec{p}}\sum_{\substack{\vec{q}, \vec{r} \\ \vec{q} + \vec{r}+ \vec{p} = \vec{0}}}
 \frac{p_j p_a p_b}{p^2}
 \hat{u}_a(\vec{q}, t_m)
 \hat{u}_b(\vec{r}, t_m)
 \hat{\psi}(-\vec{p}, t_m|\vec{k}, t_\ell) \nonumber \\
&&
+ (2\pi)^3 \sum_{\vec{p}}
 \hat{F}_j(\vec{p}, t_m) \hat{\psi}(-\vec{p}, t_m |\vec{k}, t_\ell)
+ \hat{\zeta}_j(\vec{k}, t_\ell|t_m).
\label{eqvz}
\end{eqnarray}
Here $\hat{\psi}(\vec{p}, t_m| \vec{q}, t_\ell)$ is the Fourier coefficient of the
position function \citep{LRA}. See also (\ref{fpsi}).
The formal Duhamel solution to (\ref{eqvz}) is
\begin{eqnarray}
 \hat{v}_j(\vec{k}, t_\ell|t_\ell + \varepsilon) 
= 
\hat{v}_j(\vec{k}, t_\ell|t_\ell)
 + 
\int_{t}^{t_\ell + \varepsilon} 
  [\tilde{\Lambda}_j(\vec{k}, t_\ell|s) + \hat{\zeta}_j(\vec{k}, t_\ell|s)] ~\d s,
\label{vfs}
\end{eqnarray}
where $\tilde{\Lambda}_j(\vec{k}, t_\ell|t_m)$ denotes 
the first three terms on the right hand side of (\ref{eqvz}).
Putting the solution to the velocity-noise correlation (\ref{lgm}), we have
\begin{align}
 \langle 
  \hat{v}_j(\vec{k}, t_\ell|t_\ell + \varepsilon) 
  \hat{\zeta}_j(-\vec{k}, t_\ell|t_\ell) 
  \rangle
  = 
  6\tilde{\sigma}^2(k) T 
+ 
  \int_{t_\ell}^{t_\ell + \varepsilon} 
  \langle 
  \tilde{\Lambda}_j(\vec{k}, t_\ell|s)  
  \hat{\zeta}_j(-\vec{k}, t_\ell|t_\ell) 
  \rangle 
  ~\d s,
\label{vzetacor}
\end{align}
where we use 
$\langle  \hat{v}_j(\vec{k}, t_\ell|t_\ell) \hat{\zeta}_j (-\vec{k}, t_\ell|t) \rangle = 0$
and the noise variance (\ref{zetavar}) as we did in the Eulerian case.

To further calculate the correlation functions in (\ref{vzetacor}), 
we need the evolution 
equation of the Eulerian velocity since $\tilde{\Lambda}_j(\vec{k}, t_\ell|s)$ 
are written in terms of the Eulerian velocity.
The important point here is how the noise in (\ref{eqvz})
is transformed in the equation of the Eulerian velocity. 
That can be obtained by multiplying (\ref{eqvz}) by the position function and 
using (\ref{v2u}) as
\begin{align}
 \partial_{t_m} \hat{u}_j(\vec{k}, t_m)
  =  
  \left(-\frac{{\rm i}}{2}\right)
  P_{jlm}(\vec{k}) 
  \sum_{\substack{\vec{p}, \vec{q} \\ \vec{p} + \vec{q} + \vec{k} = \vec{0}}}
  \hat{u}_l(-\vec{p}, t_m) \hat{u}_m(-\vec{q}, t_m)     
  - \nu k^2 \hat{u}_j(\vec{k}, t_m)  \nonumber \\
  + \hat{F}_j(\vec{k}, t_m)
  + (2\pi)^3 P_{j l}(\vec{k}) \sum_{\vec{p}} \hat{\zeta}_l(\vec{p}, t_\ell|t_m) \hat{\psi}(\vec{k}, t_m|-\vec{p}, t_\ell).
\label{nszeta}  
\end{align}
Here the factor $(2\pi)^3$ is due to our normalisation of the Fourier modes of the position function.

From now on we ignore the large-scale forcing $\hat{\vec{F}}$ and the 
viscous term to concentrate on the inertial-range scaling.
We put the formal Duhamel solution of (\ref{nszeta})
into the integrand of (\ref{vzetacor}). We also use the formal Duhamel solution 
of (\ref{eqpsi}) for the position function. Whenever $\hat{\vec{u}}(\vec{k}, s)$ or
$\hat{\psi}(\vec{p}, s|\vec{q}, t_\ell)$ with $s \ne t_\ell$ appear in correlations, 
we replace them by the Duhamel expressions in order to express them with 
the equal-time correlations at time $t_\ell$.
We utilise the equal-time expression of the position function 
$\hat{\psi}(\vec{p}, t_\ell|\vec{q}, t_\ell) = \delta_{\vec{p}, -\vec{q}}/(2\pi)^3$
as well.
Most of the terms at $\varepsilon^2$ cancel due to the projection operator in 
the noise term in (\ref{nszeta}). 
We then arrive at a rather simple result up to the second order ($\varepsilon^2$):
\begin{align}
 \langle 
  \hat{v}_j(\vec{k}, t_\ell|t + \varepsilon) 
  \hat{\zeta}_j(-\vec{k}, t_\ell|t) 
  \rangle
&  = 
  6\tilde{\sigma}^2(k) T \nonumber \\
& \quad - 2\tilde{\sigma}^2(k) T \varepsilon^2
  P_{jb}(\vec{k})
 \sum_{\vec{p}}
 \frac{p_j p_a p_b p_c}{p^2}
 \langle \hat{u}_c(\vec{p} + \vec{k}, t_\ell) \hat{u}_a(-\vec{p}-\vec{k}, t_\ell) \rangle \nonumber \\
& \quad + O(\varepsilon^3).
\end{align} 
This yields the desired expression of the Lagrangian response function 
\begin{align}
 G^{(L, T)}_{jj}(\vec{k}, -\vec{k}, t_\ell |t_\ell + \varepsilon, t_\ell)
&=
 1
- 
\frac{\varepsilon^2}{3}
k_a k_b
\sum_{\vec{p}}
\left(
 1 - \frac{(\vec{k}\cdot\vec{p})^2}{k^2 p^2}
\right)
 \langle 
 \hat{u}_a^*(\vec{p}+\vec{k}, t_\ell) 
 \hat{u}_b  (\vec{p}+\vec{k}, t_\ell)
 \rangle \nonumber \\
& \quad + O(\varepsilon^3).
 \label{lgsum}
\end{align}

Now we consider the leading scaling behaviour of the coefficient of the quadratic term of $\varepsilon$.
To do this,
we assume first that the diagonal components in (\ref{lgsum}) are dominant,
next that the Kolmogorov energy spectrum holds,
$\langle |\hat{\vec{u}}(\vec{k}, t_\ell)|^2 \rangle = C_K/(2\pi)\epsilon^{2/3} k^{-11/3}$ for $k_0 \le |\vec{k}| \le k_1$.
and finally that the summation can be approximated by an integral.
These assumptions lead to the following leading behaviour for $k_0 \ll k \ll k_1$
\begin{align}
& k_a k_b
\sum_{\vec{p}}
\left(
 1 - \frac{(\vec{k}\cdot\vec{p})^2}{k^2 p^2}
\right)
 \langle 
 \hat{u}_a^*(\vec{p}+\vec{k}, t_\ell) 
 \hat{u}_b  (\vec{p}+\vec{k}, t_\ell)
\rangle \nonumber \\
& \simeq
\frac{C_K}{6\pi} \epsilon^{2/3} k^2
\sum_{\vec{p}}
\left(
 1 - \frac{(\vec{k}\cdot\vec{p})^2}{k^2 p^2}
\right)
 |\vec{k}+\vec{p}|^{-11/3}  \nonumber \\
& \simeq
\frac{C_K}{6\pi} \epsilon^{2/3} k^2
\left(\frac{L}{2\pi}\right)^3 
\int_{k_0}^{k_1} \d p 
\int_0^{\pi} \d \theta 
\int_0^{2\pi} \d \varphi~ p^2 \sin^3\theta (k^2 + p^2 + 2kp \cos\theta)^{-11/6}\nonumber \\
& =
\frac{6\sqrt{3}}{35}
 \pi C_K
\epsilon^{2/3}
 k^{4/3}
 + \mbox{(subleading terms)},
\end{align}
where the integral can be calculated analytically.

Finally, the Lagrangian response function at short time is expressed as
\begin{align}
 G^{(L, T)}_{jj}(\vec{k}, -\vec{k}, t_\ell |t_\ell + \varepsilon, t_\ell)
 \simeq
 1
-
\varepsilon^2  
\frac{2\sqrt{3}}{35}
 \pi
 C_K
\epsilon^{2/3}
 k^{4/3}
 + O(\varepsilon^3).
 \label{leexp}
\end{align}
Therefore the time scale of the Lagrangian response function at short time 
is given by the Kolmogorov temporal scaling $\epsilon^{-1/3}k^{-2/3}$ as expected.
The result (\ref{leexp}) can be compared with the result of the LRA in \citet{kig},
$G^{(L, T=0)} = 1 - 0.530 C_K \epsilon^{2/3}k^{4/3} \varepsilon^2  + O(\varepsilon^3)$.
The numerical constant in (\ref{leexp}), namely $(2\sqrt{3}/35) \pi \simeq 0.311$, is 
smaller than that of the LRA result.

Next, let us consider the Lagrangian response function with the ordering $t_\ell \ge t_m$,
which is the same ordering considered in our numerical study in section \ref{s:l}.
Specifically, setting $t_\ell = t_m + \varepsilon ~(\varepsilon \ge 0)$, 
the short-time expansion is
\begin{eqnarray}
 G^{(L, T)}_{jj}(\vec{k}, -\vec{k}, t_m + \varepsilon |t_m + \varepsilon, t_m)
 &=& \frac{1}{4 \tilde{\sigma}^2(k)T}
 \langle \hat{v}_j(\vec{k}, t_m + \varepsilon|t_m + \varepsilon) \hat{\zeta}_j(-\vec{k}, t_m + \varepsilon|t_m) \rangle \nonumber \\
&=& 
\frac{1}{4\tilde{\sigma}^2(k)T}
 \langle \hat{u}_j(\vec{k}, t_m + \varepsilon) \hat{\zeta}_j(-\vec{k}, t_m + \varepsilon|t_m) \rangle
\nonumber \\
 &=& \tilde{b}_0 + \tilde{b}_1 \varepsilon + \tilde{b}_2 \varepsilon^2 + \ldots.
\label{gltexpand}
\end{eqnarray}
The factor $4$ in the denominator is due to the incompressibility as we will see.
We wish to write the coefficients $\tilde{b}_0, \tilde{b}_1$ and $\tilde{b}_2$ in terms
of the Eulerian velocity modes.

To evaluate the velocity-noise correlation, we need 
the labelling-time evolution of the random forcing, which 
is described by the passive vector equation. Hence its Duhamel solution is
\begin{align}
 \hat{\zeta}_j(\vec{k}, t_m + \varepsilon|t_m) = 
  \hat{\zeta}_j(\vec{k}, t_m |t_m)
  - \i k_a \int_{t_m}^{t_m + \varepsilon} 
  \sum_{\substack{\vec{p}, \vec{q} \\ \vec{p} + \vec{q} + \vec{k} = \vec{0}}}
  \hat{u}_a(-\vec{p}, s) \hat{\zeta}_j(-\vec{q}, s|t_m) \d s.
\label{zetaevo}
\end{align}
Similarly, we use the formal solution of the Eulerian velocity 
to the equation (\ref{nszeta}),
\begin{align}
 \hat{u}_j(\vec{k}, s)
=  \hat{v}_j(\vec{k}, s|s) 
&=  \hat{u}_j(\vec{k}, t_m) 
     + 
\int_{t_m}^{s}
ds_1
\big[
\Lambda_j(\vec{k}, s_1) \nonumber \\
&\quad
  + (2\pi)^3
  P_{ja}(\vec{k})
  \sum_{\vec{p}}
  \hat{\zeta}_a(\vec{p}, s|s_1)
  \hat{\psi}(\vec{k}, s_1|-\vec{p}, s)
\big].
 \label{duhamelrevisit2}
\end{align}
Notice that we set the labelling time appearing on the right hand side of (\ref{duhamelrevisit2})
to the same time on the left hand side. Here again we ignore the viscous term and the large-scale
forcing. We also need the formal solution of the labelling time evolution
of the position function
\begin{align}
 \hat{\psi}(\vec{k}', t_m|\vec{k}, t_m + \varepsilon)
 =
 \hat{\psi}(\vec{k}', t_m|\vec{k}, t_m )
 - \i k_a
 \int_{t_m}^{t_m + \varepsilon} ds
 \sum_{\substack{\vec{p} ,\vec{q} \\ \vec{p} + \vec{q} + \vec{k} = \vec{0}}} 
 \hat{u}_a(-\vec{p}, s)  \hat{\psi}(\vec{k}', t_m|-\vec{q}, s).
 \label{evolpsi}
\end{align}

Using (\ref{zetaevo})--(\ref{evolpsi}) successively, the velocity-noise
correlation in (\ref{gltexpand}) can be calculated as
\begin{align}
 \langle 
 \hat{v}_j(\vec{k}, t_m + \varepsilon | t_m + \varepsilon)
 \hat{\zeta}^*_j(\vec{k}, t_m + \varepsilon | t_m)
 \rangle
&=
4\sigma^2(k) T
- \sigma^2(k) T \varepsilon^2
\bigg[
   6 k_b k_d 
   \sum_{\vec{p}}
   \langle \hat{u}^*_b(\vec{p}, t_m) \hat{u}_d(\vec{p}, t_m) \rangle     \nonumber \\
&+ k_b \langle \hat{u}_j(\vec{k}), t_m) \hat{u}^*_c(\vec{k}, t_m) \rangle
       \sum_{\vec{p}} P_{bcd}(-\vec{p}) P_{dj}(\vec{k} + \vec{p}) \nonumber \\
& + P_{jab}(\vec{k})
    \sum_{\vec{p}}
    \left\{P_{bcj}(\vec{p} + \vec{k}) - 2 k_c P_{bj}(\vec{p}+ \vec{k})\right\} \nonumber \\
& \qquad \times  \langle \hat{u}^*_a(\vec{p}, t_m) \hat{u}_c(\vec{p}, t_m) \rangle \bigg] \nonumber \\
& + O(\varepsilon^3).
 \label{glt3}
\end{align}

Now we again assume that the dominant part is the diagonal components of the second-order
velocity correlations. This leads to the expansion of the response function
\begin{align}
 G^{(L, T)}_{jj}(\vec{k}, -\vec{k}, t_m + \varepsilon |t_m + \varepsilon, t_m)
\simeq
1
- \frac{\varepsilon^2}{6}
&\bigg[
   4 k^2 \langle E \rangle  \nonumber \\
& +  \frac{1}{2}\langle |\hat{\vec{u}}(\vec{k}, t_m)|^2 \rangle
      \sum_{\vec{p}} 
      \frac{\{\vec{k}\cdot(\vec{k} + \vec{p})\}
            \{\vec{p}\cdot(\vec{p} + 2 \vec{k})\}}
           {|\vec{k} + \vec{p}|^2} \nonumber \\
& - \sum_{\vec{p}} 
   (\vec{p}\cdot\vec{k})
   \langle |\hat{\vec{u}}(\vec{p}, t_m)|^2 \rangle
\bigg] 
+ O(\varepsilon^3).
\label{gltnear}
\end{align}
The last term in the $\varepsilon^2$ coefficient vanishes. The second term
can be estimated by approximating it with the integral
\begin{align}
 \sum_{\vec{p}} 
 \frac{\{\vec{k}\cdot(\vec{k} + \vec{p})\}
       \{\vec{p}\cdot(\vec{p} + 2 \vec{k})\}}
        {|\vec{k} + \vec{p}|^2}
&\simeq
 \left(\frac{L}{2\pi}\right)^3
 \int d\vec{p}
 \frac{(k^2 + \vec{k}\cdot\vec{p}) (p^2 + 2 \vec{k}\cdot\vec{p})}
      {|\vec{k} + \vec{p}|^2} \nonumber \nonumber \\
& = 2\pi k 
\int_{k_0}^{k_1} dp~ p^3
\int_{0}^{\pi} d\theta~\sin\theta
 \frac{(k + p \cos\theta) (p + 2 k \cos\theta)}
      {k^2 + p^2 + 2kp \cos\theta} \nonumber \\
& = \frac{2}{3}\pi k^2 k_1^3 + 2\pi k^3 k_1 + \mbox{(subleading terms)}.
\end{align}
Here we have dependence on the high-wavenumber cut-off $k_1 ~(\gg k)$.
Therefore, the linear response function is written as
\begin{eqnarray}
 G^{(L, T)}_{jj}(\vec{k}, -\vec{k}, t_m + \varepsilon |t_m + \varepsilon, t_m)
 &\sim&
 1
 -\frac{\varepsilon^2}{6}
 \left[
  4 k^2 \langle E \rangle 
 + \frac{1}{6} C_K \epsilon^{2/3}k^{-5/3} k_1^3
 + \ldots
 \right]
  + O(\varepsilon^3). \nonumber \\
\end{eqnarray}

In this case, the dominant time scale of the linear response function 
is $\epsilon^{-1/3} k^{5/6} k_1^{-3/2} = \epsilon^{-1/3}k_1^{-2/3} (k/k_1)^{5/6}$, 
which is an unexpected result.
So far, we do not have a clear interpretation of this time scale.
Nevertheless, this manifests non-locality, which can be ascribed to the position function.
In fact, the $k_1$ dependence is coming from the correlation between the velocity and 
the advection term in (\ref{zetaevo}).
Although the wavenumber $k$ which we are now probing is much less than $k_1$,
the time scale determined by $k_1$ (the highest active wavenumber of the velocity)
is a reminiscent of the viscous-convective-range picture of the passive scalar 
transport at high Schmidt numbers, see, e.g., \citet{Davidson}.  

Here we notice that the scaling behaviour obtained above cannot be compared 
with our DNS result described in section \ref{s:l}, since 
our simulation did not have sufficient scale separation, $k_0 \ll k \ll k_1$, between the beginning and the ending wavenumbers of 
the inertial range.
Nevertheless, we comment on 
the short-time behaviour of the Lagrangian response function observed in DNS shown in section \ref{s:l}.
As shown in figure \ref{lcg}, the Lagrangian response functions for large wavenumbers ($k_\eta/16$ and $k_\eta/32$)
at short times are so flat that the parabolic decrease does not fit well. If we plot $1 - G^{L}(k, t_\ell - t_m)$
as a function of $t_\ell - t_m$ in the log-log coordinates,
the short-time part is almost flat for $k_\eta/32$ and is close to $(t_\ell - t_m)^{0.50}$ for $k_\eta/16$.
We speculate that these apparent scaling behaviours are caused by the large-scale forcing (\ref{nsf}) and
by some competition between the $\varepsilon$ and $\varepsilon^2$ terms of the short-time Taylor expansion.

\section{\label{s:pdf}Probability density functions of Eulerian and Lagrangian velocity Fourier modes}

In this section with our numerical data, we show that probability density functions (PDFs) of
the velocity Fourier modes are self-similar and close to Gaussian. 
In particular, this holds not only for the Eulerian velocity, but also for
the Lagrangian history velocity. For the Eulerian velocity, we consider both cases
with and without the random noise $\vec{\xi}(\vec{k}, t)$ in (\ref{nsr}). The
setting of the noise is the same as in section \ref{s:NUMFRRE}, namely $\sigma(k)= k^{-1}$
and $T = 10^{-6}$.
For the Lagrangian velocity, we consider only the case without the noise. 

\begin{figure}
\centerline{\includegraphics[scale=0.7]{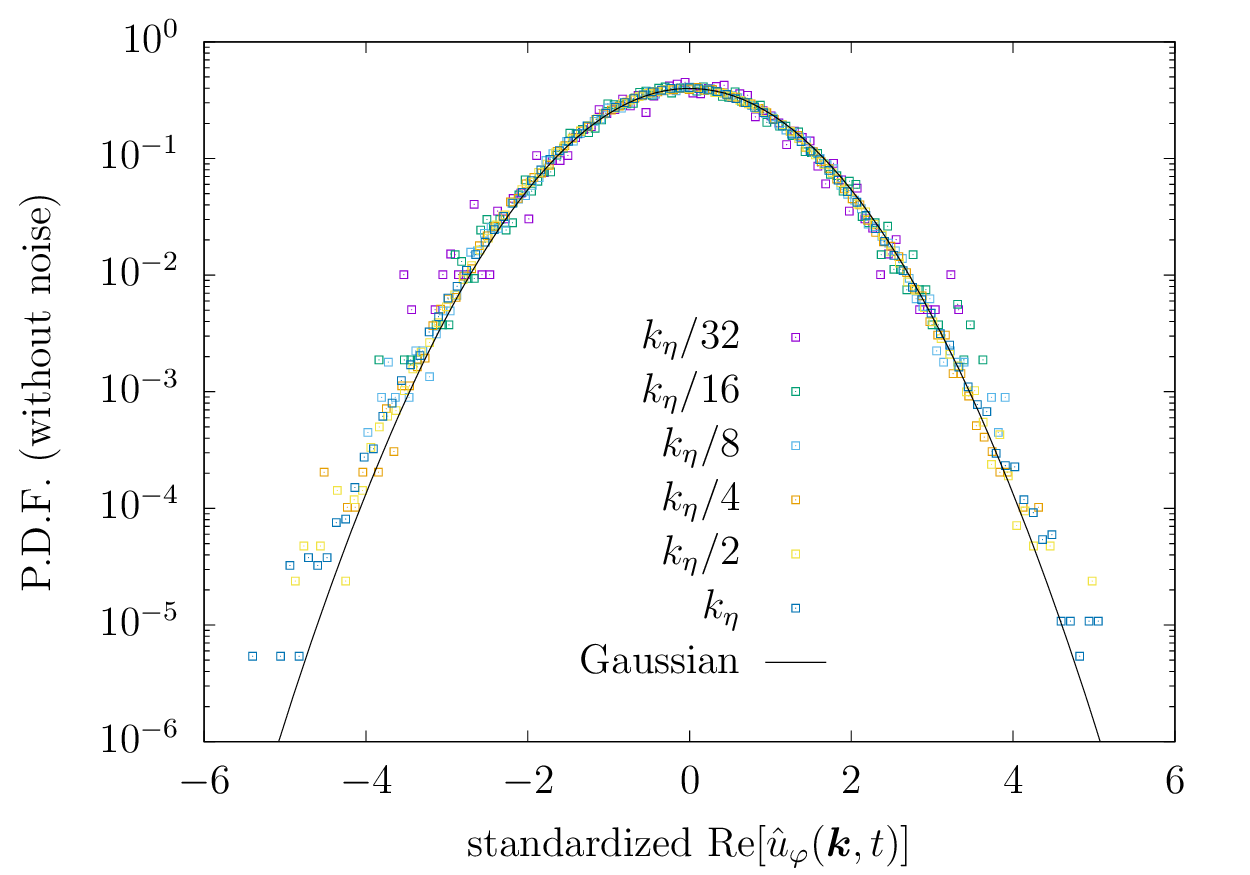}}
\centerline{\includegraphics[scale=0.7]{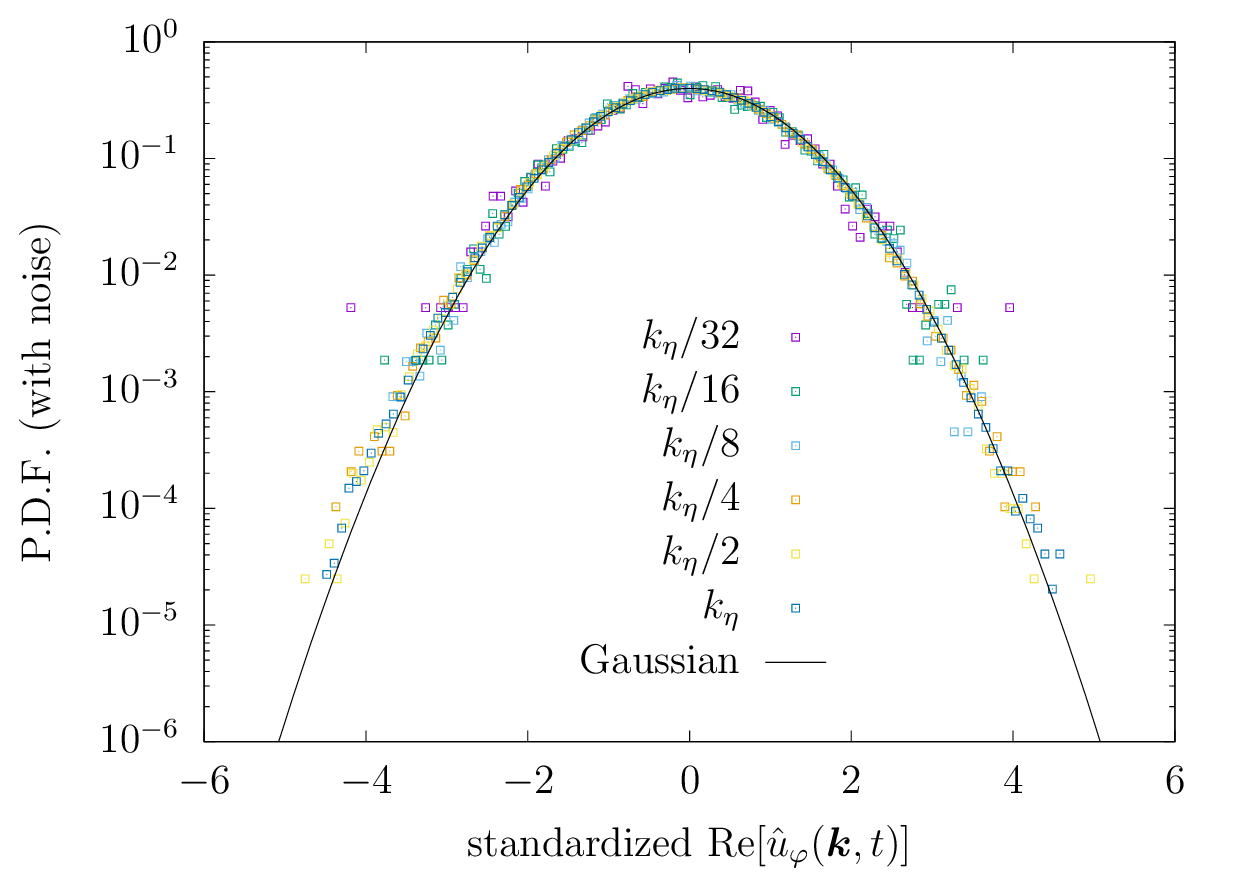}}
 \caption{\label{pdfe} Probability density functions of the real parts of the Eulerian velocity Fourier modes, ${\rm Re}[\hat{u}_\varphi(\vec{k}, t)]$, for $R_\lambda = 210 ~(k_\eta = 160)$. Top:  without the random noise, or equivalently $T = 0$. Bottom: with the random noise.}
\end{figure}

To calculate the PDF of the Eulerian velocity modes, we consider a shell in the wavenumber
space, $k \le |\vec{k}| < k + \Delta k$ with $\Delta k = 1$, as we do in calculating
the energy spectrum. Within this shell characterised with $k$, 
we calculate the PDF of the real part of the $\varphi$-component of the Fourier mode, 
${\rm Re}[\hat{u}_\varphi(\vec{k}, t)]$. We take 10 snapshots in the statistically steady state,
starting from different initial conditions.
The real parts are then standardised to have zero mean and unit standard deviation. 
The resultant PDFs shown in figure \ref{pdfe} for five different wavenumber shells indicate
that they are self-similar and close to Gaussian. This is in stark contrast to behaviour of PDFs for 
the Eulerian velocity increments in the physical space, which are not self-similar nor Gaussian, 
see, e.g., \citet{UF}.

To estimate quantitatively how close they are to Gaussian, we calculate the skewness and kurtosis, which 
are defined respectively as the 3rd and 4th moments of the standardised variables.
The results listed in table \ref{tab:ske} are indeed around those of the Gaussian 
distribution, although the smallest and largest wavenumber cases in table \ref{tab:ske} have
somewhat larger kurtosis. 
Similar results are obtained for the imaginary parts of the $\varphi$-components and both parts of the $\theta$-components
of the Eulerian velocity modes (figure not shown).
\begin{table}
  \begin{center}
\def~{\hphantom{0}}
  \begin{tabular}{ccccc}                     
      shell wavenumber & \multicolumn{2}{c}{skewness}                 &  \multicolumn{2}{c}{kurtosis}  \\
                       &  (without noise)         &   (with noise)           & (without noise)    & (with noise) \\ 
       $k_\eta / 32$ & $-1.09 \times 10^{-1}$   &   $-5.70\times10^{-2}$ & $3.19$ & $3.16$\\
       $k_\eta / 16$ & ~~$ 1.75 \times 10^{-2}$ & ~~$ 4.30\times10^{-2}$ & $3.08$ & $3.00$\\
       $k_\eta / 8$  & $-7.89 \times 10^{-3}$   & ~~$ 1.06\times10^{-2}$ & $3.05$ & $3.02$\\
       $k_\eta / 4$  & $-1.42 \times 10^{-2}$   &   $-9.03\times10^{-3}$ & $3.04$ & $3.03$\\
       $k_\eta / 2$  & ~~$ 2.50 \times 10^{-3}$ & ~~$ 3.06\times10^{-4}$ & $3.05$ & $3.05$\\
       $k_\eta$      & ~~$ 9.80 \times 10^{-4}$ &   $-3.98\times10^{-3}$ & $3.11$ & $3.05$\\
  \end{tabular}
  \caption{Skewness and kurtosis of the real part of the Eulerian velocity Fourier mode,  ${\rm Re}[\hat{u}_\varphi(\vec{k}, t)]$,
 for $R_\lambda = 210~(k_\eta = 160)$ with and without the random noise.
For Gaussian random variables, the skewness is zero and the kurtosis is $3$.}
  \label{tab:ske}
  \end{center}
\end{table}

For the Lagrangian history velocity,
we use the same method to calculate PDFs as 
in the Eulerian case. However, it should be noticed that the Lagrangian history 
velocity, $\hat{\vec{v}}(\vec{k}, t_\ell|t_m) ~(t_\ell \ge t_m)$, is not 
statistically steady  and that it is not solenoidal in general, 
$\vec{k}\cdot\hat{\vec{v}}(\vec{k}, t_\ell|t_m) \ne 0$.
Hence the decomposition of the Lagrangian mode has three components as
\begin{equation}
 \hat{\vec{v}}(\vec{k}, t_\ell|t_m)
  =  \hat{v}_\varphi(\vec{k}, t_\ell|t_m) \vec{e}_\varphi
   + \hat{v}_\theta(\vec{k}, t_\ell|t_m) \vec{e}_\theta
   + \hat{v}_c(\vec{k}, t_\ell|t_m) \hat{\vec{k}}
\end{equation}
in contrast to the Eulerian mode given in (\ref{edeco}) with two components.
Here $\hat{\vec{k}} = \vec{k}/ |\vec{k}|$ and
$\hat{v}_c$ is the compressible component which is zero at $t_\ell = t_m$.

\begin{figure}
\centerline{\includegraphics[scale=0.7]{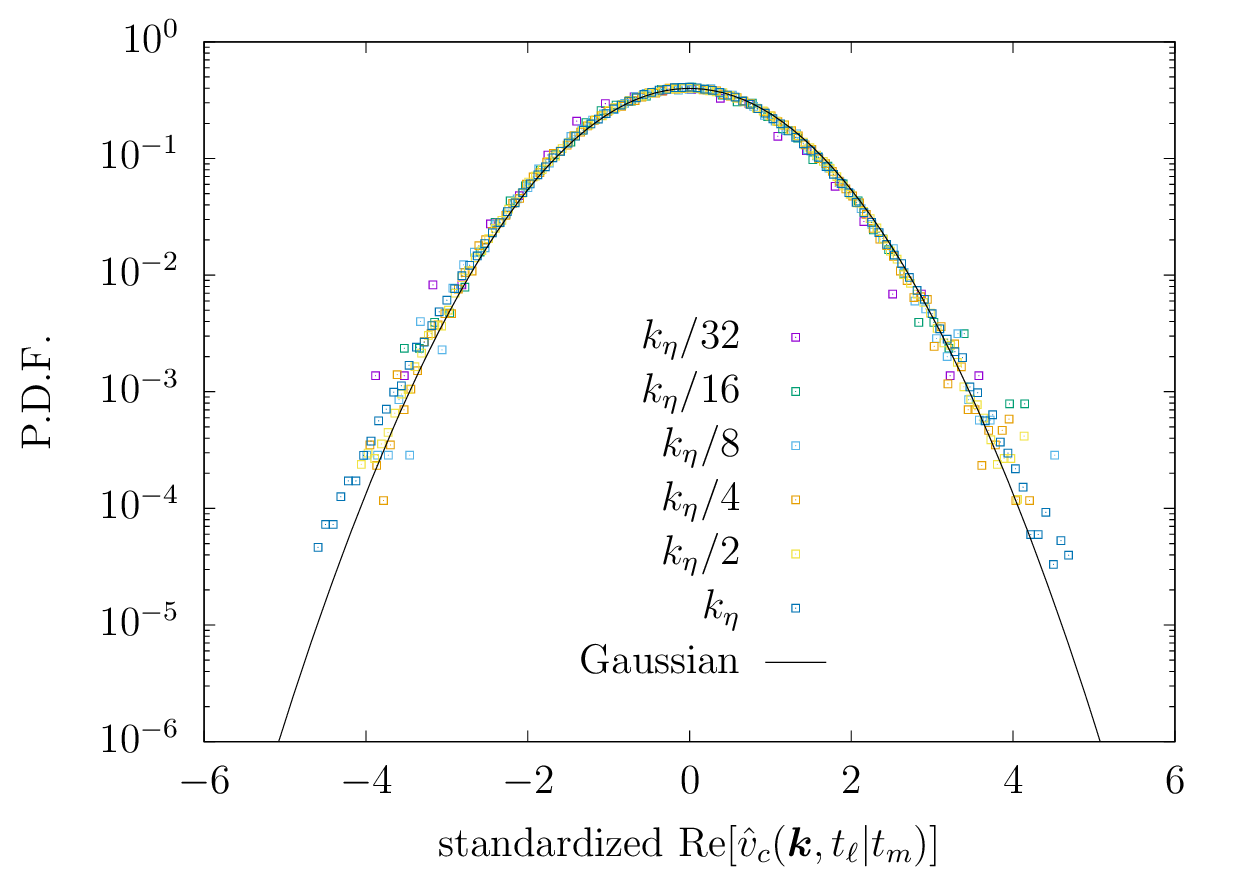}}
\centerline{\includegraphics[scale=0.7]{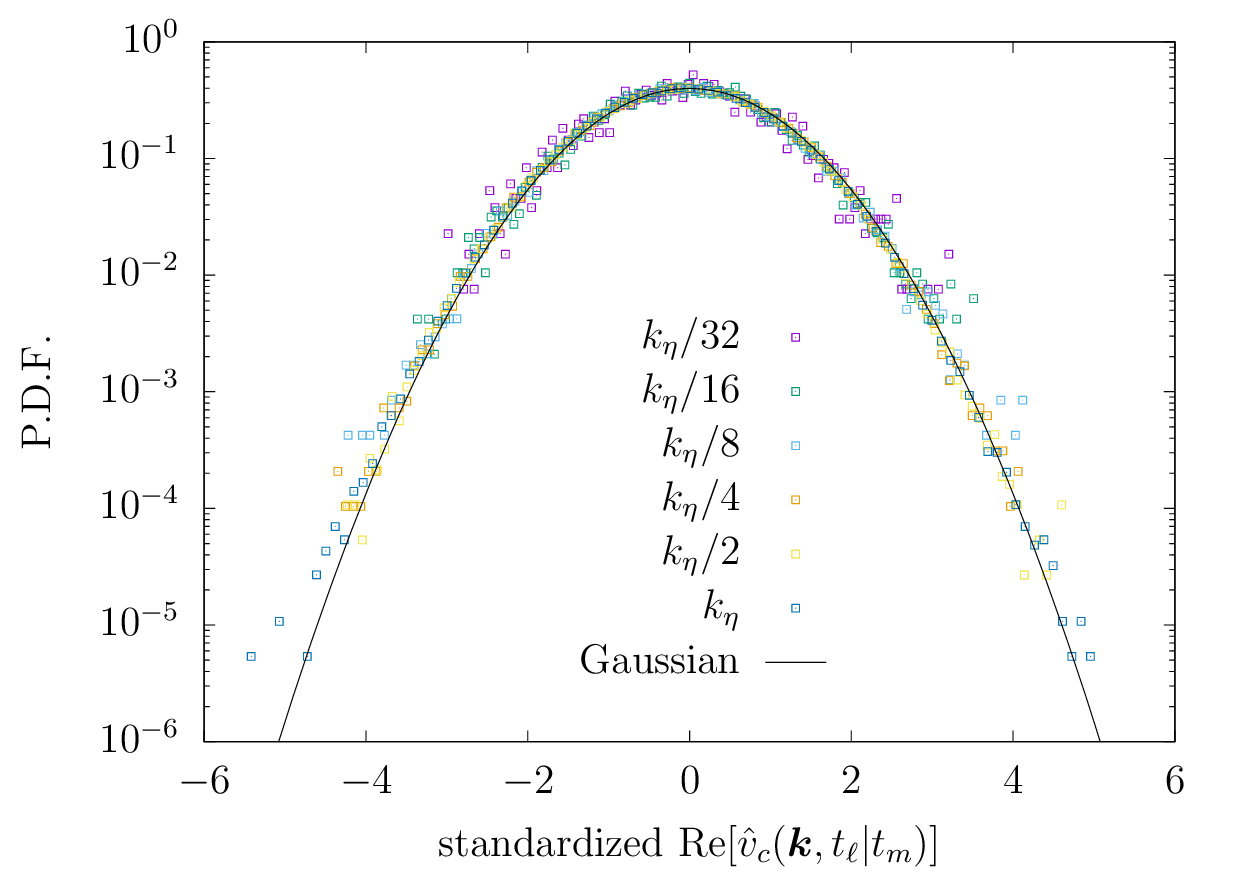}}
 \caption{\label{pdfl} Probability density functions of the real parts of the Lagrangian velocity Fourier modes, ${\rm Re}[\hat{v}_c(\vec{k}, t_\ell|t_m)]$, for $R_\lambda = 210 ~(k_\eta = 160)$. Top: $t_\ell = t_m + 0.05 \tau_{to}$. Bottom: $t_\ell = t_m + 0.50\tau_{to}$.}
\end{figure}

We calculate PDFs of the real and imaginary parts of the three components
at 11 different instances, $t_\ell = t_m + 0.05 \tau_{to}, ~t_m + 0.1\tau_{to}, ~t_m + 0.2\tau_{to}, \ldots, ~t_m + \tau_{to}$.
We use the same 10 snapshots (as we used in the Eulerian case) as the initial Lagrangian
velocity fields ($t_\ell = t_m$) for the passive vector equations (\ref{pv}).
In figure \ref{pdfl}, we plot PDFs of the real parts of the compressible components 
at two particular instances $t_\ell = t_m + 0.05\tau_{to}$ and $t_m + 0.50\tau_{to}$.
We choose this component since it may behave differently from the solenoidal Eulerian modes, 
$\hat{u}_\varphi$ and $\hat{u}_\theta$.
Although the compressible components are zero at $t_\ell = t_m$, they quickly develop
and their PDFs become  self-similar and close to Gaussian at $t_\ell = t_m + 0.05\tau_{to}$ as shown 
in figure \ref{pdfl} (how they develop to Gaussian from zero is beyond the scope of this paper).
We observe as small differences between the two times shown in figure \ref{pdfl}
that the earlier time PDFs have less developed tails and less fluctuations.
The PDFs of the imaginary parts of the compressible component behave in a similar manner.
Both real and imaginary parts of the $\varphi$- and $\theta$-components at both times are also similar 
to those at $t_\ell = t_m + 0.50\tau_{to}$ shown in figure \ref{pdfl}.
To observe how close the PDFs are to Gaussian, we list skewness and kurtosis of 
${\rm Re}[\hat{v}_c(\vec{k}, t_\ell|t_m)]$ in table \ref{tab:skl}.
Similar results are obtained for another part of the same components and both parts of the other components 
at other instances.
In conclusion, we observe that the PDFs of the Eulerian and Lagrangian history velocity Fourier modes are close 
to Gaussian in the inertial and dissipation ranges. Thus we have verified, 
for $R_\lambda = 210$, the same Gaussianity of the Eulerian velocity modes
as have been numerically found for $R_\lambda = 80$ by \cite{BrunPumir}.

\begin{table}
  \begin{center}
\def~{\hphantom{0}}
  \begin{tabular}{ccccc}                     
      shell wavenumber & \multicolumn{2}{c}{skewness}                 &  \multicolumn{2}{c}{kurtosis}  \\
                       &  ($t_\ell = t_m + 0.05\tau_{to}$)  &  ($t_m + 0.5\tau_{to}$) & ($t_m + 0.05\tau_{to}$)  & ($t_m + 0.50\tau_m$) \\ 
       $k_\eta / 32$ & ~~$3.77\times10^{-2}$  &   $-2.92\times10^{-2}$ & $3.08$ & $3.05$\\
       $k_\eta / 16$ & ~~$3.64\times10^{-2}$  &   $-7.28\times10^{-3}$ & $3.07$ & $3.13$\\
       $k_\eta / 8$  & ~~$7.60\times10^{-3}$  & ~~$2.57\times10^{-2}$  & $3.04$ & $3.07$\\
       $k_\eta / 4$  & $-1.14\times10^{-2}$   &   $-8.68\times10^{-3}$ & $3.02$ & $3.01$\\
       $k_\eta / 2$  & ~~$2.82\times10^{-3}$  &   $-1.03\times10^{-2}$ & $3.05$ & $3.02$\\
       $k_\eta$      & $-1.35\times10^{-3}$   &   $-1.85\times10^{-3}$ & $3.16$ & $3.04$\\
  \end{tabular}   
  \caption{Skewness and kurtosis of the real part of the Lagrangian velocity Fourier mode, ${\rm Re}[\hat{v}_c(\vec{k}, t_\ell|t_m)]$
   at $t_\ell = t_m + 0.05\tau_{to}$ and $t_m + 0.50\tau_{to}$ for $R_\lambda = 210~(k_\eta = 160)$.}
  \label{tab:skl}
  \end{center}
\end{table}

\bibliographystyle{jfm}
\bibliography{ns3d_fdt}

\end{document}